\documentclass[prd,aps,longbibliography,twocolumn,nofootinbib,superscriptaddress,showkeys,reprint]{revtex4-1}

\usepackage[linktocpage=true]{hyperref}
\hypersetup{colorlinks=true, linkcolor=blue, filecolor=magenta,  urlcolor=blue, citecolor=red,}

\usepackage{soul}

\usepackage{macros}
\graphicspath{{./figs/}}

\renewcommand{\EK}{E_{\rm kin}}
\newcommand{\kK}{k_{\rm kin}}
\newcommand{\KK}{K_{\rm kin}}

\newcommand{\zetaK}{\zeta_{\rm kin}}

\usepackage{ragged2e}

\begin{document}

\title{Characterization of the gravitational wave spectrum from sound waves within the sound shell model}

\author{Alberto Roper Pol}
\email{alberto.roperpol@unige.ch}
\affiliation{D\'epartement de Physique Th\'eorique, Universit\'e de Gen\`eve,
CH-1211 Geneva, Switzerland}

\author{Simona Procacci}
\email{procacci@itp.unibe.ch}
\affiliation{AEC, Institute for Theoretical Physics, University of Bern, CH-3012 Bern, Switzerland}

\author{Chiara Caprini}
\email{chiara.caprini@cern.ch}
\affiliation{D\'epartement de Physique Th\'eorique, Universit\'e de Gen\`eve,
CH-1211 Gen\`eve, Switzerland}
\affiliation{Theoretical Physics Department, CERN, CH-1211 Geneva, Switzerland}

\begin{abstract}
We compute the gravitational wave (GW) spectrum sourced by the sound waves produced during a first-order phase transition in the radiation-dominated epoch.
The correlator of the velocity field
perturbations is evaluated in accordance with the sound shell model.
In our derivation we include the effects of the expansion of the Universe, which are relevant in particular for sourcing processes whose time duration is comparable with the Hubble time.
Our results show a causal growth of the GW
spectrum at small frequencies, $\OmGW \sim k^3$, possibly followed by a linear regime $\OmGW \sim k$ at intermediate $k$, depending on the phase transition parameters.
Around the peak, we find a steep growth that approaches the
$\sim k^9$ scaling previously found within the sound shell model.
The resulting bump around the peak of the GW spectrum may represent a distinctive feature of GWs produced
from acoustic motion. Nothing similar has been observed for 
vortical (magneto)hydrodynamic turbulence.
Nevertheless, we find that the $\sim k^9$ scaling is less extended than expected in the literature, and it does not necessarily appear.
The dependence on the duration of the source, $\delta\tfin$, is quadratic at small frequencies $k$, and proportional to $\ln^2 (1 + \delta\tfin \HH_*)$ for an
expanding Universe.
At frequencies around the peak, the growth
is suppressed by a factor $\Upsilon = 1 - 1/(1 + \delta\tfin \HH_*)$
that becomes linear when the GW source is short.
We discuss in which cases the dependence on the source duration is linear or quadratic for
stationary processes.
This affects the amplitude of the GW spectrum, both in the causality
tail and at the peak, showing that the assumption of stationarity is a very relevant one, as far as the GW spectral shape is concerned.
Finally, we present a general semi-analytical template of the resulting GW spectrum,
as a function of the parameters of the phase transition.
\end{abstract}

\maketitle

\tableofcontents

\section{Introduction}
A first-order thermal phase transition can be parameterized in terms of a scalar field, whose vacuum state is degenerate at a given critical temperature $T_c$ \cite{Kirzhnits:1976ts,Coleman:1977py,Linde:1981zj}. 
According to the Standard Model (SM), both the electroweak \cite{Kajantie:1996qd} and the QCD \cite{Stephanov:2006dn} phase transitions have occurred as crossovers in
the early Universe.
However,
extensions of the SM that provide the required conditions for baryogenesis at the electroweak scale can also lead to first-order phase transitions (see Ref.~\cite{Caprini:2019egz} for a review, and references
therein).
Moreover, a large lepton asymmetry or a primordial magnetic
field may affect the QCD phase diagram, potentially
leading to a first-order QCD phase transition \cite{Schwarz:2009ii,Wygas:2018otj,Middeldorf-Wygas:2020glx,Vovchenko:2020crk,Cao:2022fow}.

We assume that, for a specific model, $T_c$ is reached while the early Universe is cooling down in the radiation-dominated era. Part of the potential energy in the unstable vacuum is then transferred to the surroundings as kinetic energy, through the nucleation and expansion
of bubbles of the broken phase \cite{Steinhardt:1981ct,Enqvist:1991xw,Ignatius:1993qn}.

The resulting shear stress of the fluid
can have 
anisotropies of the tensor type and, hence, source gravitational waves (GWs) that propagate in the homogeneous and isotropic background \cite{Witten:1984rs,Hogan:1986qda}. 
To study the power spectrum of these GWs, the shear stress from a first-order phase transition
can be decomposed into different contributions:
bubble collisions \cite{Kosowsky:1991ua,Kosowsky:1992rz,Kosowsky:1992vn,Caprini:2007xq,Huber:2008hg,Jinno:2017fby,Cutting:2018tjt}, sound waves \cite{Hindmarsh:2013xza,Hindmarsh:2015qta,Hindmarsh:2016lnk,Hindmarsh:2017gnf,Hindmarsh:2019phv,Jinno:2020eqg,Jinno:2022mie}, and turbulence \cite{Kosowsky:1992rz,Kosowsky:2001xp,Gogoberidze:2007an,Caprini:2009fx,Caprini:2009yp,Niksa:2018ofa,RoperPol:2018sap,RoperPol:2019wvy,Kahniashvili:2020jgm,Brandenburg:2021tmp,Brandenburg:2021bvg,RoperPol:2021xnd,RoperPol:2022iel,Auclair:2022jod,Sharma:2022ysf};
for
reviews see Refs.~\cite{Caprini:2018mtu,Caprini:2019egz} and references therein.

The dynamics of the expanding bubbles of the broken phase is determined
by the interaction of the plasma particles with the scalar field, which
is commonly modeled as a friction term
\cite{Ignatius:1993qn,Huber:2008hg, Espinosa:2010hh, Hindmarsh:2013xza}.
If the friction
is strong enough, we expect the
expanding bubble walls to reach a terminal velocity $\xi_w$,
which depends on the specific value of the friction term.
On the contrary, the bubbles
may run away when the friction is not sufficiently strong \cite{Bodeker:2009qy}.
However, first-order electroweak phase transitions are expected
to rarely reach this regime \cite{Bodeker:2017cim}.
If the bubbles do not run away, the long-lasting nature of the sound waves
promotes them as the dominant source of GWs.
Only if the phase transition is supercooled, it effectively
occurs in a vacuum and hence the production of sound waves
is negligible \cite{Caprini:2015zlo,vonHarling:2017yew,Kobakhidze:2017mru,Caprini:2019egz}.

The development of turbulence can occur due to the interaction
of the scalar field and the plasma \cite{Witten:1984rs,Kosowsky:1994cy}, or in the presence
of a primordial magnetic field \cite{Quashnock:1988vs,Brandenburg:1996fc},
due to the extremely
high conductivity and Reynolds number in the early Universe \cite{Ahonen:1996nq,Arnold:2000dr}.
The production of GWs from vortical turbulence
has been found to be subdominant with respect to the one from acoustic turbulence
\cite{RoperPol:2019wvy}.
However, it is not clear how much energy is converted from sound waves into
turbulence once this regime takes over, or if vortical motions can be
directly sourced from bubble collisions \cite{Cutting:2019zws}.
Moreover,
the time scales corresponding to each production mechanism are not well understood. 
This information determines the resulting GW amplitudes,
see, e.g., Refs.~\cite{Caprini:2019egz,RoperPol:2023bqa}.

In the current work, we focus on the production
of GWs from sound waves.
A semi-analytical description of the velocity spectrum originating from sound waves is provided by the sound shell model, put forward in the seminal work \cite{Hindmarsh:2016lnk}. 
The corresponding GW spectrum has been studied in detail in Ref.~\cite{Hindmarsh:2019phv} for a non-expanding Universe, and extended in Ref.~\cite{Guo:2020grp} to an expanding Universe. 
These results feature a steep growth at small frequencies, $\OmGW \sim k^9$.
The latter, however, has not been found in other numerical \cite{Hindmarsh:2013xza,Hindmarsh:2015qta,Jinno:2022mie} or
analytical \cite{Cai:2023guc} works, which are, instead, consistent with the $\sim k^3$ low-frequency tail typically expected outside the zone of both spatial and temporal correlation of the GW source \cite{Caprini:2009fx}.

The goal of this work is to generalize the results of Refs.~\cite{Hindmarsh:2019phv,Guo:2020grp} to provide
a semi-analytical template that is accurate and applicable to
the full range of frequencies of the GW spectrum.

We confirm the presence of a steep growth of the GW spectrum (cf.~Ref.~\cite{Hindmarsh:2019phv}) that, however, only appears around the peak and for certain values of the phase transition parameters. 
In particular, it depends simultaneously on the duration of the GW sourcing and the mean size of the bubbles.
The steep growth extends for a short range
of frequencies
around the peak, leading to a bump in the GW spectral shape.
At lower frequencies,
the GW power spectrum can develop an intermediate linear growth, $\OmGW \sim k$.
At even smaller frequencies,
i.e., below the inverse duration of the GW sourcing, the causal tail, $\OmGW \sim k^3$, takes over.
We also find that the bump around the GW peak is less pronounced
when one takes into account the expansion of the Universe.

With the detection of a stochastic GW background (SGWB) from the early Universe becoming conceivable in the near future, it is important to crosscheck and validate accurate theoretical templates for
the signal of the different
contributions.
The predicted spectral shape of the GW signal, in fact, strongly affects forecast observational constraints on the phase transition parameters.

The current observations by pulsar timing arrays (PTA)
have reported a SGWB at nano-Hertz
frequencies that could be compatible with sourcing anisotropic stresses produced
around the QCD scale \cite{EPTA:2023fyk,EPTA:2023xxk,NANOGrav:2023gor,NANOGrav:2023hvm,Reardon:2023gzh,Xu:2023wog}.
PTA observations have been extensively used in the literature to report constraints on the
phase transition parameters from the GW production due to sound waves \cite{Madge:2023dxc,Bringmann:2023opz,Zu:2023olm,Addazi:2023jvg,Han:2023olf,Megias:2023kiy,Li:2023bxy,DiBari:2023upq,Bai:2023cqj,Ghosh:2023aum,Figueroa:2023zhu}.
The space-based GW detector Laser Interferometer Space Antenna (LISA), planned to be launched in the early 2030s, will be sensitive to GWs with a peak sensitivity of around 1 mHz \cite{LISA:2017pwj}.
Signals produced at the electroweak phase transition are
expected to peak around these frequencies.
Several studies have used the expected sensitivity of LISA to forecast the
potential detectability of the SGWB produced by sound waves \cite{Caprini:2015zlo,Caprini:2019egz,Caprini:2019pxz,Gowling:2021gcy,Giese:2021dnw,Boileau:2022ter,Gowling:2022pzb,RoperPol:2023bqa}.
First-order phase transitions at higher energy scales, e.g., at temperatures $T > 10^8\,$GeV have been
constrained by the results of the third observing run of the LIGO--Virgo collaboration \cite{Romero:2021kby} and can be further probed
by the next generation of
ground-based GW detectors, like the Einstein Telescope
or the Cosmic Explorer.

This paper is organized as follows.
In \Sec{section1}, we provide
general formulae for the production of GWs
during
the radiation-domination era, and we introduce
the unequal-time correlator (UETC) of the anisotropic stresses originating
from sound waves.
\SSec{sw_sec} deals with the velocity field within
the framework of the sound shell model.
We provide new results regarding the
causality bounds on the velocity field and its UETC spectrum.
Being the focus of the current work on GW production, we briefly discuss
a theoretical interpretation of 
the causality argument for the initial conditions used in the sound shell model \cite{Hindmarsh:2016lnk, Hindmarsh:2019phv}, and we extend the discussion in an accompanying paper \cite{kin_sp_SSM}.

In \Sec{stationary_UETC}, we study specific features of the
GW spectrum in the sound shell model, both analytically and numerically. In particular, we discuss the occurrence of the 
$k^3$ causal tail at small frequencies. We investigate its dependence on the duration
of the source, identifying the cases in which the assumptions of Refs.~\cite{Hindmarsh:2019phv,Guo:2020grp} do not apply. 
The dependence of the GW amplitude on the duration of
the source is the topic of \Sec{linear_vs_quadratic}.
We study the GW production for stationary processes by comparing the results obtained within the sound shell model 
with those obtained for a velocity field with Gaussian (cf.~Kraichnan) decorrelation.

Numerical results for the GW spectrum are presented in \Sec{numerical_GW}. 
We show that a steep $\OmGW \sim k^7$ growth may appear below the peak under
certain circumstances, leading to a bump in the spectral shape. 
A linear growth $\OmGW \sim k$ can also develop between
the causal $\OmGW \sim k^3$ and the steep bump.
Studying the dependence of the amplitude on the duration of the source $\delta\tfin$, 
we find
that the causality tail is always quadratic in $\delta\tfin$,
while the peak may present a quadratic or a linear dependence,
with the latter being the one obtained in Refs.~\cite{Hindmarsh:2019phv,Guo:2020grp}.

We provide a template for the current-day observable $\OmGW$, as a function of the parameters that describe the phase transition. 
In \Sec{conclusions}, we discuss the implications and conclude.

In the following, the notation is such that the characteristic scales and time intervals are physical and therefore time dependent.
They are normalized
by the conformal Hubble factor
${\cal H}_* \equiv (a_*/a_0)\, H_*$, where
throughout this paper, an asterisk subscript indicates a quantity
evaluated at the initial time of GW generation, and a zero subscript indicates today's values.


\section{GW production during radiation domination}
\label{section1}

\subsection{Tensor-mode perturbations}

We consider tensor-mode perturbations $\ell_{ij}$ in an expanding Universe, described by conformal coordinates
\begin{equation}
    \dd s^2 = a^2(\tau) \left[ -\dd \tau^2 + (\delta_{ij} + \ell_{ij}) \dd x^i \dd x^j \right] \,,
\end{equation}
where $a$ is the scale factor.
The perturbations are
traceless and transverse (TT): $\ell^i_i=0$ and $\partial^i\ell_{ij}=0$.
Assuming radiation domination, the scale factor
$a$ evolves linearly with conformal time. 
We set $a(\tini)=1$
at the starting time of GW generation, such that
$a(\tau) = \HH_* \tau$, 
where $\mathcal{H}_*\equiv a'/a(\tau_*)$ is the conformal
Hubble parameter evaluated at $\tau_*$, and
a prime
denotes the derivative with respect to conformal time, $a' \equiv \partial_\tau a$.

The dynamics of small perturbations is described by the linearized Einstein equations. 
In comoving momentum space,\footnote{For a generic function $f(\xx)$, we use the Fourier convention
\begin{equation}
    f (\kk) = \int\!\!\dd^3 \xx\, f(\xx)\, e^{i \kk \cdot \xx}, \quad
    f (\xx) =  \int\!\!\frac{\dd^3 \kk}{(2\pi)^3}\, f(\kk) \, e^{-i\kk \cdot \xx} \, . \nonumber
\end{equation}
Fourier-transformed quantities are distinguished only by their argument $\kk$.} $\kk$,
the tensor-mode perturbations are governed by the GW equation
\begin{align}
    \bigl(\partial_\tau^2 +  2\mathcal{H}\partial_\tau +   k^2\bigr)&\, \ell_{ij} (\tau,\kk) \nonumber \\ &\,
    = 16 \pi G a^2 \bar\rho \, \Pi_{ij} (\tau,\kk) \,,
    \label{Einstein_eq}
\end{align}
with $G$ being the gravitational constant and $k\equiv |\kk|$.
The perturbations of the stress-energy tensor $T_{ij}$ are 
denoted by $\bar \rho \,\Pi_{ij} (\tau, \kk) \equiv 
\Lambda_{ijlm}(\hat \kk) \, T_{lm} (\tau, \kk)$, where
$\bar \rho \equiv 3\mathcal{H}^2 / (8\pi G a^2)$ is the critical
energy density,
and $\Lambda_{ijlm}$ denotes the projection onto TT components,
\begin{equation}
    2\Lambda_{ijlm}
    \equiv P_{il} P_{jm} + P_{im} P_{jl} - P_{ij} P_{lm}\,,
    \label{Lambda_ijlm}
\end{equation}
with $P_{ij} (\hat \kk) =  \delta_{ij} - \hat k_i \hat k_j$
and $\hat k_i = k_i/k$.
Rewriting \Eq{Einstein_eq} for $h_{ij}\equiv a \ell_{ij}$ during radiation domination 
yields
\begin{equation}
    \left(\partial_\tau^2 + k^2 \right)\, h_{ij} (\tau,\kk) =
    \frac{6 \, \HH_* \, \Pi_{ij} (\tau,\kk)}{\tau} \,.
    \label{GW_eq}
\end{equation}
\EEq{GW_eq} shows that the scaled strains $h_\ij$ are sourced by the
normalized and comoving TT projection of the anisotropic
stresses, $\Pi_\ij$.

While the source is active,\footnote{Since the initial time of GW production occurs within the radiation-dominated era, $\tau_* \simeq 1/\HH_*$.}
$\tini \leq \tau \leq \tfin$, the solution to \Eq{GW_eq}
with initial conditions $h_\ij(\tini, \kk) = h_\ij' (\tini, \kk) = 0$
is the convolution of the source with the Green's function,
\begin{align}
    h_{ij} (&\,\tau_* \leq \,\tau\leq\tfin^{ },\kk) \nonumber\\
    &\, = \frac{6 \HH_*}{k} \int_{\tini}^{\tau}
    \!\!\! \dd \tau_1 \,
    \frac{ \Pi_{ij} ( \tau_1^{ }, \kk) }{\tau_1^{ }} \sin k (\tau - \tau_1^{ }) \,.
    \label{sol_GW_eq}
\end{align}
At later times, $\tau > \tfin$, the solution in the free propagation regime is
\begin{align}
    h_\ij (\tau&\,>\tfin^{ },\kk) \nonumber \\ \,
    &\, = \frac{6 \HH_*}{k} \int_{\tini}^{\tfin^{ }} 
    \!\!\! \dd \tau_1^{ } \,
    \frac{ \Pi_\ij (\tau_1^{ },\kk)}{\tau_1^{ }} \sin 
    k (\tau - \tau_1^{ }) \,.
    \label{sol_hij_late_t}
\end{align}

We are interested in the fractional energy
density of GWs today
\begin{align}
    \OmGW (\tau_0^{ })
    &= \int_{-\infty}^\infty \, \OmGW (\tau_0^{ },
    k)
    \dd \ln k  \label{Ogw_def_eq} \\
    &\,\equiv \frac{1}{32\pi G \bar \rho_0 }
    \bra{\dot{\ell}_\ij (\tau_0^{ },\xx) \,
    \dot{\ell}_\ij (\tau_0^{ },\xx)} \label{Ogw_eq1}\\
    &\approx \frac{1}{12\,
    \mathcal{H}_0^2 \, a_0^2} \,
    \bra{h'_{ij} (\tau_0^{ },\xx) \,
    h'_\ij (\tau_0^{ },\xx)} \,, \label{Ogw_eq2}
\end{align}
where a dot
denotes derivatives with respect to cosmic time
$\partial_t \equiv a^{-1} \partial_\tau$,\footnote{
The exact relation from \Eq{Ogw_eq1} to \Eq{Ogw_eq2} is 
$\dot{\ell}_\ij /H=(h'_\ij - h_\ij/\tau)/(a \HH)$, where
$H\equiv \dot a/a$ and $a \HH = \HH_*$. However, terms proportional to $1/(k\tau)$ are negligible inside the horizon at
present time, $k \tau_0 \gg 1$.}
and $\OmGW (\tau_0, k)$ is the GW spectrum today.

Following the notation of Ref.~\cite{RoperPol:2018sap}, the unequal-time correlator (UETC) spectrum $S_{h'}$ of the strain derivatives that appear in \Eq{sol_hij_late_t}, is defined as
\begin{align}
    \bra{ h_{ij}' (\tau_1^{ },\kk) &\,  {h_{ij}^*}'
    (\tau_2^{ },\kk_2)} 
    \nonumber \\ \equiv &\, \bigl(2 \pi\bigr)^6 \delta^3 (\kk^{ } - \kk_2^{ })
    \frac{S_{h'} (\tau_1^{ }, \tau_2^{ }, k)}{4\pi k^2} \,,
    \label{eq_Sh} 
\end{align}
where $S_{h'}$ only depends on the wave number $k$ for a stochastic field with a homogeneous and isotropic distribution,
and on the unequal times $\tau_1$ and $\tau_2$.\footnote{
The UETC spectrum can also be expressed in terms of the power
spectral density $P_{h'} \equiv 2 \pi^2 S_{h'}/k^2$ or the 
spectrum in units of $\ln k$, ${\cal P}_{h'} \equiv k\, S_{h'}$ in analogy to \Eq{Ogw_def_eq}; see  Ref.~\cite{Hindmarsh:2019phv}.
}

Evaluating \Eq{Ogw_eq2} with \Eq{sol_hij_late_t}
at equal times $\tau_{1,2} = \tau_0 \gg \tfin$, the GW spectrum $\OmGW$
today is
\begin{align}
    \hspace{-.2cm}
    \OmGW  (\tau_0, k) =  &\, 3k
    \, {\cal T}_{\rm GW}
    \int_{\tini}^{\tfin^{ }}\! \frac{\dd \tau_1^{ }}{\tau_1^{ }}
    \int_{\tini}^{\tfin^{ }} \!\! \frac{\dd \tau_2^{ }}{\tau_2^{ }} \,
    E_\Pi^{ } (\tau_1^{ }, \tau_2^{ }, k) \nonumber \\
    &\;\times 
    \cos k(\tau_0^{ } - \tau_1^{ }) \cos k(\tau_0^{ } - \tau_2^{ }) \,. \label{OmGW_full}
\end{align}
$E_\Pi^{ }$ is 
the UETC spectrum of the anisotropic stresses,\footnote{Note that Ref.~\cite{Hindmarsh:2019phv} uses the spectral density $U_\Pi = 2 \pi^2 E_\Pi/k^2$.} defined as
\begin{align}\label{eq_Epi}
    \bra{ \Pi_{ij}^{ }& (\tau_1^{ },\kk^{ })  \,
          \Pi^*_{ij} (\tau_2^{ },\kk_2^{ })}  \vphantom{\frac{E_\Pi^{ }}{k^2}} \nonumber\\
    &\;\equiv (2 \pi)^6 \delta^3 (\kk^{ } - \kk_2^{ }) \, \frac{E_\Pi^{ }(\tau_1^{ }, \tau_2^{ }, k) }{4 \pi k^2} \,.  
\end{align}

We call {\it transfer function}, introduced in \Eq{OmGW_full}, the prefactor that describes the redshift from GW sourcing time to today,
\begin{align}
   h^2 \, {\cal T}_{\rm GW}
       \equiv &\,  \biggl( \frac{ a_*}{a_0 } \biggr)^4 \biggl(\frac{H_*}{H_0/h} \biggr)^2
   = \biggl(\frac{g_0}{g_*}\biggr)^{4\over 3} \biggl(\frac{h\, T_0^2/H_0}{T_*^2/H_*}\biggr)^2 \nonumber \\ 
     \simeq &\, 1.6 \times 10^{-5}\,  \biggl(\frac{100}{g_*}\biggr)^{1\over3} \,,
    \label{eq:evolGW}
\end{align}
where $g_*$ and $g_0 = 3.91$ are the number of entropic degrees of
freedom at $\tau_*$
(e.g., $g_* \simeq 100$ at the electroweak phase transition) and
at present time, respectively \cite{Kolb:1990vq}.
The temperature today is taken to be $T_0 = 2.725$ K \cite{Fixsen:2009ug}.
The Hubble rate at present time is given in terms of 
$h = H_0/(100 \, {\rm km}/{\rm s}/{\rm Mpc})$,
while its value during the radiation-dominated era, $H_*$, used in \Eq{eq:evolGW}, is
$ H_*^2 = 4 \pi^3 G g_* T_*^4 / (45 \, \hbar^3)$ \cite{Kolb:1990vq}.

The product of the Green's functions in \Eq{OmGW_full} can be expressed as
\begin{align}
    2\cos k  ( & \tau_0^{ } - \tau_1^{ }) \cos k (\tau_0^{ } - \tau_2^{ }) 
    \nonumber \\ = & \cos  k(\tau^{ }_1 - \tau^{ }_2) + \cos 2k\tau_0^{ }   \cos k (\tau^{ }_1 + \tau^{ }_2) \nonumber \\
    &
    + \sin 2k\tau^{ }_0 \sin k(\tau^{ }_1 + \tau^{ }_2) \label{eq_coss} \,.
\end{align}
An average over highly oscillating modes 
$k \tau^{ }_0 \gg 1$, yields
\begin{align}\nonumber
    \OmGW (k) \approx  \frac{3 k}{2}  
    {\cal T}_{\rm GW}
    \int_{\tini^{ }}^{\tfin^{ }} \!\! \frac{\dd \tau^{ }_1}{\tau^{ }_1} & 
    \int_{\tini^{ }}^{\tfin^{ }} \!\! \frac{\dd \tau^{ }_2}{\tau^{ }_2}    
    E_\Pi^{ } (\tau^{ }_1, \tau^{ }_2, k) \\
    &\times \cos k (\tau^{ }_1 - \tau^{ }_2) \,. \label{eq_OmGW_UETC}
\end{align}
Note that the approximation in \Eq{eq_OmGW_UETC} is not valid if one is interested in computing the GW spectrum while the source is active. For this case, we provide a formula for the full time dependency of $\OmGW$ in \App{app_GW_tdep}.

\subsection{GWs sourced by sound waves}

The stress-energy tensor $\Pi_{ij} \equiv \Lambda_{ijlm} T_{lm}$ that sources GWs [see \Eq{GW_eq}] can contain
contributions from the fluid (depending on the enthalpy $w$, the pressure $p$, and on $u^i \equiv \gamma v^i$, where
$\gamma$ is the Lorentz factor and $v^i$ the velocity), and
from gradients of the scalar field, $\phi$, among other
possible contributions (e.g., gauge fields),
\begin{align}
    T_{ij} \supset 
      w \,  u_i \, u_j  
    + p \,\delta_{ij} 
    + \partial_i \phi \, \partial_j \phi 
    - \frac{1}{2} \bigl(\partial \phi \bigr)^2 \delta_{ij} \,, \label{eq_Tij}
\end{align}
where $w = p + \rho$, being $\rho$ the energy density.

In the current work, we focus on the GWs sourced by sound waves
in the aftermath of a first-order phase transition.
Hence, we only consider the GW production from the linearized fluid motion (omitting the potential development of turbulence), and
neglect the contributions from bubble collisions, as well as
the possible presence of electromagnetic
fields that would alternatively affect the fluid dynamics and
also source GWs \cite{Witten:1984rs,Quashnock:1988vs}.

Since diagonal terms in \Eq{eq_Tij} are ruled out by the TT projection, the contributing part of the energy-momentum tensor is
the convolution of the velocity field in Fourier space
\begin{align}
T_{ij} (\tau,\kk)
\supset
\bar w
\int \!\!\frac{\dd^3 \pp}{(2\pi)^3} \, u_i(\tau,\pp) \, u_j (\tau, \tilde \pp) \,,
\label{Tij_ui_uj}
\end{align}
where we have denoted $\tilde \pp \equiv \kk - \pp$.
The velocity field from sound waves corresponds to perturbations over a background at rest with mean
enthalpy $\bar w$.
Hence, fluctuations in the enthalpy field correspond
to higher-order terms in the perturbative expansion and
can be neglected at first order.
In the linear regime, we also have $\gamma \sim 1$.

If we assume that the stochastic velocity field is Gaussian,
Isserlis' (or Wick's) theorem \cite{Isserlis:1916} allows us to express the four-point correlations
as linear superposition of the product of two-point functions,
\begin{align}
    &\bra{ T_{ij}^{ } (\tau_1^{ },\kk) \, T_{lm}^* (\tau_2^{ },\kk) }
    \supset
    \bar w^2
    \int \!\! \frac{\dd^3 \pp_1}{(2\pi)^3} 
    \int \!\! \frac{\dd^3 \pp_2}{(2\pi)^3} \label{eq_EPi1}\\
    &\;\times \Bigl[ \,
    \bra{ u_i (\tau_1^{ },        \pp_1^{ }) \, u_l^* (\tau^{ }_2,        \pp_2^{ }) } \,
    \bra{ u_j (\tau^{ }_1, \tilde \pp_1^{ }) \, u_m^* (\tau^{ }_2, \tilde \pp_2^{ }) } \nonumber\\
    &\hphantom{\times} + 
    \bra{ u_i (\tau^{ }_1,        \pp_1^{ }) \, u_m^* (\tau^{ }_2, \tilde \pp_2^{ }) } \,
    \bra{ u_j (\tau^{ }_1, \tilde \pp_1^{ }) \, u_l^* (\tau^{ }_2,        \pp_2^{ }) } \, \Bigr] \,. \nonumber
\end{align}

In general, the spectrum of any statistically homogeneous and isotropic field 
can be decomposed in a spectrum proportional to the projector $P_{ij}$, given below \Eq{Lambda_ijlm}, and a
spectral function proportional to $\hat k_i \hat k_j$ \cite{MY75}.
In the particular case of irrotational fields (as it is the case for sound waves), the contribution
proportional to $P_{ij}$ is zero,
and the two-point correlation function of the velocity field is\footnote{Note that Ref.~\cite{Hindmarsh:2019phv} uses the spectral density $G = 4 \pi^2 \EK/k^2$.
We add an extra factor of $2$ in \Eq{eq_EK} such that the kinetic energy density is
$\half \bra{\uu^2 (\xx)} = \int \EK (k) \dd k$.}
\begin{align}
    \bra{ u_i (\tau^{ }_1,  &\, \kk^{ }) \, {u_{j}^*} (\tau^{ }_2,\kk_2^{ })} \nonumber \\
         = &\, (2 \pi)^6 \,
    \hat k_i \hat k_j \, \delta^3 (\kk - \kk_2^{ })
    \frac{2 \EK^{ } (\tau^{ }_1, \tau^{ }_2, k)}
     {4\pi k^2} \,. \label{eq_EK}
\end{align}
The assumption of the velocity field being irrotational
is motivated by the results of numerical simulations \cite{Hindmarsh:2013xza,Hindmarsh:2015qta,Hindmarsh:2017gnf}.

In a semi-analytical approach,
the sound shell model describes the
velocity field as the linear superposition of the single-bubble contributions until the moment of collision \cite{Hindmarsh:2016lnk,Hindmarsh:2019phv}, based on the
hydrodynamics of expanding bubbles \cite{Espinosa:2010hh}.
At later times, the velocity field is assumed to be described
by the superposition of sound waves.
Hence, the resulting velocity field is irrotational and is described by the tensor structure of \Eq{eq_EK}.

Using \Eq{eq_EK}, the TT projection of the stress tensor in 
\Eq{eq_EPi1}
acts as
\begin{align}
    \Lambda_{ijlm} (\hat \kk) &\,
\hat{p}^i \hat{\tilde p}^j \hat{p}^l \hat{\tilde p}^m =
\frac{p^2}{\tilde{p}^2}\frac{(1-z^2)^2}{2} \,,
\end{align}
where $z = \hat \kk \cdot \hat \pp$.
The UETC spectrum of the anisotropic stresses $E_\Pi^{ }$,
which sources the GW spectrum in \Eq{eq_OmGW_UETC},
becomes
\begin{align}
    E_\Pi (\tau^{ }_1,  \tau^{ }_2, k)
    = &\, 2\, k^2 \bar w^2  
    \int_{-1}^1 \!\! \dd z  \int_0^\infty \!\! \dd p \,
    \frac{p^2}{\tilde p^4} 
    (1-z^2)^2  \nonumber\\
    \times &\,
    \EK  (\tau^{ }_1, \tau^{ }_2,p) \,
    \EK (\tau^{ }_1, \tau^{ }_2,\tilde p) \,.
    \label{E_Pi}
\end{align}
Hence, under the assumption of Gaussianity of the velocity field, the UETC of the anisotropic stresses $E_\Pi$ is reduced to
a quadratic function of the UETC of the velocity field $\EK$, integrated over $p$ and $z$.

A useful alternative form of \Eq{E_Pi} is found by changing
the integration variable from $z$ to $\tilde p$ with
\begin{equation}
    \tilde p^2 \equiv |\kk - \pp|^2 = p^2 + k^2 - 2 pk z \,,
\end{equation}
yielding
\begin{align}
    E_\Pi (\tau^{ }_1, \tau^{ }_2, k) &= 
     2\, k \, \bar w^2  
    \int_0^\infty \!\!\dd p\,p\, \EK (\tau^{ }_1,\tau^{ }_2,p)  \nonumber \\ 
    \times \int_{|k - p|}^{k + p} &\, \!\!\dd \tilde p \, 
     \frac{\EK(\tau^{ }_1, \tau^{ }_2,\tilde p)}{\tilde p^3} \,
     \bigl[1 - z^2(\tilde p) \bigr]^2 \,. \label{eq_EPi}
 \end{align}
This expression is used in Ref.~\cite{Hindmarsh:2019phv} and
we use it in \App{HH19} for a comparison with their results.

In \Sec{sw_sec}, we present the computation of the UETC of the velocity field for the sound
waves produced upon collision of broken-phase bubbles, following
the sound shell model.
A detailed derivation, and theoretical aspects of the
velocity UETC are presented in an accompanying paper \cite{kin_sp_SSM}.

\section{Sound waves from first-order phase transitions in the sound shell model}
\label{sw_sec}

\subsection{Velocity field}

In a first-order phase transition, the hydrodynamic equations of the fluid around the expanding bubbles of the broken phase can be derived
imposing the conservation of energy and momentum, $\partial_\mu T^{\mu \nu}=0$, and assuming
radial symmetry around the center of bubble nucleation
\cite{Espinosa:2010hh,Hindmarsh:2019phv}.
Once the broken-phase bubbles collide, it can be assumed that
the Higgs field has reached its true vacuum state and the
fluid perturbations follow a linear hydrodynamical
description without any forcing term, leading to the development
of compressional sound waves, according to the sound shell model \cite{Hindmarsh:2016lnk,Hindmarsh:2019phv}.
Defining the energy density fluctuations $\lambda \equiv 
(\rho - \bar \rho)/ \bar w$, the linearization of the fluid equations
leads to wave equations for $\uu$ and $\lambda$,
\begin{align}
    \lambda' (\tau,\kk) - &\, i k_i \, u_i (\tau,\kk) = 0 \,, \label{eq_hydro_e} \\
    u_i' (\tau,\kk) - &\,  i k_i \, \cs^2 \, \lambda (\tau,\kk) = 0 \,. \label{eq_hydro_p}
\end{align}
The equation of state  $c_s^2 \equiv \dd \bar p / \dd \bar \rho $ relates the background fluid
pressure $\bar p$ and energy density $\bar \rho$. 
The solution is a longitudinal velocity field, $u_i = \hat k_i u$,
\begin{align}
    u (\tau,\kk) =
    \sum_{s=\pm} A_s (\kk)\,
    e^{i s \omega (\tau - \tini)},
    \label{vel_sw}
\end{align}
where the dispersion relation is $\omega = c_s k$.
The coefficients $A_\pm$ depend on the velocity and energy density fields at the
time of collisions
\cite{Hindmarsh:2019phv,kin_sp_SSM},
\begin{equation}
    A_\pm (\kk) = \frac{1}{2} \Bigl[u (\tini, \kk) \pm \cs \lambda(\tini, \kk) \Bigr].
    \label{init_cond}
\end{equation}
Alternatively, as initial conditions, we could use the velocity $u$ and acceleration $u'$
fields, as done in Ref.~\cite{Hindmarsh:2013xza}.
Reference~\cite{Hindmarsh:2019phv} suggests the use of 
$\lambda$ in \Eq{init_cond} to respect the causality condition
of irrotational fields when $k \to 0$ \cite{MY75,Caprini:2003vc}.
We show in an accompanying paper that the causal limit does not depend on this choice, however the latter 
is required
to avoid discontinuities on $u$ and $\lambda$ at
$\tini$ \cite{kin_sp_SSM}.

According to the sound shell model, 
the velocity and energy density fields are the linear superposition of the fields produced by the expansion of each 
of the $N_b$ single bubbles \cite{Hindmarsh:2013xza,Hindmarsh:2019phv},
\begin{equation}
    A_\pm (\kk) = \sum_{n = 1}^{N_b} {\cal A}_\pm (\chi) \, T_n^3 \,
    e^{i \kk \cdot \xx_0^{(n)}},
    \label{coeff_A}
\end{equation}
where, for the $n$-th bubble, $T_n = \tini - \tau_0^{(n)}$ is its lifetime, $\tau_0^{(n)}$ is its time
of nucleation, and $\xx_0^{(n)}$ is its nucleation location.
The functions ${\cal A}_\pm (\chi)$, where $\chi \equiv k \, T_n$, are
\begin{equation}
    {\cal A}_\pm (\chi) = - \frac{i}{2} \bigl[f'(\chi) \pm i \cs l(\chi) \bigr],
    \label{coeff_AA}
\end{equation}
being
$f(\chi)$ and $l(\chi)$ integrals of the single-bubble radial profiles $v_\ip (\xi)$ and $\lambda_\ip (\xi)$
over a normalized radial coordinate $\xi$,
\begin{align}
    f(\chi) &\, =
    \frac{4\pi}{\chi} \int_0^\infty \dd \xi \, v_\ip (\xi) \, \sin (\chi\xi)  \ , \label{eq_f} \\ 
     l(\chi) &\, = 
    \frac{4\pi}{\chi} \int_0^\infty \dd \xi \, \xi \, \lambda_\ip (\xi) \, \sin (\chi\xi)  \ . \label{eq_l}
\end{align}
We follow Refs.~\cite{Espinosa:2010hh,Hindmarsh:2019phv} to
compute the single-bubble profiles, and present the detailed
calculation in an accompanying paper \cite{kin_sp_SSM}.

\subsection{UETC of the velocity field}
\label{UETC_velfield}

The UETC of the velocity field in \Eq{eq_EK} can be computed
from the resulting velocity field given in \Eq{vel_sw},
\begin{align}
    \EK (\tau_1, \tau_2, k) = &\, \EK^{(1)}(k) \cos \omega (\tau_1 - \tau_2) \nonumber \\ + &\, \EK^{(2)} (k)
    \cos \omega (\tau_1 + \tau_2 - 2 \tini)
    \nonumber \\ + &\, \EK^{(3)} (k)
    \sin \omega (\tau_1 + \tau_2 - 2 \tini),
    \label{eq_EK_Eu}
\end{align}
whose coefficients $\EK^{(n)}(k)$
are given as \cite{Hindmarsh:2019phv,kin_sp_SSM}
\begin{align}
    \EK^{(n)} &\, (k) = \nonumber \\ &\, \frac{k^2}{2 \pi^2 \beta^6 R_*^3} \int_0^\infty
    \!\!\dd \tilde T \, \nu(\tilde T)\, \tilde T^6  \, {\cal E}^{(n)} (\tilde T k/\beta) \,,
    \label{EK_sw_HH}
\end{align}
where $\beta$ denotes the inverse duration of the phase transition
and $\tilde T \equiv T \beta$ is the normalized bubble lifetime.
The mean bubble separation, $R_* \equiv (8\pi)^{1/3} \xi_w/\beta$ \cite{Enqvist:1991xw},
corresponds to the characteristic length scale of the fluid motion.
The distribution of the bubbles' lifetime, $\nu(\tilde T)$,
is considered in Ref.~\cite{Hindmarsh:2019phv} for
the scenarios of exponential and simultaneous nucleation,
\begin{equation}
    \nu_{\rm exp} (\tilde T) = e^{-\tilde T}, \quad 
    \nu_{\rm sim} (\tilde T) = \half \tilde T^2 e^{-{1\over 6}\tilde T^3} \,.
\end{equation}
\begin{figure*}
    \centering
    \includegraphics[width=.32\textwidth]{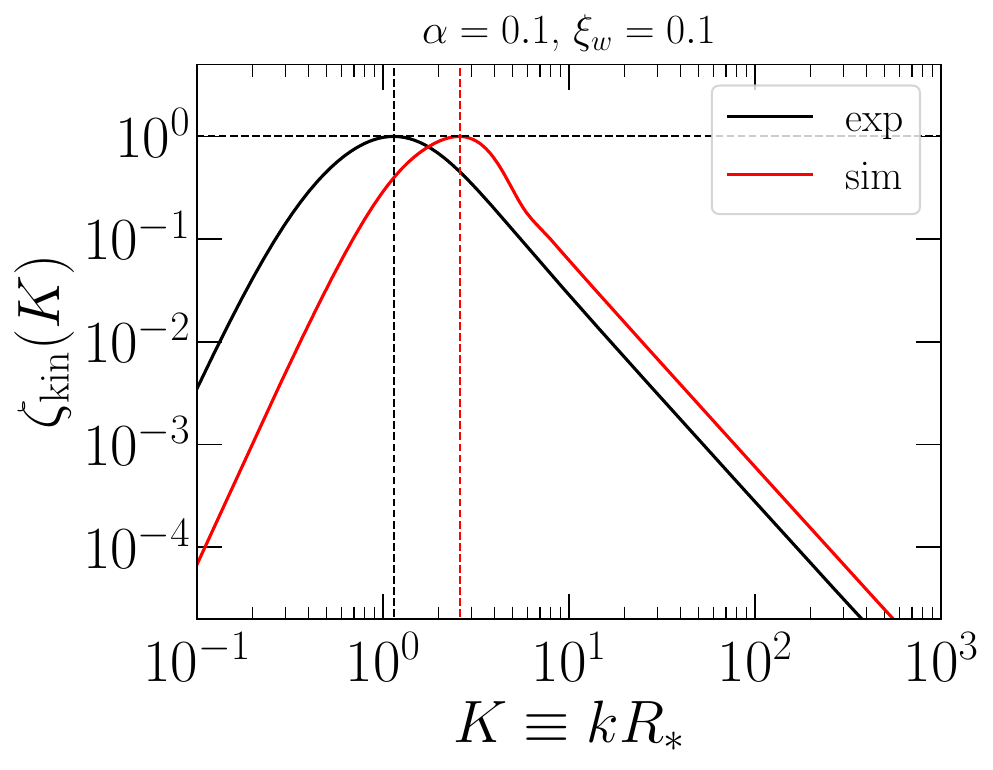}
    \includegraphics[width=.32\textwidth]{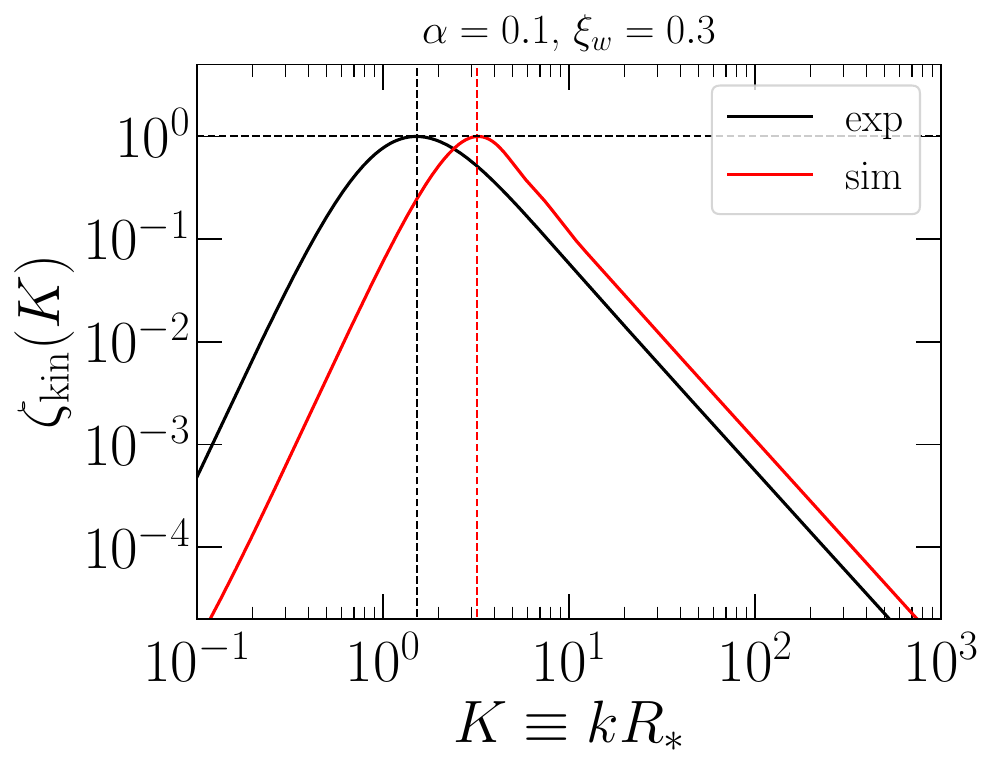}
    \includegraphics[width=.32\textwidth]{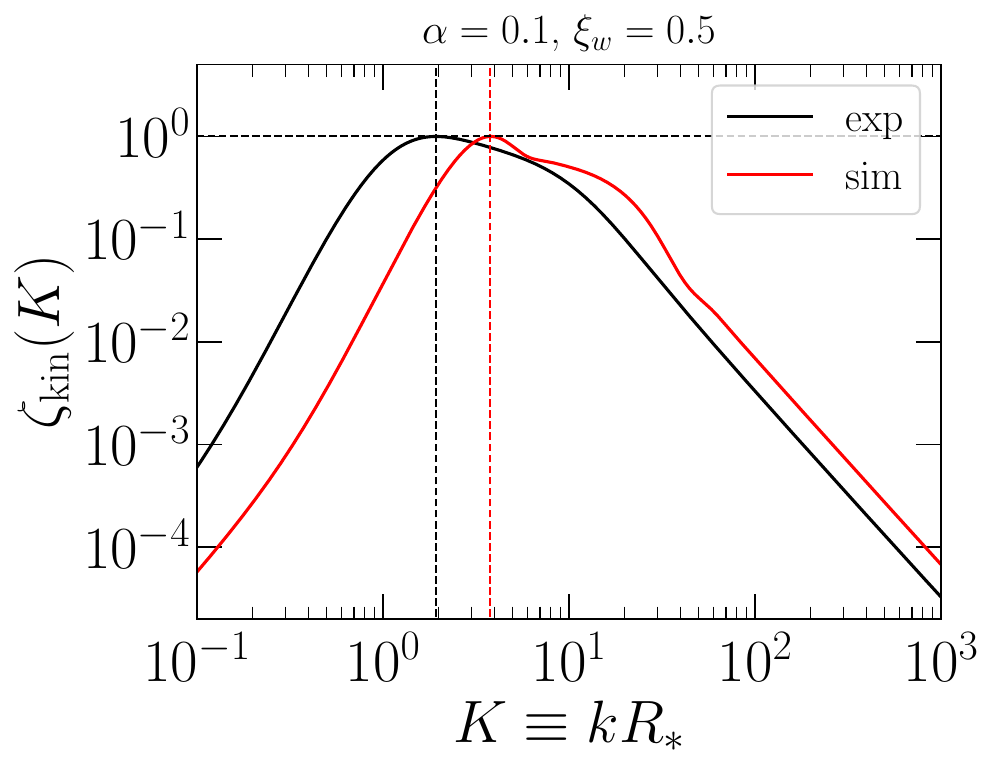}
    \includegraphics[width=.32\textwidth]{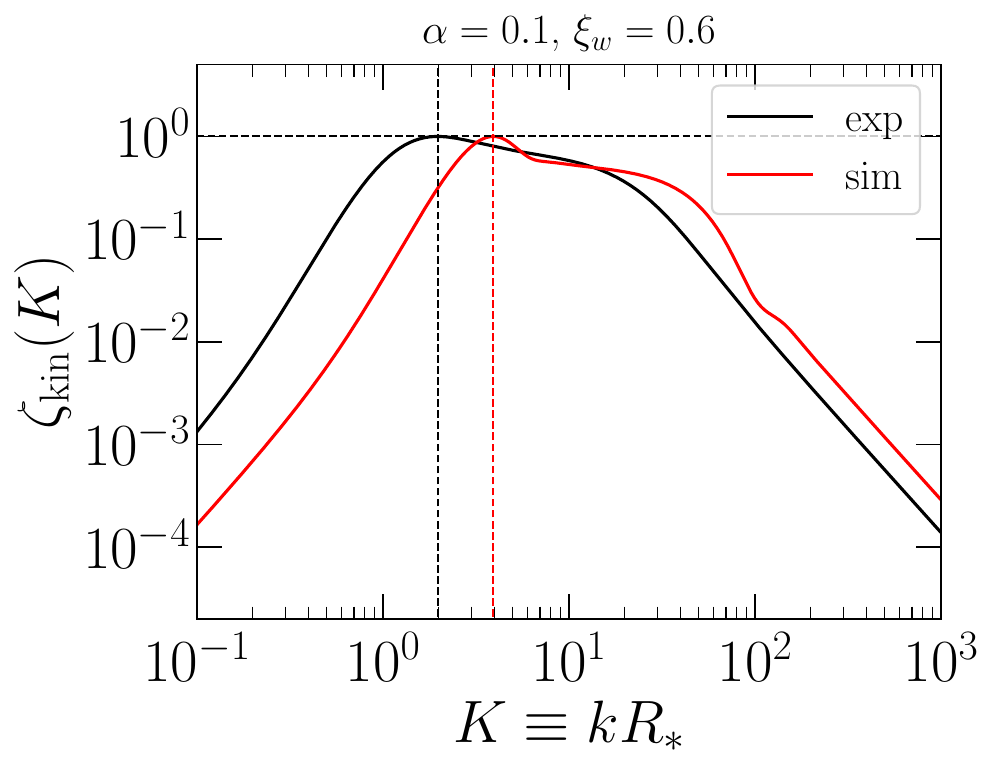}
    \includegraphics[width=.32\textwidth]{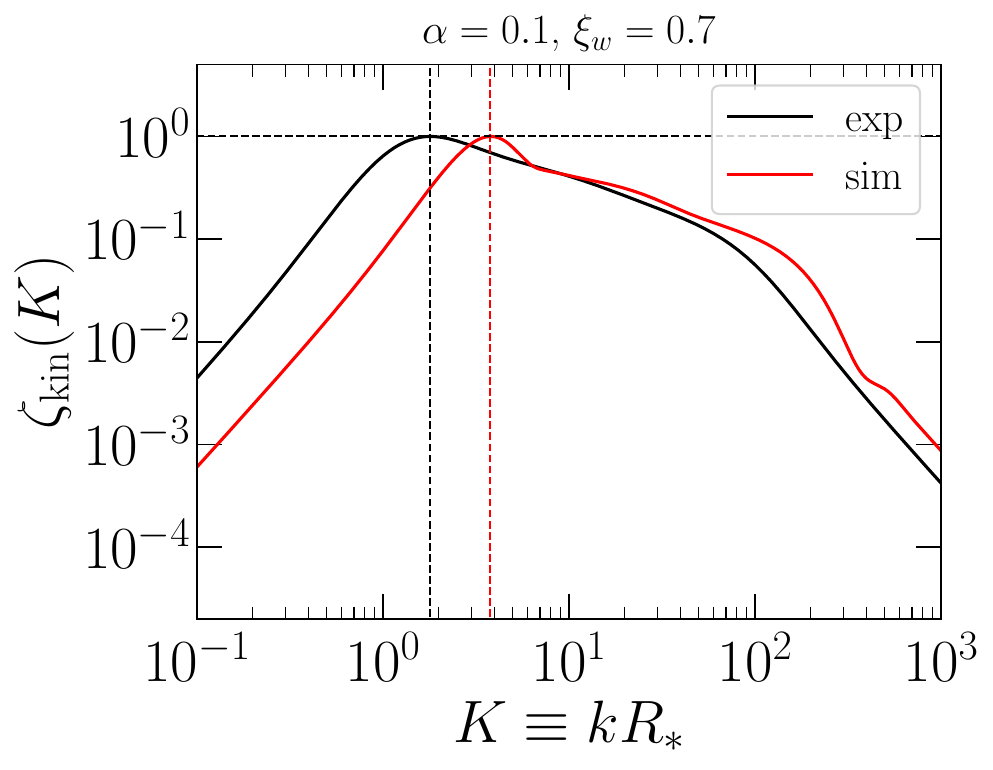}
    \includegraphics[width=.32\textwidth]{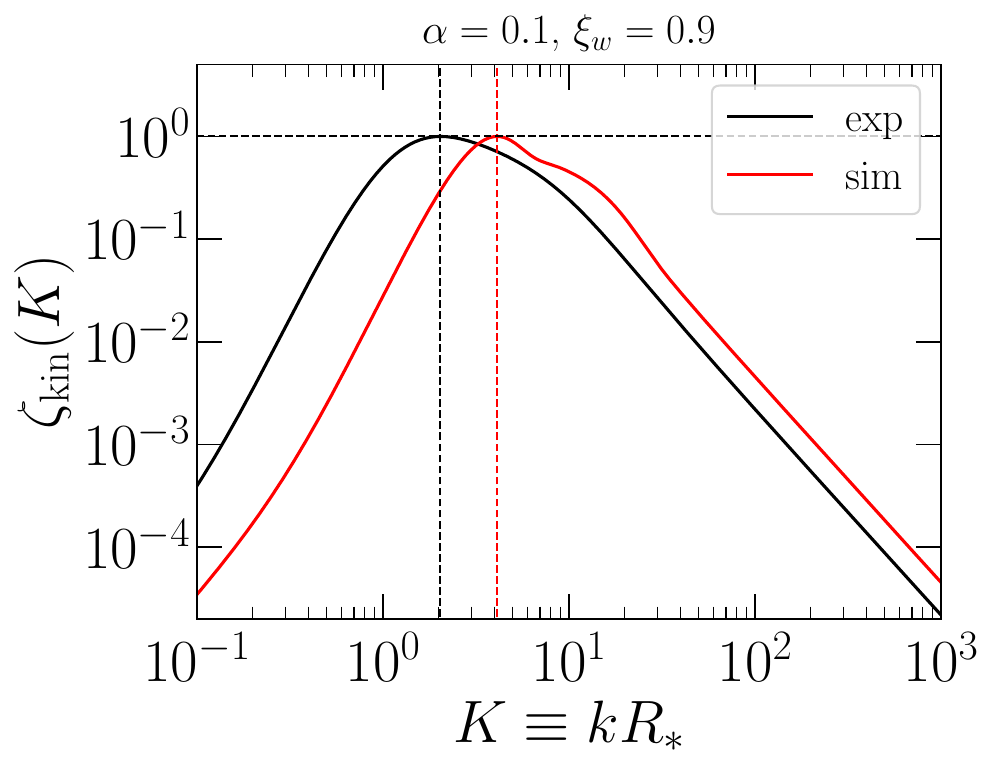}
    \caption{\justifying Time-independent component of the normalized velocity field UETC spectrum $\zetaK (k) \equiv \EK(k)/\EK^*$ [see \Eq{eq_EK_Eu_st}],
    $\EK^*$ being the maximum value of the spectrum.
    Numerical results are obtained according to the sound shell model \cite{Hindmarsh:2019phv},
    as described in Ref.~\cite{kin_sp_SSM},
    for the phase transition strength parameter $\alpha = 0.1$ and a range of wall velocities
    $\xi_w \in [0.1, 0.9]$.
    The results are computed in the cases of exponential (black) and simultaneous (red) bubble nucleations \cite{Hindmarsh:2019phv}.
    Vertical dashed lines indicate the wave
    numbers, $\kK^{\rm peak}$, where the maxima, $\EK^*$, are reached.
    Their numerical values are given in \Tab{tab:my_label}.
    }
    \label{kinetic_spectra}
\end{figure*}
The functions ${\cal E}^{(n)}$ in \Eq{EK_sw_HH} are
\begin{align}
    {\cal E}^{(1)} (\chi) = &\, |{\cal A}_+|^2 = \frac{1}{4} \Bigl[{f'}^2 (\chi) + \cs^2 l^2(\chi) \Bigr], \label{EE_1} \\ 
    {\cal E}^{(2)} (\chi) = &\, {\rm Re}\bigl({\cal A}_+ {\cal A}_-^* \bigr) = \frac{1}{4} \Bigl[{f'}^2 (\chi) - \cs^2 l^2(\chi) \Bigr],  \label{EE_2} \\
    {\cal E}^{(3)} (\chi) = &\, {\rm Im}\bigl({\cal A}_+ {\cal A}_-^* \bigr) = \frac{1}{2} \cs f'(\chi) \, l(\chi),
\end{align}
where ${\cal A}_\pm (\chi)$ are defined in \Eq{coeff_AA}.

Following Ref.~\cite{Hindmarsh:2019phv}, we expect
the amplitude of the oscillatory contributions corresponding to $\EK^{(1)}$ in \Eq{eq_EK_Eu} to be larger than those
from $\EK^{(2)}$ and $\EK^{(3)}$.
This is a consequence of the inequalities among their amplitudes,
\begin{equation}
    {\cal E}^{(1)} (\chi) \geq {\cal E}^{(2)} (\chi) \geq {\cal E}^{(3)} (\chi).
\end{equation}
However, when the term proportional to $\EK^{(2)}$ is not highly oscillating,
it cannot be neglected with respect to the one proportional to $\EK^{(1)}$.
This occurs in the limit $\omega \equiv k \cs \ll (2 \delta \tfin)^{-1}$, since $0\leq \tau_1 + \tau_2 - 2 \tini \leq 2 \delta \tfin$, where we denote the duration of the source as $\delta \tfin \equiv \tfin^{ } - \tini^{ }$.

Let us first focus on the case $k \gg 1/(2 \cs \delta \tfin)$. Then, we find a stationary UETC \cite{Hindmarsh:2019phv},
\begin{equation}
    \EK (k, \tau_1, \tau_2)
    \approx
    \EK(k) \cos (k \cs \tau_-) \,, \label{eq_EK_Eu_st}
\end{equation}
where $\EK(k) = \EK^{(1)} (k)$ and $\tau_- = \tau_2 - \tau_1$.
\FFig{kinetic_spectra} shows benchmark results for the normalized $\zetaK (k) = \EK (k)/\EK^*$, $\EK^*$ denoting the maximum value of $\EK (k)$, obtained for a
benchmark phase transition strength $\alpha = 0.1$ and a range of broken-phase bubble wall speeds $\xi_w \in [0.1, 0.9]$.
We present the details of these calculations in an accompanying
paper \cite{kin_sp_SSM}.

Since the resulting velocity field due to the superposition of sound
waves is irrotational, the causality condition requires $\EK (\tau_1, \tau_2, k) \sim k^4$ in the limit $k \rightarrow 0$ \cite{MY75,Caprini:2003vc,Hindmarsh:2019phv}.
We note that, since $\EK (k)$ is an integral over $\tilde T$ of  
${\cal E}^{(1)} (\chi)$, the limit of $\EK (k)$ when $k \to 0$ is equivalent to the limit of ${\cal E}^{(1)} (\chi)$
when $\chi \to 0$.
The integrand is then proportional to $f'^2 (\chi) + \cs^2 l^2 (\chi)$ [see \Eq{EE_1}].

As mentioned above, Ref.~\cite{Hindmarsh:2019phv} justifies the choice of $\lambda$ (which leads to the $\cs^2 l^2$ contribution in $\EK$)
in \Eq{coeff_AA} for the initial conditions,
instead of $u'$, to ensure the causality condition.
However,
the function $l^2 (\chi)$ in \Eq{eq_l}
leads to the asymptotic limits $l^2 (\chi) \sim \chi^0$ when $\chi \rightarrow 0$,
and $\EK (k) \sim k^2$ when $k \to 0$, as we show in an 
accompanying paper \cite{kin_sp_SSM}.
This naively seems
to violate causality. 
The same is true when one chooses $u'$ to impose the initial conditions.
The key point to recover the causality condition is to note that the assumption in \Eq{eq_EK_Eu_st} is not valid in
the limit $k \ll 1/(2 \cs \delta \tfin)$.
In this limit, one finds from \Eq{eq_EK_Eu},
\begin{equation}
    \lim_{k \rightarrow 0} \EK (\tau_1, \tau_2, k) = \EK^{(1)} (k) + 
    \EK^{(2)} (k).
    \label{lim_k0_UETC}
\end{equation}
The UETC of the velocity field in the $k \to 0$ limit
is then proportional to ${f'}^2 (\chi)$ [see \Eqs{EE_1}{EE_2}], and not
to $l^2 (\chi)$, as previously found using \Eq{eq_EK_Eu_st}.
Then the $\chi \to 0$ limit is indeed ${f'}^2 \sim \chi^2$,
such that $\EK \sim k^4$, as expected
from causality.

In the following, we take \Eq{eq_EK_Eu_st} to describe the
UETC spectrum and will refer to $\EK$ as the kinetic spectrum.
Even though $\EK$ does not describe the UETC
in the limit $k \to 0$, it does for all the scales that are relevant
for the study of GW production (see \Fig{kinetic_spectra}).

\begin{table*}[t]
    \centering
    \begin{tabular}{|l|c|c|c|c|c|c|c|c|c|c|c|c|c|}
    \hline type &
     $\xi_w$ & $10^4 \EK^*/R_*$ &
     $\KK^{\rm peak}$ & ${\cal K}$ & $ 10^2 \OmK$ & ${\cal C}$  & $K_{\rm GW}$ &
     $(K^3 \zeta_\Pi)_{\rm peak}$ & $K_1$ & $K_2$ & $b$ & $\alpha_1$ & $\alpha_2$ \\  \hline
     exp &
    0.1 & $26.8$ & 1.15 & 7.49 & 2.0 & 1.21 & 2.03 & 0.90 & 1.18 & 2.39 & 0.34 & 0.76 & 1.22  \\
     exp &
    0.2 & $25.7$& 1.28 & 5.42 & 1.4 & 0.94 & 2.39 & 1.36 & 1.59 & 2.39 & 1.06 & 0.66 & 1.33  \\
     exp &
    0.3 & $21.6$ & 1.53 & 5.52 & 1.2 & 0.80 & 3.01 & 2.41 & 1.98 & 3.00 & 0 & 0.67 & 1.10  \\
     exp &
    0.4 & $16.0$& 1.80 & 8.43 & 1.4 & 0.76 & 4.01 & 4.93 & 1.99 & 6.70 & 0 & 0.70 & 1.30  \\
     exp &
    0.5 & $10.3$ & 2.02 & 21.95 & 2.3 & 0.75 & 8.44 & 13.97 & 2.26 & 12.81 & 0.36 & 0.73 & 1.31 \\
     exp &
    0.6 & $5.3$ & 2.28 & 58.88 & 3.1 & 0.75 & 20.33 & 68.80 & 2.79 & 26.94 & 0.78 & 0.68 & 1.08  \\
     exp &
    0.7 & $3.2$ & 2.21 & 88.82 & 2.9 & 0.73 & 56.63 & 102.37 & 3.42 & 91.89 & 0.42 & 0.49 & 1.27 \\
     exp &
    0.8 & $5.3$  & 2.05 & 22.98 & 1.2 & 0.72 & 9.53 & 14.09 & 1.63 & 11.94 & 1.47 & 1.81 & 0.39 \\
     exp &
    0.9 & $5.1$ & 2.04 & 10.90 & 0.6 & 0.68 & 6.36 & 9.09 & 2.33 & 10.66 & 0 & 0.69 & 1.38 \\
     exp &
    0.99 & $4.5$  & 2.04 & 7.95 & 0.4 & 0.66 & 4.99 & 7.36 & 2.20 & 7.82 & 0 & 0.73 & 1.49 \\   \hline
    sim &
    0.1 &  $18.2$  & 2.59 & 10.43 & 1.9 & 0.44 & 3.42 & 7.51 & 1.74 & 3.81 & 0.92 & 1.41 & 2.34 \\
     sim &
    0.2 & $19.0$ & 2.82 & 7.08 & 1.3 & 0.31 & 4.03 & 10.85 & 2.09 & 4.04 & 0.93 & 1.34 & 2.13 \\
     sim &
    0.3 & $16.1$ & 3.29 & 7.29 & 1.2 & 0.26 & 5.07 & 18.45 & 2.56 & 4.77 & 1.29 & 1.39 & 1.25 \\
     sim &
    0.4 & $11.6$ & 3.64 & 11.11 & 1.3 & 0.25 & 6.76 & 35.29 & 3.53 & 10.50 & 0 & 0.92 & 2.39 \\
     sim &
    0.5 & $7.2$ & 3.85 & 30.66 & 2.2 & 0.25 & 16.95 & 102.40 & 4.20 & 20.63 & 0.33 & 0.85 & 2.95 \\
     sim &
    0.6 & $3.7$ & 4.16 & 84.35 & 3.1 & 0.25 & 40.79 & 528.01 & 5.36 & 44.62 & 0.75 & 0.71 & 2.15 \\
     sim &
    0.7 & $2.3$ & 4.20 & 123.32 & 2.8 & 0.24 & 113.65 &  718.45 & 7.15 & 154.60 & 0.29 & 0.45 & 2.86 \\
     sim &
    0.8 & $3.8$  & 4.06 & 31.14 & 1.2 & 0.23 & 16.07 & 97.56 & 3.14 & 23.38 & 1.02 & 1.70 & 0.71 \\
     sim &
    0.9 &  $3.6$  & 4.12 & 14.92 & 0.5 & 0.22 & 10.72 & 63.57 & 4.15 & 16.72 & 0 & 0.88 & 2.74 \\
     sim &
    0.99 & $3.2$ & 4.13 & 10.84 & 0.4 & 0.22 & 8.41 & 52.26 & 3.96 & 12.37 & 0 & 0.96 & 2.66 \\\hline
    \end{tabular}
    \caption{\justifying Numerical values of
    the amplitudes and peak frequencies that characterize the spectra of the
    velocity field (columns $3$ to $6$) and of GWs (columns $7$ to $9$), within the sound shell model for
    exponential (``exp'') and simultaneous (``sim'') types of nucleation \cite{Hindmarsh:2019phv,kin_sp_SSM}. 
    The bubble wall velocities $\xi_w$ correspond to the benchmark phase transitions shown in \Fig{kinetic_spectra}, with $\alpha = 0.1$. 
    The parameters in the last five columns determine the fit of $K^3\zeta_\Pi(K)$ in \Eq{zeta_Pi_fit}.}
    \label{tab:my_label}
\end{table*}

Following the normalization of Ref.~\cite{RoperPol:2022iel},
we define a characteristic amplitude $\EK^*$ and
wave number $k_*$.
For the kinetic spectrum corresponding to sound waves,
we set $k_* = 1/R_*$ and $\EK^*$ to be the maximum amplitude,
which is located at $\KK^{\rm peak} = \kK^{\rm peak} R_* \sim {\cal O} (1)$
(see \Fig{kinetic_spectra} and values in \Tab{tab:my_label}).
Then, the kinetic spectrum can be expressed as
\begin{equation}
    \EK(k) = \EK^* \, \zetaK (K),
    \label{dec_EK}
\end{equation}
where $K = k/k_* = kR_*$ and $\zetaK$ determines the
spectral shape of the kinetic spectrum.
The spectral shape found within the sound shell
model (see \Fig{kinetic_spectra}) is proportional to $k^4$ at low $k$, as
discussed in \Sec{UETC_velfield}, and follows a $k^{-2}$ decay at large $k$.
At intermediate scales, $\zetaK$ can present an additional intermediate power law, especially for
values
of the wall velocity $\xi_w$ close to the speed of sound $\cs = 1/\sqrt{3}$, and
develop a double peak structure, as can be seen in \Fig{kinetic_spectra} and shown in 
Refs.~\cite{Hindmarsh:2019phv,Gowling:2021gcy}.

The total kinetic energy density $\OmK$,
expressed as a fraction of the critical energy density,
is computed from \Eq{eq_EK_Eu_st} at equal times $\tau_1 = \tau_2 = \tau$,
\begin{equation}
    \OmK = \int_0^\infty \EK(\tau, \tau, k) \dd k = \frac{\EK^*}{R_*} \, {\cal K} ,
    \label{OmK_vs_EK}
\end{equation}
where we have used \Eq{dec_EK}, and\footnote{As discussed above, $\zetaK$ is not a valid description of the
UETC spectrum at small $K$. 
However, the effect on
${\cal K}$ is negligible, since $\zetaK$ 
becomes very small
in this range of $K$, and it does not contribute appreciably to the integral.}
\begin{equation}
    {\cal K}  \approx  \int_0^\infty
    \zetaK ( K) \dd K,
    \label{KK}
\end{equation}
only depends on the spectral shape, characterizing how broad
is the spectrum around $K = 1$.
The values of ${\cal K}$ are listed in \Tab{tab:my_label} for the benchmark phase
transitions shown in \Fig{kinetic_spectra}.
The kinetic energy density $\OmK$ is estimated by
the single-bubble profiles in Ref.~\cite{Espinosa:2010hh}
as $\OmK \equiv \kappa \alpha/(1 + \alpha)$, where $\kappa$ is
an efficiency factor that depends on $\alpha$ and $\xi_w$.
We omit the comparison of $\OmK$ found in the sound shell model with that
of Ref.~\cite{Espinosa:2010hh} since we focus on the GW production in the current work.
This relation will be explored in an accompanying paper \cite{kin_sp_SSM} (see also the discussion of
Refs.~\cite{Hindmarsh:2013xza,Hindmarsh:2016lnk,Hindmarsh:2017gnf,Hindmarsh:2019phv,Jinno:2020eqg,Jinno:2022mie}).

\subsection{UETC of the anisotropic stress}
\label{sec_norm}

We consider the UETC of the anisotropic stresses $E_\Pi$, defined in \Eq{E_Pi}, under the stationary assumption of 
\Eq{eq_EK_Eu_st}. Introducing the normalization
of \Eqss{dec_EK}{KK} one obtains,
\begin{align}
    k \, E_\Pi^{ } &(\tau_1, \tau_2, k) \nonumber\\
    & \simeq 2 \, \bar w^2 \, K^3 \, \biggl(\frac{\OmK}{{\cal K}}\biggr)^2\,  {\cal C} \, \zeta_\Pi^{ } (\tau_-, K),
\end{align}
where, following Ref.~\cite{RoperPol:2022iel}, we have defined
\begin{align}
    {\cal C}\, \zeta_\Pi^{ }(\tau_-, K) = &\, \int_0^\infty P^2 \zetaK (P) \, \cos (P \cs \, k_* \, \tau_-) \, \dd P
    \nonumber \\
    \int_{-1}^1 (1 -  z^2&)^2 \, \frac{\zetaK(\tilde P)}{\tilde P^4}
    \cos (\tilde P \cs \, k_* \, \tau_-) \dd z \,,
    \label{p_Pi}
\end{align}
and used the notation
$P \equiv p/k_* = p R_*$ and $\tilde P \equiv \tilde p/k_* = \tilde p R_*$.
The constant ${\cal C}$ is defined such that $\zeta_\Pi (K) \to 1$, in the $K \to 0$ limit, at equal times, i.e., $\tau_- = 0$
(see \Tab{tab:my_label} for values of ${\cal C}$ of the benchmark phase transitions),
\begin{equation}
    {\cal C} = \frac{16}{15} \int_0^\infty \frac{\zetaK^2 (K)}{K^2} \dd K \,.
    \label{CC_alp}
\end{equation}
The spectral shape is therefore encoded in $\zeta_\Pi$.
We note that, as discussed in the previous section, the UETC of the velocity field in this limit
should be taken from \Eq{lim_k0_UETC}, so it does not 
only depend on the time
difference $\tau_-$ when $k \ll 1/(2 \cs \delta \tfin)$.

\begin{figure}
    \centering
    \includegraphics[width=\columnwidth]{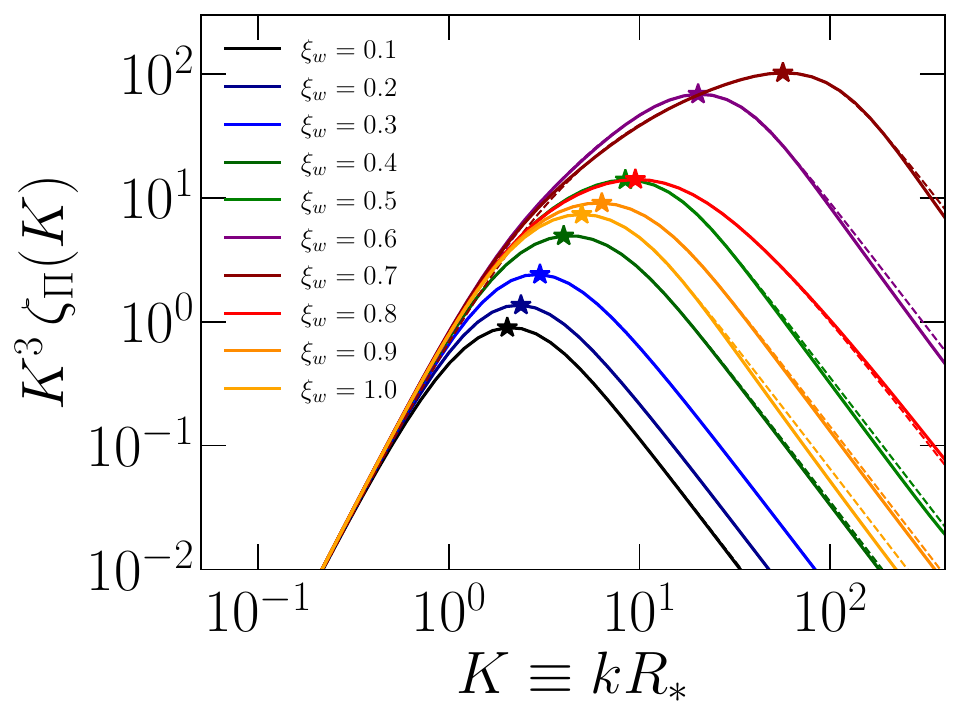}
    \caption{\justifying Normalized spectrum of the anisotropic stresses, $\zeta_\Pi$, for the
    kinetic spectra of the benchmark phase transitions shown in \Fig{kinetic_spectra},
    in the case of exponential nucleation, computed numerically (solid lines) compared to the fit from \Eq{zeta_Pi_fit} (dashed lines).
    $\zeta_\Pi$ is multiplied by $K^3$ since this
    is the relevant contribution to the resulting GW spectrum (see \Sec{numerical_GW}).
    The stars correspond to $K_{\rm GW}$, where $K^3 \zeta_\Pi$ is maximum (see values in \Tab{tab:my_label}).}
    \label{zeta_Pi}
\end{figure}

At equal times, \Eq{p_Pi} becomes
\begin{align}
    {\cal C}\,  \zeta_\Pi^{ }(K) =  \int_0^\infty &\, P^2 \zetaK (P) \dd P
    \nonumber \\  &  \times
    \int_{-1}^1 (1 -  z^2)^2 \, \frac{\zetaK(\tilde P)}{\tilde P^4}
    \dd z,
    \label{p_Pi_equal_times}
\end{align}
where $\zeta_\Pi (K) \leq 1$ is a monotonically
decreasing function, shown in
\Fig{zeta_Pi} for the benchmark 
phase transitions of \Fig{kinetic_spectra}.
This condition can be understood from the derivative of ${\cal C} \zeta_\Pi$ with respect to $K$,
\begin{align}
    {\cal C}\, \partial_K \zeta_\Pi  & (K) = \int_0^\infty P^2 \zetaK (P) \dd P \int_{-1}^1 (1 -  z^2)^2 \,
    \nonumber \\  & \times \biggl[
    \zetaK' (\tilde P) - \frac{4 \zetaK (\tilde P)}{\tilde P}
    \biggr] \frac{K - Pz}{\tilde P^4} \dd z.
    \label{derivative_zetaPi}
\end{align}
We find that the term in square brackets is always negative if
$\zetaK(\tilde P)\propto \tilde P^n$
with $n \leq 4$ at all $\tilde P$, which is indeed the case.
The second term $K - Pz$ is positive for most of the
integration range since it becomes negative only when $z > K/P$.
Since $1 - z^2$ is symmetric in $z$,
then the final integral is {\em almost} always negative, unless
the term in the square bracket, once multiplied by $P^2 \zetaK(P)$, has a larger contribution when $K/P < 1$ and $K/P < z < 1$ than in the rest of the range, but
this is not the case for any of the evaluated spectra (see 
\Fig{zeta_Pi}).

At intermediate $K$, $\zeta_\Pi$ strongly depends on the
specific spectral shape of the velocity power spectrum $\zetaK (K)$, and it requires numerical evaluation
of the integral in \Eq{p_Pi_equal_times}.
However, in the asymptotic limit $K \to \infty$, 
indicated by a $\infty$ superscript,
\Eq{p_Pi_equal_times} becomes
\begin{equation}
    \zeta_\Pi^\infty = \frac{\zetaK^\infty}{K^4} 
    \frac{\displaystyle\int_0^\infty P^2 \zetaK (P) \dd P}
    {\displaystyle\int_0^\infty \frac{\zetaK^2 (P)}{P^2} \dd P} \ .
\end{equation}
Therefore, if the kinetic spectrum decays as $\zetaK^\infty \sim K^{-b}$, then
$\zeta_\Pi^\infty$ decays as $K^{-b - 4}$.
However, since $P$ is integrated from $0$ to $\infty$,
it can become of the same order as $K$ and
the power law decay $K^{-b - 4}$ might not be reached exactly.
In particular, we find $\zeta_\Pi^\infty \sim K^{-5}$, which is close to the estimated $K^{-6}$ slope, for the
benchmark kinetic spectra (see \Fig{zeta_Pi},
where dashed lines correspond to the fit in \Eq{zeta_Pi_fit}, with an exact $K^{-5}$ decay).

We find in \Sec{numerical_GW} that the final GW spectrum
is proportional to $K^3 \zeta_\Pi$ in the limit of short
duration of the GW sourcing, $\delta \tfin/R_* \ll 1$.
For longer duration, the GW spectrum can deviate with respect to $K^3 \zeta_\Pi$ by a factor $\tilde \Delta$ (see \Sec{numerical_GW}).
In any case, the GW spectrum
approximately peaks at $K_{\rm GW}$, defined as the wave number
where $K^3 \zeta_\Pi$ takes its maximum value $(K^3 \zeta_\Pi)_{\rm peak}$.
The value of $K_{\rm GW}$ depends on how steep is the
negative slope of $\zeta_\Pi$ when it starts to decay
around $K \sim {\cal O} (1)$;
it therefore requires numerical
evaluation of $\zeta_\Pi$ using \Eq{p_Pi_equal_times}.
We give in \Tab{tab:my_label} the numerical values of $K_{\rm GW}$ and $(K^3 \zeta_\Pi)_{\rm peak}$.

Due to the double peak structure of $\zetaK (K)$, which 
appears when the wall velocity $\xi_w$ approaches $c_s$,
an appropriate fit for $K^3 \zeta_\Pi$ is a smoothed
double broken power law,
\begin{equation}
    K^3 \zeta_{\Pi} (K) = \frac{K^3 \bigl[1 + (K/K_1)^{(3 - b) \alpha_1} \bigr]^{-{1 \over \alpha_1}}}{\bigl[1 + (K/K_2)^{(2 + b)\alpha_2} \bigr]^{{1 \over \alpha_2}}} \ ,
    \label{zeta_Pi_fit}
\end{equation}
where $K_{1,2}$ are the wave number breaks,
$b$ is the intermediate slope, and $\alpha_{1,2}$ are
parameters that determine the smoothness of the transition between
slopes.
At low $K$, we fix $\zeta_\Pi = 1$, as desired, and at large
$K$, we fix $\zeta_\Pi^\infty \sim K^{-5}$.
We note that, in general, $K_2$ is not necessarily equal to $K_{\rm GW}$.
We show the corresponding values of $K_{1,2}$, $b$, and $\alpha_{1,2}$, found for the benchmark
phase transitions of \Fig{kinetic_spectra}, in \Tab{tab:my_label}.
We note that some $\zeta_\Pi$ are already well approximated by a single
broken power law since they do not present a double peak structure,
especially for $\xi_w \lesssim 0.5$ and $\xi_w \gtrsim 0.8$.

The exact values of the amplitude $(K^3 \zeta_\Pi)_{\rm peak}$, the frequency breaks $K_{1,2}$, and the intermediate
slopes
highly depend on the specific
spectral shape of the velocity power spectrum $\zetaK$.
According to the sound shell model, Refs.~\cite{Hindmarsh:2016lnk,Hindmarsh:2019phv} proposed that
the two peaks are determined by the inverse mean size of the bubbles, $1/R_*$, and the inverse sound shell
thickness, $1/(R_* \Delta_w)$, where $\Delta_w = |\xi_w - \cs|/\cs$.
Similar dependencies
are found in numerical simulations \cite{Hindmarsh:2013xza,Hindmarsh:2015qta,Hindmarsh:2017gnf,Jinno:2020eqg,Jinno:2022mie}.
We explore the relations between the phase transition parameters
and the shape of the anisotropic stresses,
which will ultimately impact the GW spectrum,
in an accompanying paper \cite{kin_sp_SSM}.
In the following, we study the spectral shape of GWs once
we know the spectral shape of $\zeta_\Pi$, shown in \Fig{zeta_Pi}
for a set of benchmark phase transitions.

\section{Low wave number tail of the GW spectrum from sound waves}

\label{stationary_UETC}

In this section, we study the amplitude of the GW spectrum analytically, by evaluating its low-frequency limit $k \to 0$. We do not assume flat space-time but consider an expanding Universe.
Following the sound shell model \cite{Hindmarsh:2019phv}, we adopt the stationary assumption of \Eq{eq_EK_Eu_st}, assuming its validity down to $k\rightarrow 0$ (see discussion in \Sec{UETC_velfield}).
The
source is assumed to be stationary but still characterized by a finite lifetime, $\delta \tfin$. 
Note that this might introduce a spurious effect in the final GW spectrum due to the sharp cutoff of the integrals in time \cite{Caprini:2009yp}. However, we deem this not important, given the good agreement of the GW spectrum evaluated semi-analytically following the sound shell model with the one from numerical simulations \cite{Hindmarsh:2016lnk,Hindmarsh:2019phv}.
The study of the GW spectrum at all $k$ is presented in \Sec{numerical_GW}.

In \Sec{GW_SSM}, we start by collecting the results of \Sec{sw_sec} to evaluate the GW spectrum, and comment on the consequences of the expansion of the Universe.
We find in \Sec{k_to_0_sec} that the GW spectrum follows a
$k^3$ scaling at low $k$:
this is expected from previous analyses, both analytical \cite{Cai:2023guc} and numerical \cite{Jinno:2022mie,Sharma:2023mao}, but it is in contradiction with the findings of the original sound shell model of Ref.~\cite{Hindmarsh:2019phv}, which obtains instead that, at scales larger than the peak, the GW spectrum goes as $k^9$.
In \Sec{k3_k9}, we reproduce the calculation of Ref.~\cite{Hindmarsh:2019phv} and show that the $k^9$ behavior is recovered only when one makes an
assumption for the UETC that is, however, only justified under certain conditions that do not
hold in the $k \to 0$ limit. We therefore claim that the $k^3$
scaling is the correct one in the low-$k$ limit.

Moreover, we find in \Sec{k_to_0_sec} that the GW
amplitude in the $k \to 0$ limit is proportional to $\ln^2 (1+ \delta\tfin\HH_*)$. 
This factor becomes quadratic in the source duration parameter $\delta\tfin\HH_*$
when one ignores the expansion of the Universe, i.e., in the limit $\delta \tfin \HH_* \ll 1$.
A similar quadratic dependence has also been found in the numerical analysis of Ref.~\cite{RoperPol:2019wvy} for acoustic turbulence, as well as for (magneto)hydrodynamical [(M)HD] vortical turbulence, both analytically \cite{Caprini:2009fx,Caprini:2009yp,RoperPol:2022iel} and
numerically \cite{RoperPol:2019wvy,Kahniashvili:2020jgm,Brandenburg:2021bvg,Brandenburg:2021tmp,RoperPol:2021xnd,RoperPol:2022iel,Auclair:2022jod}.
However, this result is in contradiction with the
linear dependence in the source duration usually assumed for stationary UETCs \cite{Kosowsky:2001xp,Gogoberidze:2007an,Caprini:2009fx,Huber:2007vva,Hindmarsh:2013xza,Hindmarsh:2019phv,Guo:2020grp}.
In particular, a linear growth is assumed for sound waves
in analytical (see, e.g., Refs.~\cite{Caprini:2015zlo,Weir:2017wfa,Caprini:2019egz,Hindmarsh:2019phv,Hindmarsh:2020hop,Gowling:2021gcy,Gowling:2022pzb,RoperPol:2023bqa}) and numerical (see, e.g., Refs.~\cite{Hindmarsh:2013xza,Hindmarsh:2015qta,Hindmarsh:2017gnf,Cutting:2019zws,Jinno:2020eqg,Jinno:2022mie}) studies.
We investigate this issue in \Sec{linear_vs_quadratic}.
We show that
the linear growth of Ref.~\cite{Hindmarsh:2019phv}, and the suppression factor $\Upsilon = 1 - 1/(\tfin\HH_*)$ of Ref.~\cite{Guo:2020grp} for an expanding Universe, are
valid for stationary processes {\em only} under specific assumptions
\cite{Caprini:2009fx}, which are equivalent
to those used in Refs.~\cite{Hindmarsh:2019phv,Guo:2020grp}.
We show that these assumptions do not hold in the $k \to 0$ limit.
Therefore, the causality tail,
proportional to $k^3$, is also proportional to $\ln^2 (1+ \delta\tfin\HH_*)$.

In \Sec{sec_kraichnan}, we extend our analysis
to a stationary Gaussian UETC
(cf.~Kraichnan decorrelation \cite{Kraichnan:1965zz,Kosowsky:2001xp,Gogoberidze:2007an,Caprini:2009yp,Auclair:2022jod}) to show, within a general framework,
when the aforementioned assumptions hold.
We find that this occurs
when $k \tau_c \gg 1$, where $\tau_c$ is
a characteristic time of the process (e.g., $\delta \tfin$ in
the sound shell model).
Hence, if $\delta \tfin/R_* \gg 1$, 
the slope of the GW spectrum around its spectral peak, $\kK^{\rm peak} \sim 1/R_*$,
is well described under these assumptions.
As discussed in \Sec{sec:timenonlin}, this limit corresponds to low fluid velocities and correspondingly weak first-order phase transitions.

Indeed, in \Sec{numerical_GW}, we extend our analysis to all $k$ and we show that, even though the causality
tail is proportional to $k^3$ and follows a quadratic growth with $\delta \tfin$, the
amplitude around the peak can present
a steep slope approaching the $k^9$ scaling, and can follow a linear
growth with $\delta\tfin$, when $\delta \tfin/R_* \gg 1$.
Hence, at frequencies $k \gg 1/\delta \tfin$, and when $\delta \tfin/R_* \gg 1$, the GW spectrum
can be approximately described by the calculation of Refs.~\cite{Hindmarsh:2019phv,Guo:2020grp}, reproduced in
\App{HH19}.
Including the expansion of the Universe, the quadratic $(\delta\tfin \mathcal{H}_*)^2$ and linear $\delta\tfin \mathcal{H}_*$ dependencies
become respectively $\ln^2 (1+\delta\tfin \HH_*)$ and $\Upsilon$.

\subsection{GW spectrum in the sound shell model}
\label{GW_SSM}

We adopt the stationary assumption of \Eq{eq_EK_Eu_st} and combine Eqs.~\eqref{eq_OmGW_UETC} and \eqref{E_Pi} to find the GW spectrum today. After averaging over fast
oscillations in time, it becomes
\begin{align}
    &\, \OmGW  ( \delta \tfin, k) = 
    3 \, \bar w^2 \, k^3 \, {\cal T}_{\rm GW} \int_{-1}^1 \bigl(1 - z^2 \bigr)^2 \, \dd z
      \nonumber \\ 
    &\, \hspace{2mm}
    \times 
    \int_0^\infty \!\! \dd p \, \frac{p^2}{\tilde p^4} \, \EK (p)
    \EK (\tilde p) \, 
    \Delta (\delta\tfin, k, p, \tilde p) \, .
    \label{OmGW_GG}
\end{align}
Note that
\Eq{OmGW_GG}
gives the present-day GW spectrum, i.e., the observable we are generally interested in. 
While the source is still active, the GW spectrum would depend not only on the source duration $\delta \tfin \equiv \tfin^{ } - \tini^{ }$, but also on the absolute time $\tau$.  
During the production phase in the early Universe, in fact, the dependence on $\tau$ cannot be averaged out.
We present this case in \App{app_GW_tdep},
which
is particularly relevant when one compares it with the results of 
numerical simulations: depending on the wave number span and on the duration of the simulation, it is often required to take into account the residual dependence on $\tau$ of the GW spectrum, instead of on $\delta\tfin$ only.

\begin{figure*}[t]
    \centering
    \includegraphics[width=.85\columnwidth]{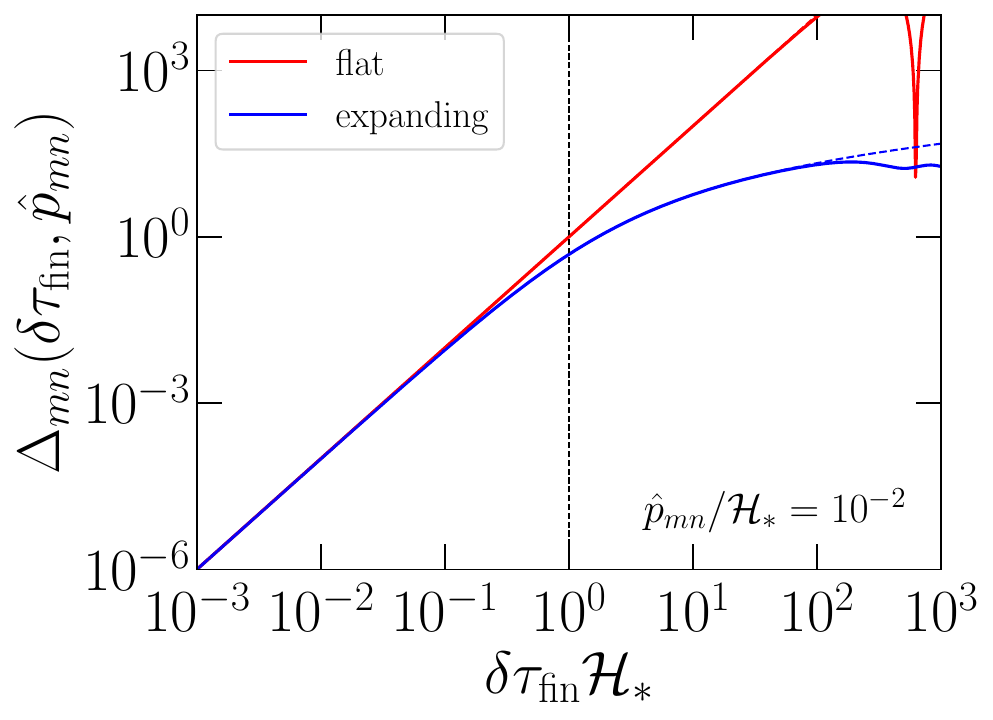}
    \includegraphics[width=.85\columnwidth]{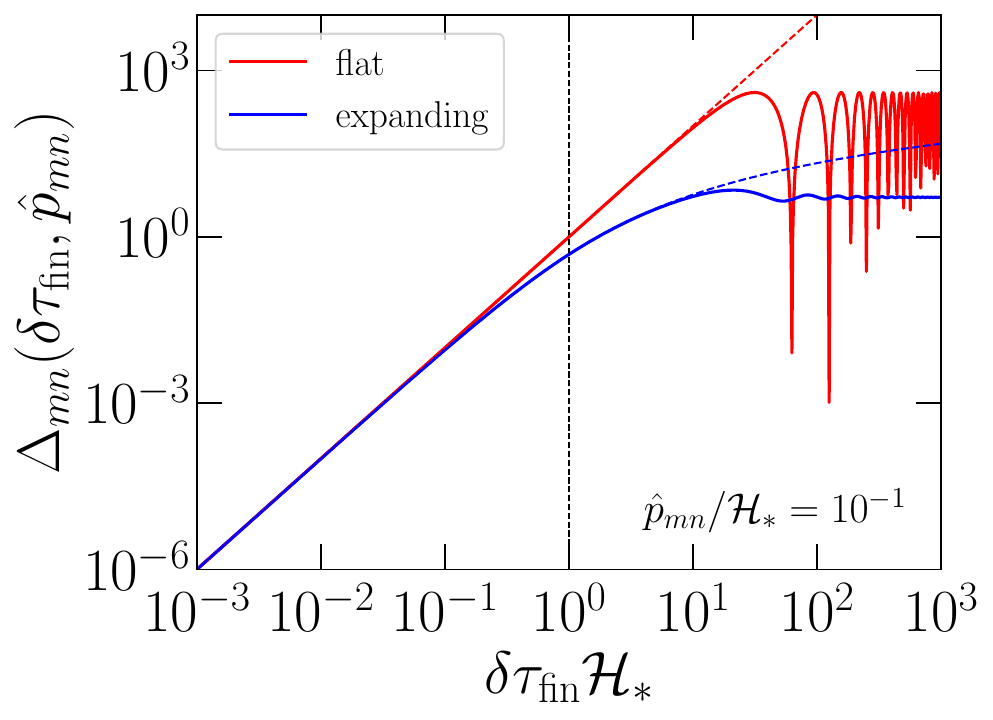}
    \includegraphics[width=.85\columnwidth]{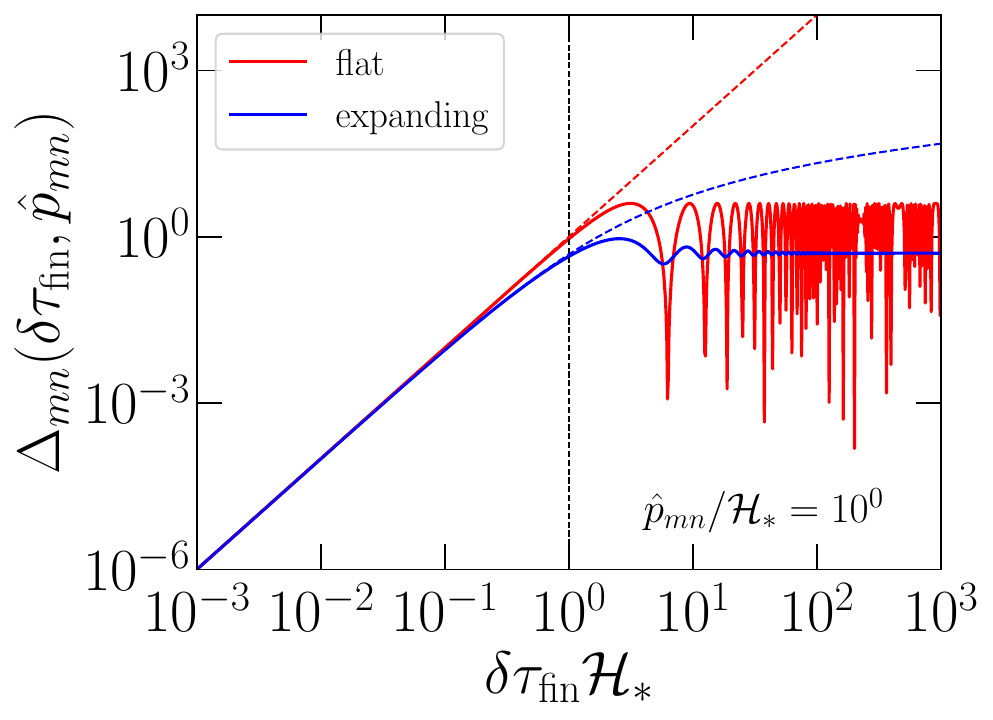}
    \includegraphics[width=.85\columnwidth]{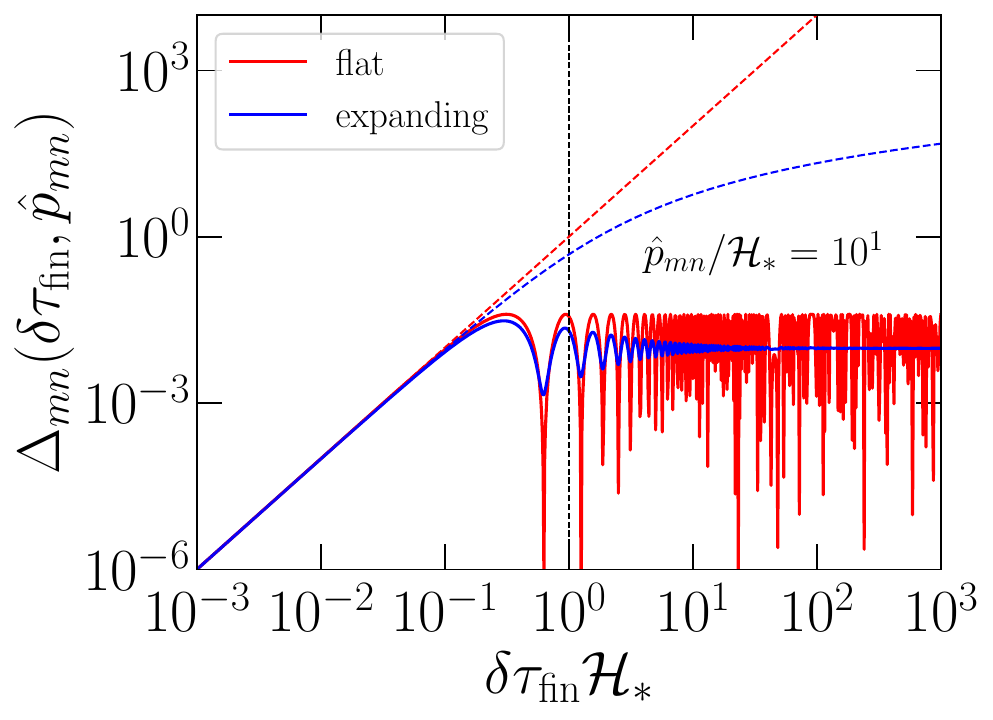}
    \caption{\justifying $\Delta_{mn}$ as a function of the duration of GW production $\delta\tfin \HH_* \equiv \tfin \HH_* - 1$, given in units of the Hubble time $\tini = \HH_*^{-1}$, for an expanding [blue, \Eq{GG_kp_stat}] and a flat [red, \Eq{G_func_flat}] Universe.
    The four panels show different values of the combined momenta $\hat p_{mn}^{ }/\HH_* \in \{10^{-2}, 10^{-1}, 1, 10\}$.
    The dashed lines correspond to the asymptotic limits: $(\delta \tfin \HH_*)^2$ when $\delta \tfin \HH_* \ll 1$ (red), and $\ln^2 (\tfin \HH_*)$ when $\hat p_{mn} \ll \HH_*$ for an expanding Universe (blue). Both asymptotic limits are equivalent when $\delta \tfin \HH_* \ll 1$.}
    \label{GG_vs_t}
\end{figure*}

The function $\Delta$ in \Eq{OmGW_GG} contains the integral over times $\tau_1$ and $\tau_2$ of the Green's functions and the time dependence of the stationary UETC,
\begin{align}
    \Delta (\delta \tfin, &\,k, p, \tilde p) \equiv
    \int_{\tini^{ }}^{\tfin^{ }} \!\! \frac{\dd \tau^{ }_1}{\tau^{ }_1} 
    \int_{\tini^{ }}^{\tfin^{ }} \!\! \frac{\dd \tau^{ }_2}{\tau^{ }_2} \nonumber \\
    &\, \times
    \cos (p \cs \tau^{ }_-)
    \cos (\tilde p \cs \tau^{ }_-)
    \cos (k\tau^{ }_-) \,.
    \label{GG_function}
\end{align}
The product of cosines can be expressed as
\begin{align}
     \cos (p  \cs \tau^{ }_-) 
    \cos (\tilde p \cs  & \tau^{ }_-)
    \cos (k \tau^{ }_-) \nonumber \\ &\, = \frac{1}{4} \sum_{m, n = \pm 1}
    \cos (\hat p_{mn} \tau^{ }_-) \,,
\end{align}
where we have defined
$\hat p_{mn} \equiv (p + m \tilde p) \, \cs + nk$.
We
separate the time dependencies using
\begin{align}
    \cos (\hat p_{mn} \tau_-) = &\, \cos (\hat p_{mn} \tau_2) \cos (\hat p_{mn} \tau_1) \nonumber \\ &\, + \sin (\hat p_{mn} \tau_2)
    \sin (\hat p_{mn} \tau_1),
\end{align}
so that the integrals over $\tau^{ }_1$ and $\tau^{ }_2$ yield
\begin{align}
    \Delta (\delta \tfin, k, p, \tilde p) = \sum_{m, n = \pm 1} \Delta_{mn} (\delta \tfin, \hat p_{mn}),
    \label{GG_kp_stat0}
\end{align}
where we have defined the function
\begin{align}
    \Delta_{mn} & (\delta \tfin, \hat p_{mn}) 
    \nonumber \\ = &\, \frac{1}{4} \Bigl[ \Delta \Ci^2 (\tfin, \hat p_{mn}) + \Delta \Si^2 (\tfin, \hat p_{mn}) \Bigr],
    \label{GG_kp_stat}
\end{align}
and
\begin{align}
    \Delta \Ci (\tau, p) &\equiv 
        \Ci \bigl(p \tau^{ }\bigr) - \Ci \bigl(p\tini^{ }\bigr) \,, \label{Ci_Delta}\\
    \Delta \Si (\tau, p) &\equiv 
        \Si \bigl(p \tau^{ }\bigr) - \Si \bigl( p \tini^{ }\bigr) \,. \label{Si_Delta}  
\end{align}
Even though $\Delta_{mn}$ is an intermediate function, which needs to be integrated over $p$ and $z$ to obtain the GW spectrum [see \Eq{OmGW_GG}], it is still very useful to study its behavior as a function of both $\delta \tfin \HH_*$ and $\hat p_{mn}/\HH_*$. 
In \Fig{GG_vs_t}, we show $\Delta_{mn}$ as a function of $\delta \tfin \HH_*$
for different fixed values of $\hat p_{mn}/\HH_*$.

In the limit $\hat p_{mn} \ll \HH_*$, the functions $\Delta \Ci
\rightarrow
\ln (\tfin^{ } \HH_*)$ and $\Delta \Si
\rightarrow 0$, such that
$\Delta \rightarrow \ln^2 (\tfin^{ } \HH_*)=\ln^2 (1+\delta\tfin \HH_*)$ (see \Fig{GG_vs_t}).
This limit is very relevant: we show in \Sec{k_to_0_sec} that, indeed, this logarithmic scaling with the
source duration holds also for the GW spectrum in the
$k \to 0$ limit. 

If the duration of the production of GWs from sound waves is short, $\delta \tfin \HH_*  \ll 1$, the expansion of the Universe can be neglected.
As a consequence,
$\tau \approx 1/\HH_*$ in \Eq{GW_eq}, and the
factor $1/(\tau_1 \tau_2)$ in the integrand of \Eq{GG_function} becomes constant, $\HH_*^2$.
In this case, we obtain the solution for a flat (non-expanding)
Universe,
\begin{align}
    \Delta_{mn}^{\rm flat} (\delta \tfin, &\, \hat p_{mn}) =  \nonumber\\
   & \frac{1 - \cos \bigl[(\hat p_{mn}^{ }/\HH_*) (\HH_*\delta \tfin) \bigr]}
    {2\, \bigl(\hat p_{mn}/\HH_*\bigr)^2} \ .
    \label{G_func_flat}
\end{align}
Since $\delta \tfin \HH_*  \ll 1$, one has $\Delta^{\rm flat}
\rightarrow (\delta \tfin \HH_*)^2_{ }$ from \Eq{G_func_flat},
suggesting that the GW spectrum grows quadratically in
$\delta \tfin$.
This quadratic scaling also holds for an expanding Universe,
since the same limit can be found from \Eq{GG_kp_stat}:
for $\hat p_{mn} \ll \HH_*$ and $\delta \tfin \HH_*  \ll 1$, one has $\Delta \to \ln^2 (1+\delta\tfin \HH_*)\rightarrow  (\delta \tfin \HH_*)^2_{ }$.
These behaviors for a flat and an expanding Universe are shown in \Fig{GG_vs_t} and are due to the asymptotic limits of the cosine
and sine integral functions, as pointed out in Refs.~\cite{Caprini:2009fx,RoperPol:2022iel}.

\subsection{Low-frequency limit}
\label{k_to_0_sec}

In the previous section, we have shown that the
function $\Delta_{mn}$, given in \Eqs{GG_kp_stat}{G_func_flat} respectively for an expanding and a flat Universe,
depends logarithmically
on the duration of the source, $\ln^2 (\tfin \HH_*)$,
for small values of $\hat p_{mn}/\HH_*$.
In this section, 
we compute explicitly the GW spectrum in the limit $k \to 0$, and confirm that the GW spectrum inherits the same logarithmic dependence at large scales.
We also show how the $k^3$ scaling, expected
from causality \cite{Caprini:2003vc},
appears in this limit, instead of the $k^9$
scaling found in Ref.~\cite{Hindmarsh:2019phv}.

In the low-frequency limit $k \to 0$, $\tilde p \to p$ and
$\hat p_{mn} \rightarrow (p + m \tilde p)\, \cs$. The latter becomes
$0$ for $m = -1$ and $2 p \cs$ for $m = 1$.
Therefore, the $z$-dependence in \Eq{OmGW_GG} is reduced only to the function $(1 - z^2)^2$, and the GW spectrum becomes
\begin{align}
     &\, \lim_{k\rightarrow 0} \OmGW(\delta\tfin, k) =\nonumber \\ 
     &\,  3\,  \bar w^2 \, k^3 \, \frac{16}{15} \, {\cal T}_{\rm GW}
     \int_0^\infty
    \frac{\EK^2(p)}{p^2} \Delta_0(\delta \tfin, p) \dd p.
    \label{GW_spec_k0}
\end{align}
This expression already shows an important result: the GW spectrum scales with $k^3$ in the limit $k \rightarrow 0$,
since the integral in \Eq{GW_spec_k0} does not depend on $k$.
We defer the comparison of this result to the
$k^9$ scaling found in Ref.~\cite{Hindmarsh:2019phv} to \Sec{k3_k9}.
There, we demonstrate that
a simplifying approximation of $\Delta$ used in Ref.~\cite{Hindmarsh:2019phv} leads to an additional
dependence of $\Delta_0$ on $k$. 
However, this approximation does not apply in the $k \to 0$
limit, invalidating the $k^9$ behavior at large scales.
In the following,
we rather focus on the dependence of $\OmGW$ with the source duration $\delta \tfin$.

The $\Delta_0$ function that appears in \Eq{GW_spec_k0} corresponds
to $\Delta$, given in \Eq{GG_kp_stat0}, in the $k \to 0$ limit,
\begin{align}
    \Delta_0 (\delta \tfin, p) =  \lim_{k\rightarrow 0} \Delta (\delta \tfin, k,  & \,  p, \tilde p)   \nonumber \\ =  \frac{1}{2} 
    \Bigl[  \ln^2 (\tfin \HH_*) +   & \, \Delta \Ci^2 (\tfin, 2 p \cs) \nonumber \\ +   & \, \Delta \Si^2 (\tfin, 2 p\cs) \Bigr],
    \label{Delta_0_p_exp}
\end{align}
which, for a flat (non-expanding) Universe, reduces to
\begin{align}
    \Delta_0^{\rm flat} (\delta \tfin, p) = &\, \lim_{k\rightarrow 0} \Delta^{\rm flat} (\delta \tfin, k, p, \tilde p) \nonumber \\  = \frac{1}{2} \Biggl[ &\, \bigl(\delta \tfin \HH_*\bigr)^2  + \frac{\sin^2 \bigl(p \cs \delta \tfin\bigr)}{\bigl(p \cs/\HH_*\bigr)^2} \Biggr].
    \label{Delta_0_p_flat}
\end{align}

We find in \Eq{Delta_0_p_exp} a first term, $\half \ln^2 (\tfin \HH_*)$, independent of $p$, and
a second term that depends on $p$ and will enter the integral over $p$ in \Eq{GW_spec_k0}. 
We can parameterize the 
dependence of the GW amplitude
with $\delta \tfin$ by defining a weighted average of the function $\Delta_0$ with the spectral function $\zetaK$,
\begin{align}
    \tilde \Delta_0  (\delta \tfin, & \, R_*)  \nonumber \\ =  &\, \frac{\displaystyle\int_0^\infty \frac{\zetaK^2 (K)}{K^2} \, \Delta_0 (\delta \tfin,
    K /R_*) \, \dd K}{\displaystyle\int_0^\infty \frac{\zetaK^2 (K)}{K^2} \dd K} \ ,
    \label{tilde_Delta0}
\end{align}
where we have used the normalized quantities $\zetaK (K) = \EK (K)/\EK^*$ and $K \equiv k/k_* = k R_*$, defined
in \Sec{sec_norm}.
Introducing \Eq{tilde_Delta0} into \Eq{GW_spec_k0},
and using the normalization of \Sec{sec_norm} for the UETC
of the anisotropic stresses, we find
\begin{align}
    \lim_{K\rightarrow 0} &\, \OmGW(\delta\tfin, K) \nonumber \\ = 
    &\, 3 \, \bar w^2 \, K^3 \, {\cal T}_{\rm GW} \biggl(\frac{\OmK}{{\cal K}}\biggr)^2 \,
    {\cal C} \, \tilde \Delta_0 (\delta \tfin , R_*),
    \label{GW_spec_k0_norm}
\end{align}
where $\EK^* = \OmK R_*/{\cal K}$ [see \Eq{OmK_vs_EK}].
Since the dimensionless kinetic power spectrum $\zetaK$ is peaked 
at $\KK^{\rm peak} = \kK^{\rm peak} R_* \sim {\cal O} (1)$ (see \Fig{kinetic_spectra}), we can approximate it as $\zetaK(K) \sim \delta(K-1)$ in the integrals of \Eq{tilde_Delta0}.
Under this assumption, $\tilde \Delta_0 (\delta \tfin, R_*) \to \Delta_0 (\delta \tfin, 1/R_*)$, showing that
$\tilde \Delta_0/\Delta_0$ characterizes the deviations with
respect to a delta-peaked kinetic power spectrum.
We can now study the dependence of $\tilde \Delta_0$ with $\delta \tfin$ under
this approximation,
\begin{align}
    \tilde \Delta_0 & (\delta \tfin,  \, R_*)  \sim \frac{1}{2}\Biggl\{\ln^2 (\tfin \HH_*) \, +
    \label{eq:tildeDelta0approx} \\
    & \left[\Ci\left(\frac{2c_s}{\mathcal{H}_*R_*}+2c_s\frac{\delta\tfin}{R_*}\right)-\Ci\left(\frac{2c_s}{\mathcal{H}_*R_*}\right) \right]^2 +\nonumber \\
   &  \left[\Si\left(\frac{2c_s}{\mathcal{H}_*R_*}+2c_s\frac{\delta\tfin}{R_*}\right)-\Si\left(\frac{2c_s}{\mathcal{H}_*R_*} \right) \right]^2 
    \Biggr\}\,. \nonumber
\end{align}
If $c_s\delta\tfin/R_* \ll 1$, from the expansion of the $\Ci$ and $\Si$ functions one gets:
\begin{align}
    \tilde \Delta_0  (c_s\delta\tfin/R_* \ll 1) & \sim  \frac{1}{2}\left\{\ln^2 (\tfin \HH_*)+ (\delta\tfin \mathcal{H}_*)^2 \right\} \nonumber \\ 
    &\sim (\delta\tfin \mathcal{H}_*)^2\,,
\end{align}
where the last estimate holds when $\delta \tfin\HH_*\ll R_*\HH_*\leq 1$. 
In the opposite limit $c_s\delta\tfin/R_* \gg 1$, the contribution from the $\Ci$ and $\Si$ functions is oscillating and decaying, and therefore subdominant. 
One then expects,
\begin{align}
    \tilde \Delta_0  (c_s\delta \tfin/R_*\gg 1) & \sim  \frac{1}{2}\ln^2 (\tfin \HH_*)\,.
    \label{Delta0_limit1}
\end{align}
\begin{figure}[t!]
    \centering
    \includegraphics[width=\columnwidth]{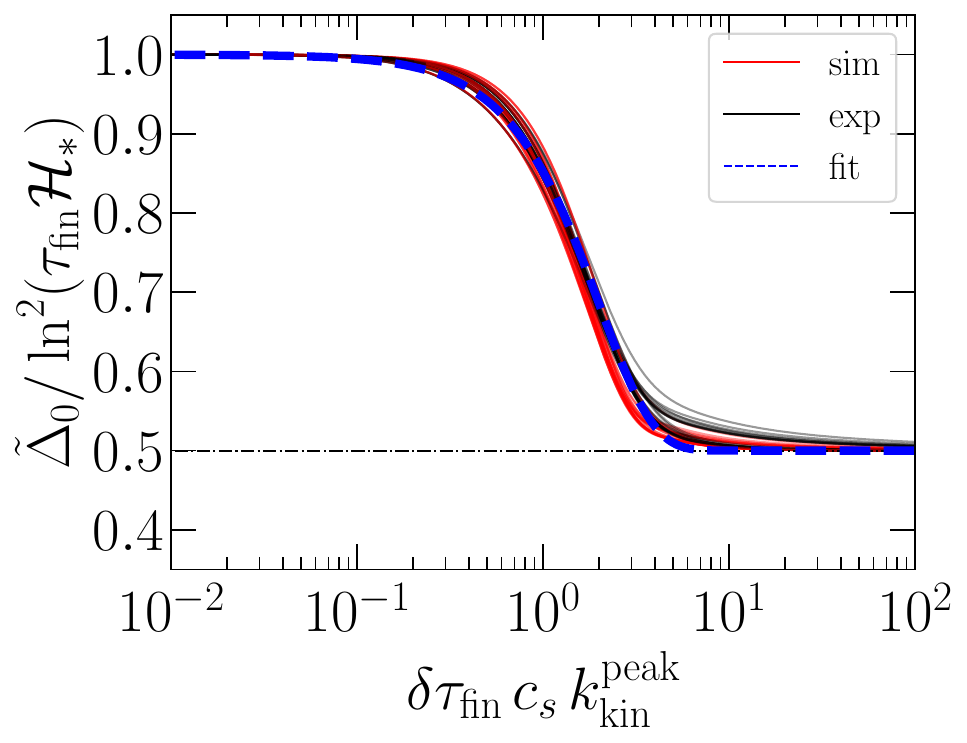}
    \caption{\justifying The function $\tilde \Delta_0$
     compensated by its short-duration limit $\ln^2 (\tfin \HH_*)$.
    $\tilde \Delta_0$ is shown as a function of $\delta \tfin \, \cs \, \kK^{\rm peak}$, where $\kK^{\rm peak} \sim {\cal O} (1/R_*)$ is the spectral peak of the kinetic spectra shown in \Fig{kinetic_spectra} (see dashed lines and values in \Tab{tab:my_label}).
    The blue dashed line corresponds to the empirical fit in \Eq{A_fit}.}
    \label{Delta_0_ratio}
\end{figure}
The asymptotic behavior
at the extremes of the quantity $c_s\delta\tfin R_*$
is confirmed by \Fig{Delta_0_ratio}, showing the function $\tilde \Delta_0$, compensated by the logarithmic dependence $\ln^2 (\tfin \HH_*)$, for the benchmark phase transitions of \Fig{kinetic_spectra}.
One can appreciate
that almost all curves collapse into one, apart from small deviations, which are due to the
specific spectral shape of the kinetic spectra $\zetaK (k)$ around
their peak $\kK^{\rm peak}$ (see~\Fig{kinetic_spectra}).
In all the cases considered, the dependence of the GW amplitude with $\delta \tfin$ given in \Eq{GW_spec_k0_norm} can be expressed as $A \ln^2 (\tfin \HH_*)$, where $A$ monotonically decreases around $\delta \tfin \sim (\cs \kK^{\rm peak})^{-1}$ between its asymptotic values, i.e., from 1 to 0.5, as it can be derived approximately from \Eqss{eq:tildeDelta0approx}{Delta0_limit1}.
The exact variation of the function $A$ at intermediate
$\delta \tfin$ requires numerical computation of \Eq{tilde_Delta0}
for the specific spectral shape
$\zetaK$.
However, we show in \Fig{Delta_0_ratio} that the empirical fit
\begin{equation}
    A \approx \frac{1}{2} \biggl[1 + \exp  \Bigl(- 0.35 \, \bigl[\delta \tfin \, \cs \, \kK^{\rm peak}\bigr]^{1.5} \Bigr) \biggr],
    \label{A_fit}
\end{equation}
gives an accurate estimate for the evaluated phase transitions.

By taking the low-frequency limit $k \to 0$ of the GW
spectrum, we have found that its amplitude 
depends quadratically on the duration of the 
GW source when $\delta \tfin$ is short, compared to the Hubble time,
and it is proportional to $\ln^2 (\tfin \HH_*)$ in general (see \Fig{Delta_0_ratio}).
As previously discussed, this result is in contradiction with the
linear dependence on the source duration usually assumed for the GW spectrum from sound waves, and from stationary processes in general.
We come back to this aspect in \Sec{linear_vs_quadratic}
and extend the discussion to a generic class of stationary
UETC.
In the next \Sec{k3_k9}, we instead analyze the $k$-dependence of the GW spectrum at large scales, and provide insight on the reasons why a $k^9$ behavior is found in Refs.~\cite{Hindmarsh:2019phv,Guo:2020grp}, as opposed to the usual causal $k^3$ scaling given in \Eq{GW_spec_k0_norm}.

\subsection{$k^3$ vs $k^9$ tilt in the low-frequency limit}
\label{k3_k9}

In \Sec{k_to_0_sec},
we have found that the GW
spectrum scales proportional to $k^3$ when $k \rightarrow 0$ [see \Eq{GW_spec_k0_norm}].
The causal $k^3$ branch is, in general,\footnote{This agreement is not always completely clear, since the
numerical studies of the IR regime of the GW spectrum
are computationally challenging, see discussion in Refs.~\cite{RoperPol:2022iel,Jinno:2022mie}.} in agreement
with numerical simulations of sound waves \cite{Hindmarsh:2013xza,Hindmarsh:2015qta,Jinno:2020eqg,Jinno:2022mie,Sharma:2023mao}
and the recent analytical derivation of Ref.~\cite{Cai:2023guc}.
However, as mentioned above, it is in contradiction with
the $k^9$ scaling reported in the sound shell model
\cite{Hindmarsh:2016lnk,Hindmarsh:2019phv,Guo:2020grp}.
To understand the $k^3$ vs $k^9$ discrepancy of the GW spectrum,
we reproduce in this section the calculation of Refs.~\cite{Hindmarsh:2019phv,Guo:2020grp}.
Since Ref.~\cite{Hindmarsh:2019phv} considers
that the duration of the phase transition is short and hence ignores the expansion
of the Universe,\footnote{We note, however, that even if
the duration of the phase transition $\beta^{-1}$ is short with respect to the Hubble time,
the duration
of the GW sourcing from sound waves can
last longer, until the plasma develops
non-linearities or until the sound waves are completely
dissipated \cite{Pen:2015qta,Caprini:2019egz}.} we will consider the limit $\delta\tfin \HH_* \ll 1$
when comparing our results to theirs.

In order to reproduce the calculations of Ref.~\cite{Hindmarsh:2019phv}, we need to
compute the growth rate of $\OmGW$ with the duration of GW production, $\delta \tfin$.
Note that in Ref.~\cite{Hindmarsh:2019phv}, the growth
rate $\dot \Delta$ [see their Eq.(3.38)] is
defined instead as the derivative of $\Delta$ with respect to cosmic time $t$.
We consider this interpretation to be
misleading since $\Delta$,
see \Eq{GG_function}, has been defined after averaging over time and
it is valid {\em only} in the free propagation regime at late times $\tau \gg \tfin$, e.g., at present time $\tau_0$
[see \Eqs{OmGW_full}{eq_OmGW_UETC}].
We show in \App{app_GW_tdep} the correct time-dependence of
$\Delta$ with conformal time during the phase of GW production, $\tau < \tfin$.
We note that using \Eq{GG_kp_stat0} during the sourcing could
lead to wrong results when comparing, for example, to the
results from numerical simulations \cite{Hindmarsh:2013xza, Hindmarsh:2015qta, Hindmarsh:2017gnf, Jinno:2020eqg, Jinno:2022mie}.

As a present-day observable,
we are then interested in the dependence of the GW
spectrum with the source duration $\delta \tfin$, so
we define $\Delta' \equiv \partial_{\tfin} \Delta$.
Note that in the current work, we distinguish $\Delta'$
from $\dot \Delta \equiv \partial_{t_{\rm fin}} \Delta$ since we take into account the expansion of the Universe.

\begin{figure}
    \centering
    \includegraphics[width=1\columnwidth]{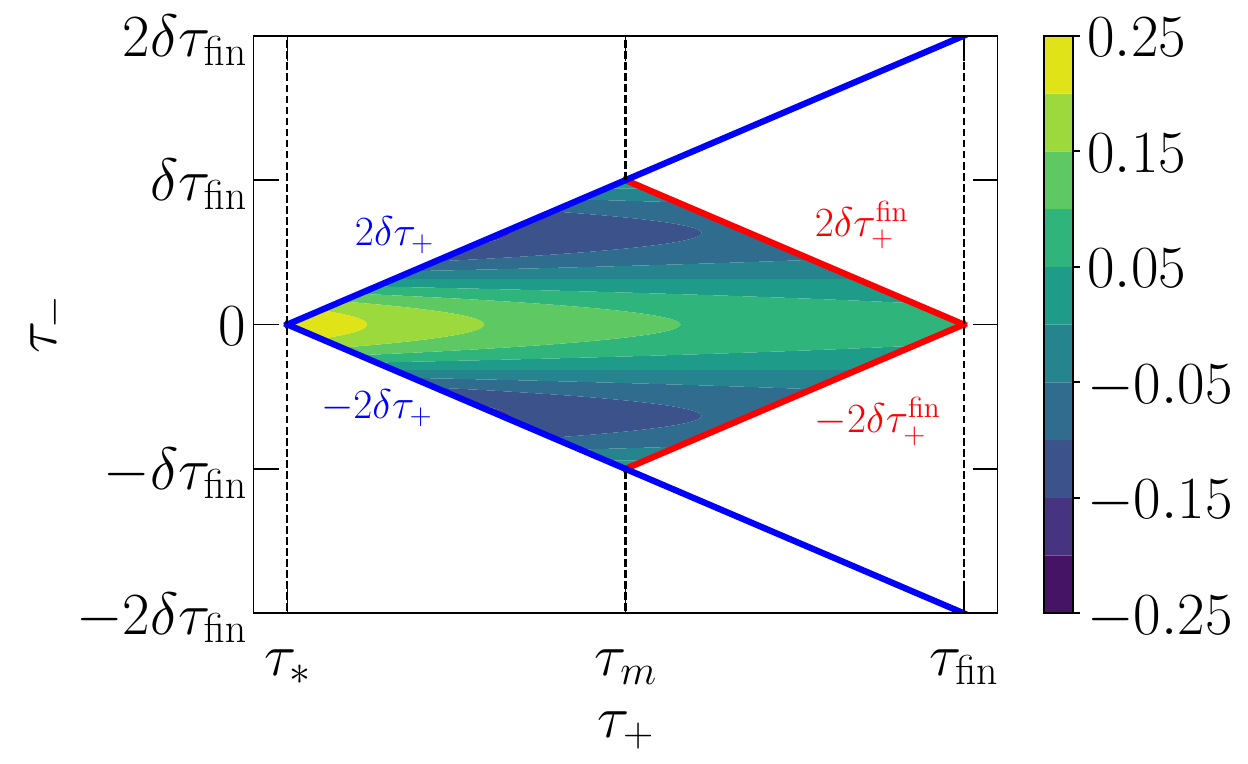}
    \includegraphics[width=1\columnwidth]{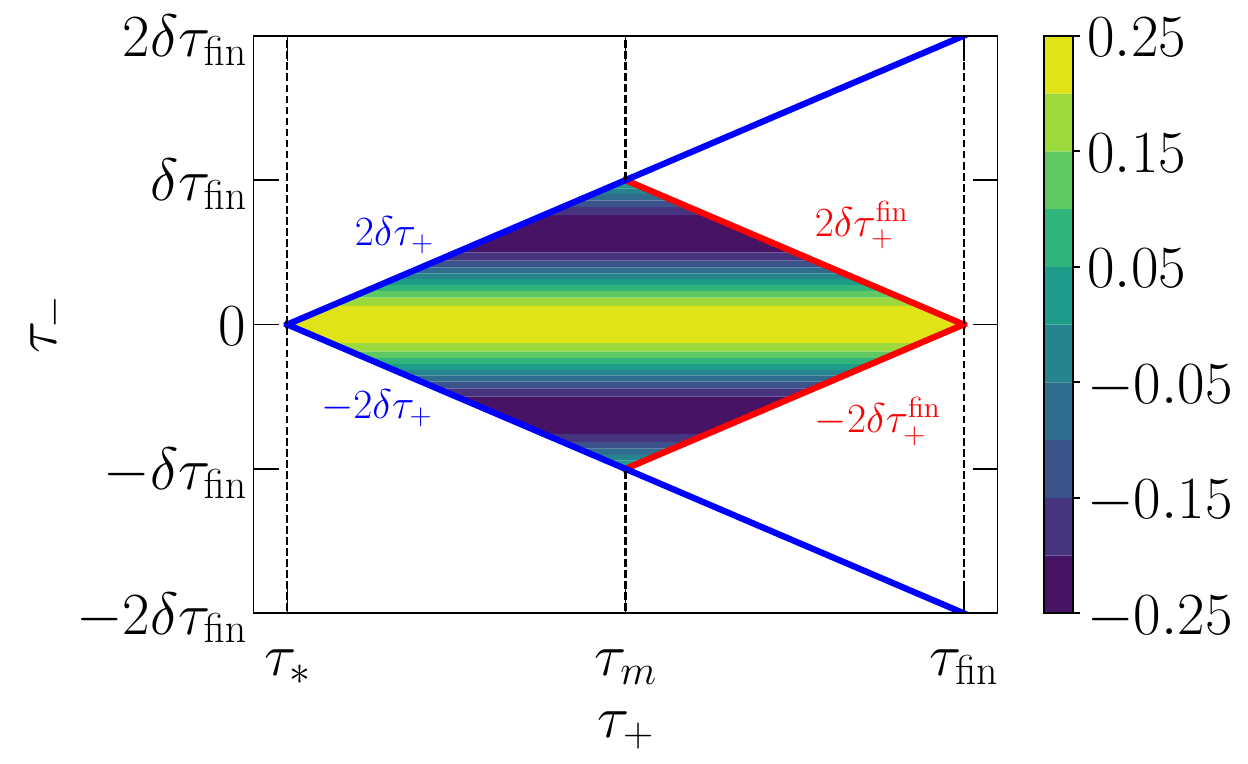}
    \caption{\justifying The limits of integration for the change of variables $\lbrace \tau^{ }_{1,2}\rbrace \to \lbrace \tau^{ }_\pm \rbrace$, with $\tau^{ }_+\equiv (\tau^{ }_1+\tau^{ }_2) / 2$ and $\tau^{ }_-\equiv \tau^{ }_2-\tau^{ }_1$ in the $\tau_{1, 2} \in [\tini, \tfin]$
    range corresponds to the region bounded by the blue and red lines.
    The extension of the blue lines up to $\tfin$ indicates the region considered for the same integration in Ref.~\cite{Hindmarsh:2019phv}, which thus ignores the
    bounds given by the red lines.
    For illustration, we show contour values of the integrand multiplied by $\HH_*^2$ for an expanding (\Eq{Delta_mn_taupm}, upper panel) and a flat (\Eq{integral_tp_tm}, lower panel) Universe, with $\hat p_{mn}/\HH_* = 5$.}
    \label{limits_integral}
\end{figure}

We start by performing the change of variables $\lbrace \tau^{ }_{1,2}\rbrace \to \lbrace \tau^{ }_\pm \rbrace$ in the integral of \Eq{GG_function}, with $\tau^{ }_+\equiv (\tau^{ }_1+\tau^{ }_2) / 2$ and $\tau^{ }_-\equiv \tau^{ }_2-\tau^{ }_1$.
The limits of integration can be found in the following way.
Since $\tau_1, \tau_2 \in [\tini, \tfin]$, one has that $\tau_+ \in [\tini, \tfin]$, and 
\begin{equation}
    \tau_- = 2 (\tau_+ - \tau_1) = 2 (\tau_2 - \tau_+),
\end{equation}
which, since $\tau_1, \tau_2 \in  [\tini, \tfin]$, leads to the limits
\begin{equation}
    \tau_- \in 2 \ [- \delta \tau_+^{\rm fin}, \delta \tau_+] \; \vee \; 
    \tau_- \in 2 \ [- \delta \tau_+, \delta \tau_+^{\rm fin}],
\end{equation}
where we have defined $\delta \tau_+ \equiv \tau_+ - \tini$ and $\delta \tau_+^{\rm fin} \equiv \tfin - \tau_+$.
Combining both limits we see that, when $\tau_+ \leq \tau_m \equiv \half(\tau_* + \tfin)$, the limits of integration for $\tau_-$ are $\tau_- \in 2\, [-\delta \tau_+, \delta \tau_+]$,
and when $\tau_+ > \tau_m$, then $\tau_- \in 2\, [-\delta \tau_+^{\rm fin},
\delta \tau_+^{\rm fin}]$ (see \Fig{limits_integral}).
Hence, the change of variables $\lbrace \tau^{ }_{1,2}\rbrace \to \lbrace \tau^{ }_\pm \rbrace$ in \Eq{GG_function} yields
\begin{align}
    \Delta_{mn}  (\delta & \tfin, \hat p_{mn}) = 
    \int_{\tini^{ }}^{\tfin^{ }} \!\! \frac{\dd \tau^{ }_1}{2 \tau^{ }_1} 
    \int_{\tini^{ }}^{\tfin^{ }} \!\! \frac{\dd \tau^{ }_2}{2 \tau^{ }_2} \cos (\hat p_{mn} \tau_-) \nonumber \\
    = &\, \int_{\tini}^{\tau_m} \dd \tau_+ \int_{-2 \delta \tau_+}^{ 2\delta \tau_+}
    \frac{\cos (\hat p_{mn} \tau_-)}{4 \tau_+^2 - \tau_-^2} \dd \tau_-
     \nonumber \\ 
     + &\, \int_{\tau_m}^{\tfin} \dd \tau_+ \int_{-2\delta \tau_+^{\rm fin}}^{2 \delta \tau_+^{\rm fin}}
    \frac{\cos (\hat p_{mn} \tau_-)}{4 \tau_+^2 - \tau_-^2} \dd \tau_- \ .
    \label{Delta_mn_taupm}
\end{align}

In particular, if we ignore the expansion of the Universe, we can set $\tau_* \to 0$ and take $\tau_1 \tau_2 \approx 1/\HH_*^2$ in \Eq{Delta_mn_taupm}, such that $\Delta_{mn}$
becomes
\begin{align}
     4\, \Delta_{mn}^{\rm flat}  & (\delta \tfin, \hat p_{mn})  = \nonumber \\
     &\, \HH_*^2  \int_{0}^{\half \tfin} \dd \tau_+ \int_{-2 \tau_+}^{2\tau_+} \hspace{-2mm}
    \cos (\hat p_{mn} \tau_-)
    \dd \tau_- \nonumber \\
    + &\, \HH_*^2\int_{\half \tfin}^{\tfin}  \dd \tau_+ \int_{-2\delta\tau_+^{\rm fin}}^{ 2\delta\tau_+^{\rm fin}} \cos (\hat p_{mn} \tau_-)
    \dd \tau_-.
    \label{integral_tp_tm}
\end{align}

\FFig{limits_integral} shows the values of the integrand in \Eqs{Delta_mn_taupm}{integral_tp_tm} as a function of $\tau_\pm$.
Ignoring the expansion of the Universe, the integrand is
constant in $\tau_+$ and only depends on $\tau_-$ as $\cos(\hat p_{mn} \tau_-)$.
 
If we compare this integral with the one computed in Ref.~\cite{Hindmarsh:2019phv} [see their Eq.~(3.36)], we find that the limits
of the integral are taken to be $\tau_+ \in [0, \tfin]$
and $\tau_- \in [-2\tau_+, 2\tau_+]$.
This corresponds to integrating over $\tau_-$ according to the
blue limits in \Fig{limits_integral} in all the range $\tau_+ \in [0, \tfin]$, hence including the areas of
integration that are not allowed, limited by the red lines.
The inclusion of the upper and lower right triangles, out
of the limits denoted by the red lines, leads to $\tau_2 > \tfin$ and $\tau_1 > \tfin$, respectively.
Using these limits of integration,
the explicit dependence of the limits of the integral
over $\tau_-$ on the source duration $\tfin$ is ignored,
leading to the wrong value of $\Delta'$, as we show below.

We now compute the growth rate $\Delta'$ from \Eq{Delta_mn_taupm},\footnote{The derivative can be taken from the
the integral over $\tau_{1,2}$ or from the integral over $\tau_{\pm}$.
The dependence of the integration
limits on $\tfin$ is simpler in the former case after using the correct limits (see \Fig{limits_integral}) but both computations
lead to the same result.}
\begin{align}
    \Delta'_{mn}  (\delta \tfin, \hat p_{mn}) &\,
    \nonumber \\ =
    \frac{1}{2 \,\tfin}  \Bigl[&\,\cos \bigl(\hat p_{mn} \tfin
    \bigr) \Delta\Ci (\tfin, \hat p_{mn})
    \nonumber \\
     + &\, \sin \bigl(\hat p_{mn} \tfin
    \bigr) \Delta \Si (\tfin, \hat p_{mn}) \Bigr] \ ,
    \label{GG_dot_general}
\end{align}
which can also be directly found from \Eq{GG_kp_stat}.
Ignoring the expansion of the Universe we get, from 
either \Eq{G_func_flat} or \Eq{integral_tp_tm},
\begin{equation}
    \Delta_{mn}^{\rm flat '} (\delta \tfin, \hat p_{mn}) =
    \HH_* \, \frac{\sin \bigl(\hat p_{mn} \delta \tfin \bigr)}{2 \, (\hat p_{mn}/\HH_*)}.
    \label{Delta_prime}
\end{equation}
If one omits the dependence on $\tfin$ in the integration limits over $\tau_-$ in \Eq{integral_tp_tm}, the solution to Eq.~(3.38)
of Ref.~\cite{Hindmarsh:2019phv} is found, which is equivalent
to \Eq{Delta_prime} with an extra factor of 2 in the $\sin$ function, $\sin(2 \hat p_{mn} \delta \tfin)$.

\begin{figure*}[t]
    \centering
    \includegraphics[width=.75\columnwidth]{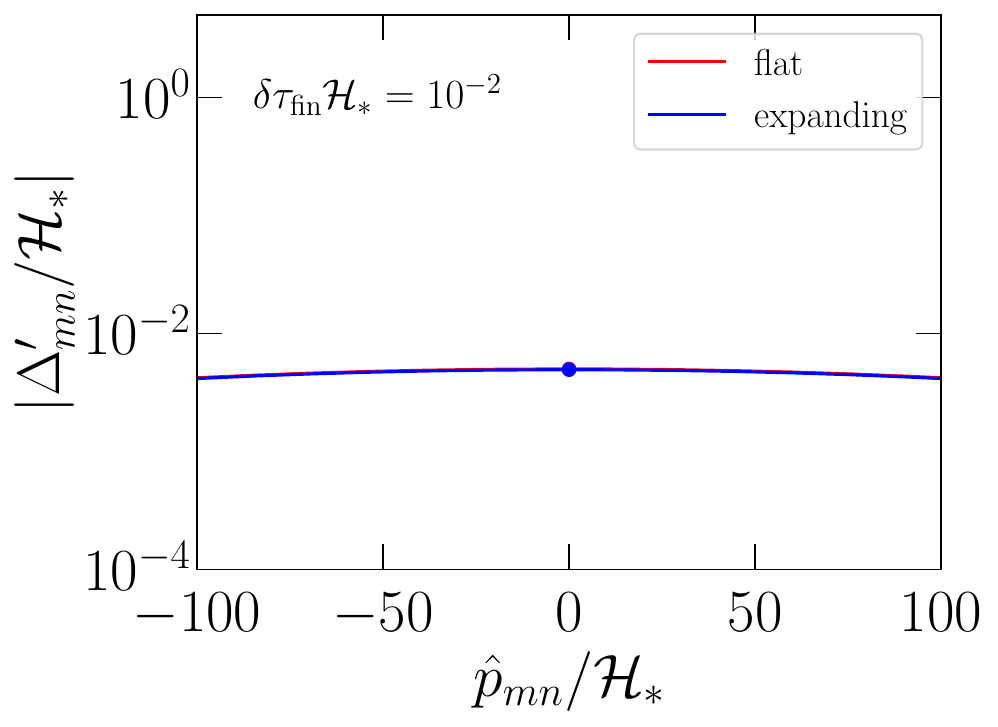}
    \includegraphics[width=.75\columnwidth]{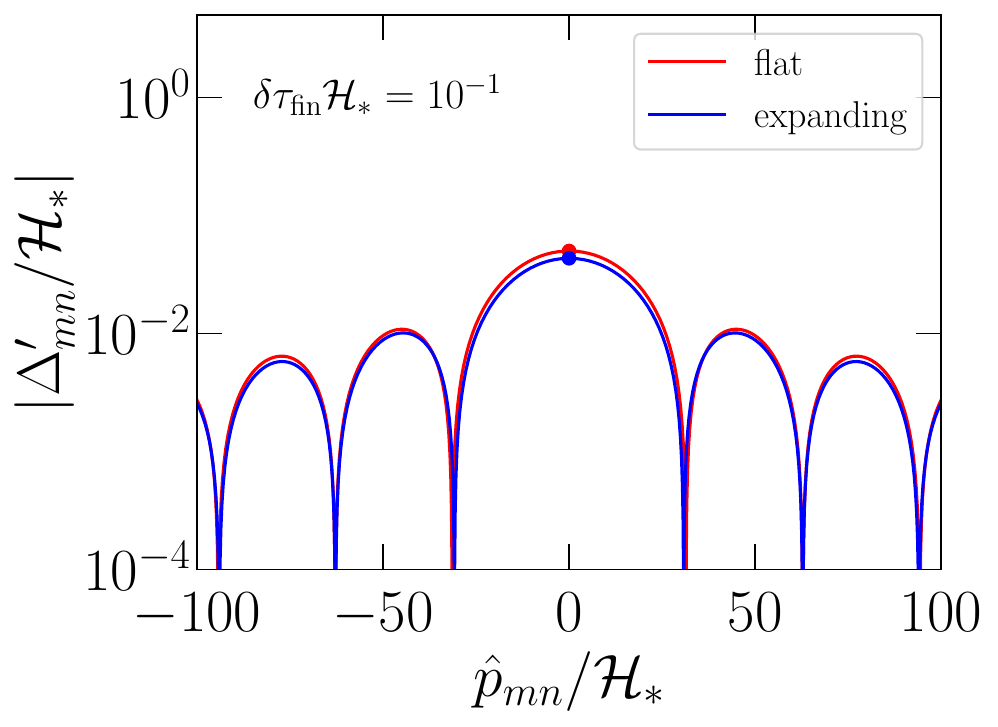}
    \includegraphics[width=.75\columnwidth]{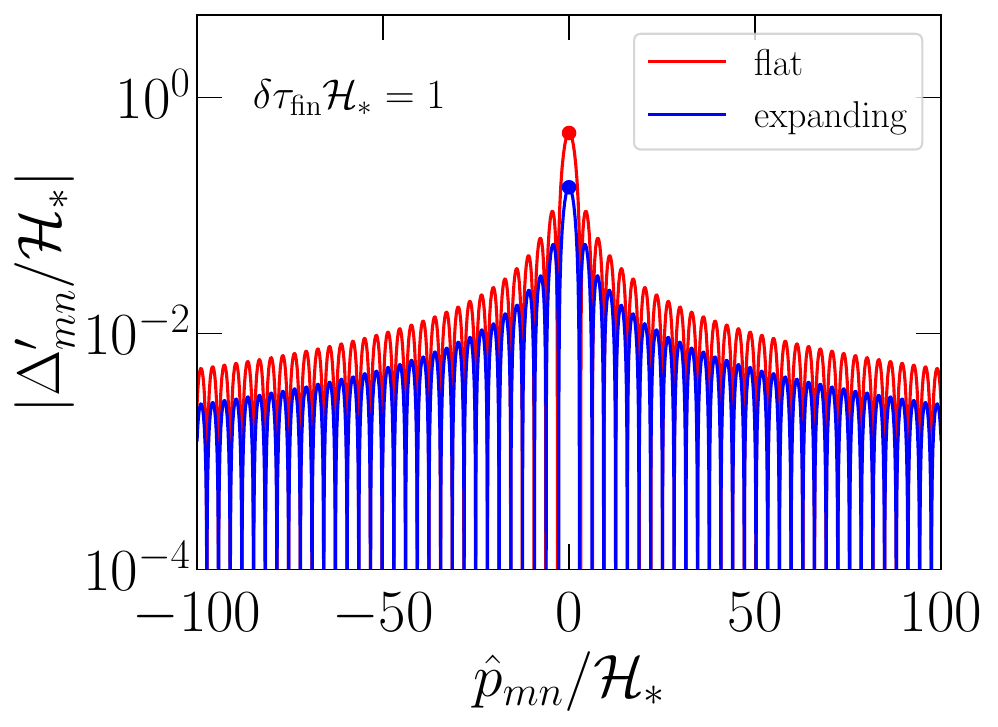}
    \includegraphics[width=.75\columnwidth]{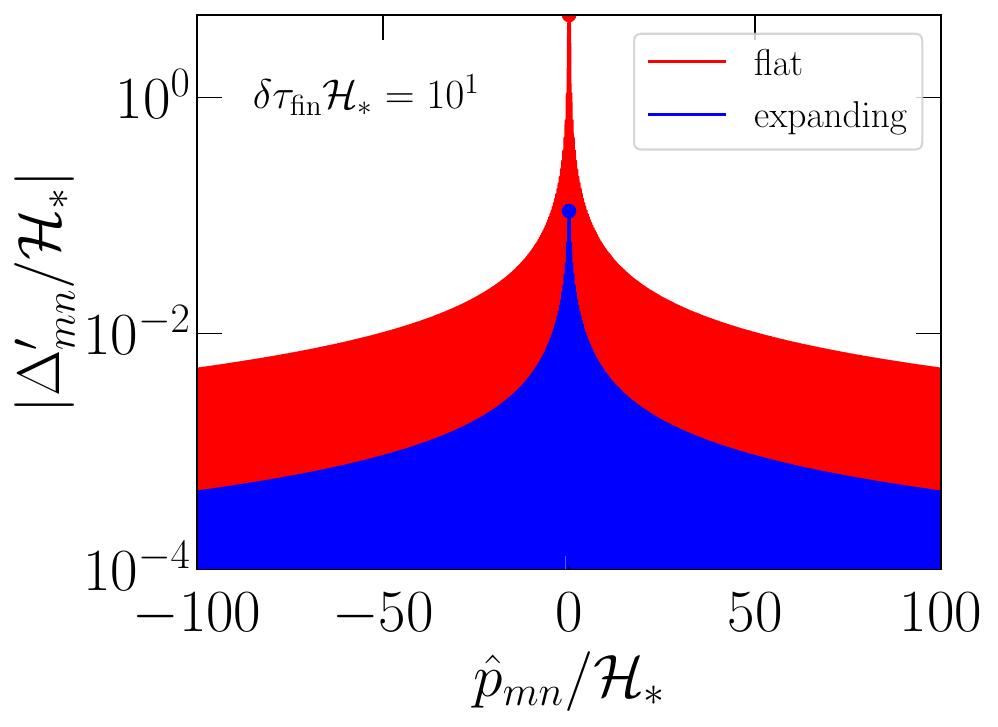}
    \caption{\justifying Function $|\Delta'_{mn}|/\HH_*$ for an expanding [blue, \Eq{GG_dot_general}] and a flat [red, \Eq{Delta_prime}] Universe
    as a function of $\hat p_{mn}/\HH_*$ for different GW sourcing
    duration
    $\delta \tfin \HH_* \in \{10^{-2}, 10^{-1}, 1, 10\}$.
    }
    \label{dotGG_vs_t}
\end{figure*}

\FFig{dotGG_vs_t} shows the dependence of the growth rate $\Delta'_{mn}$, given in \Eqs{GG_dot_general}{Delta_prime},
on the combined momenta $\hat p_{mn}$ for
different values of the GW source duration $\delta \tfin$.
We observe that, as $\delta \tfin$ increases, $\Delta'_{mn}$ becomes
more confined around $\hat p_{mn} \rightarrow 0$.
Taking into account the relation between the sinc
and the Dirac $\delta$ function,
\begin{equation}
    \delta (x) = \lim_{a \to 0} \frac{\sin (\pi x/a)}{\pi x},
\end{equation}
Ref.~\cite{Hindmarsh:2019phv} approximates \Eq{Delta_prime}
in the $1/\delta \tfin \to 0$ limit, i.e., for large GW duration,\footnote{\EEq{Delta_prime_lim} is equivalent
to Eq.~(3.39) in Ref.~\cite{Hindmarsh:2019phv} after taking
into account the extra factor of 2 [see text below \Eq{Delta_prime}] and that their $\Delta$
is defined with an extra $\half$ factor [see their Eq.~(3.36) compared to \Eq{GG_function}].}
\begin{equation}
    \lim_{\delta \tfin \HH_* \to \infty} \Delta_{mn}^{\rm flat '} (\hat p_{mn}) = \HH_* \frac{\pi}{2} \delta \bigl(\hat p_{mn}/\HH_* \bigr).
    \label{Delta_prime_lim}
\end{equation}

This approximation is used in Refs.~\cite{Hindmarsh:2019phv,Guo:2020grp} to simplify the calculation of the
integral in \Eq{OmGW_GG}.
However, it is not required to compute the GW amplitude,
as we have done in \Sec{k_to_0_sec} in the $k \to 0$ limit and
we extend in \Sec{numerical_GW} to all $k$.
We show in the 
following that it is precisely this assumption the
one that leads to the linear growth with the source duration
and the $k^9$ scaling of the GW spectrum when $k \to 0$.
We also show in \Sec{linear_vs_quadratic} that this assumption is equivalent to the one usually taken for stationary
processes that decay very quickly with the time difference $\tau_-$ \cite{Kraichnan:1965zz,Kosowsky:2001xp,Gogoberidze:2007an,Caprini:2009fx,Caprini:2009yp,Niksa:2018ofa,Auclair:2022jod}.
However, the UETC found in the sound shell model is a periodic function in $\tau_-$ [see \Eq{eq_EK_Eu_st}] so this assumption is, in general, not justified.

On the other hand, when $k$ is large
and oscillations over $\tau_-$ become very rapid,
this assumption might become justified.
In such circumstances, 
as we show in \Sec{numerical_GW}, the expression computed in \App{HH19}, based on this approximation, can describe the GW spectrum in the regime
$k \gg 1/\delta \tfin$ and, in particular, around the spectral peak
if $\delta \tfin/R_* \gg 1$.
One can understand this by noting that the limit leading to \Eq{Delta_prime_lim}
is equivalent to considering $\hat p_{mn} \delta \tfin \to \infty$.
At low and moderate $k$, in general, this limit does not hold, since $p$
and $\tilde p$ are integrated from 0 to $\infty$.
However, when $k \delta \tfin \to \infty$, this assumption is valid,
since $\Delta_{mn}$ is symmetric in $\hat p_{mn}$ and then
$\hat p_{mn} \delta \tfin \to \infty$.

We note that this approximation is {\em only} valid when $k \delta \tfin$ becomes sufficiently large,
not when $\tau$ is large, since $\Delta'$ is the growth
with respect to $\delta \tfin$.
The assumption of asymptotically large $\delta \tfin$
is not justified for GW production from sound waves and it
is in contradiction with the assumption that the
expansion of the Universe can be ignored, so expansion becomes
relevant in this regime.

In general, we find that $\Delta'_{mn}$
is widely spread along a broad range of $\hat p_{mn} \neq 0$
for short and moderate (around one Hubble time) duration (see \Fig{dotGG_vs_t}).
Its maximum value at $\hat p_{mn} = 0$ is $\half \, \delta \tfin \HH_*$, as can be inferred from \Eq{Delta_prime}.
For longer sourcing duration, one can no longer ignore the expansion
of the Universe and we find that the growth rate at $\hat p_{mn} = 0$ decreases to $\half \, \ln (\tfin \HH_*)/\tfin$
(see blue and red dots in \Fig{dotGG_vs_t}).
Therefore, the integral over $p$ and $z$ in \Eq{OmGW_GG}
includes non-negligible contributions from $\hat p_{mn} \neq 0$
that are being ignored if one uses \Eq{Delta_prime_lim}.

We now explicitly show how this approximation affects the limit $k \to 0$ of the GW spectrum, computed in \Sec{k_to_0_sec}.
Denoting $\OmGW' \equiv \partial_{\tfin} \OmGW$ as the growth rate of the GW spectrum and
using \Eqs{GW_spec_k0_norm}{Delta_prime_lim}, we find,
\begin{align}
    \lim_{K\rightarrow 0} \OmGW' & (\delta\tfin, K) = \frac{8 \pi}{5} \,
     R_* \, \bar w^2 \, K^3 \, {\cal T}_{\rm GW} \biggl(\frac{\OmK}{\cal K}\biggr)^2
    \nonumber \\ \times &\, \int_0^\infty \frac{\zetaK^2(P)}{P^2} \delta(K - 2 P \cs) \dd P,
    \label{HH19_EK}
\end{align}
where, following Ref.~\cite{Hindmarsh:2019phv}, we have further assumed
that $\hat p_{mn}$ only cancels when $m = 1$ and $n = -1$,
and $\Delta_0' \to \half \HH_* \pi \, \delta(2 \cs p - k) = \half (\HH_* R_*) \pi \, \delta(2 \cs P - K)$.
We note that this additional assumption does not take into account
the case $m = -1$, such that $p + m \tilde p = 0$, which always holds
when $k \to 0$.
From \Eq{Delta_0_p_flat}, one can see that the $m = -1$ case
would include in $\OmGW'$ a linear term in $\delta \tfin$ that would lead to the quadratic scaling
and a function proportional to $k^3$ when $k \to 0$ that would dominate
over the $k^9$ term.
Therefore, the $k^9$ scaling appears due to the inclusion of a $k$ 
dependence in the integral over $p$ of \Eq{HH19_EK} and due to neglecting
the leading-order term when $k \to 0$.
The extension of \Eq{HH19_EK} to all values of $k$ is shown in \App{HH19}.

The integral in \Eq{HH19_EK} is directly computed by substituting $P = K/(2 \cs)$,
\begin{align}
    \lim_{K\rightarrow 0}   &\, \OmGW' (\delta\tfin, K) =
    \nonumber \\ &\, \frac{32 \pi}{5} \, \cs^2 \, R_* \, \bar w^2 \,
    K \, {\cal T}_{\rm GW} \biggl(\frac{\OmK}{\cal K}\biggr)^2  \, \zetaK^2 (K).
    \label{OmGW_prime}
\end{align}

Therefore, we find that the GW spectrum in the $k \to 0$
regime is proportional to $K \zetaK^2(K)$.
For irrotational fields, $\zetaK \sim K^a$ with $a \geq 4$ (see \Sec{UETC_velfield}) and,
for
the kinetic spectra of the benchmark phase transitions of \Fig{kinetic_spectra}, we find $a = 4$.
Therefore, one finds that the GW spectrum is proportional to $K^{2a + 1} = K^9$ in this case, as argued in Ref.~\cite{Hindmarsh:2019phv}.
As discussed above, this result is a consequence of the
assumption that the growth rate $\Delta'$ can be approximated as
a Dirac $\delta$ function [see \Eq{Delta_prime_lim}].
The calculation using the stationary UETC found in the sound shell
model [see \Eq{eq_EK_Eu_st}] in the $k \to 0$ limit
has been presented in \Sec{k_to_0_sec}, where
we recover the low-frequency scaling with $k^3$ as expected by causality [see \Eq{GW_spec_k0_norm}].
We note that this result also holds when one takes into
account the expansion of the Universe.

\section{GW production from stationary processes}
\label{linear_vs_quadratic}

In \Secs{GW_SSM}{k_to_0_sec}, we have shown that the dependence
of the GW amplitude in the $k\to 0$ limit with the source duration is $\ln^2 (\tfin \HH_*)$, which becomes
quadratic when the duration is short.
In addition, we have shown in \Sec{k3_k9} that the approximation of the growth rate $\Delta'$, given in \Eqs{GG_dot_general}{Delta_prime}, as a Dirac $\delta$
function [see \Eq{Delta_prime_lim}], taken in Refs.~\cite{Hindmarsh:2019phv,Guo:2020grp}, leads to the
conclusion that the GW spectrum is proportional to $k^9$ in the $k \to 0$ limit.
We have found that this scaling is actually $k^3$ as expected from causality and found in numerical studies.
In addition, from \Eq{OmGW_prime} we directly find that
since $\OmGW'$ does not depend on $\tfin$, then $\OmGW = \delta \tfin \, \OmGW$, which corresponds to the assumed
linear growth with the source duration.
Hence, this result is also a consequence of
the aforementioned assumption, which does not hold in the $k \to 0$
limit.
We note that this is not necessarily the case at all $k$, however,
as we show in \Sec{numerical_GW}, it can give an accurate estimate of the GW amplitude at $k \gg 1/\delta \tfin$.

To understand the transition from the quadratic
to the linear growth of $\OmGW$ with $\delta \tfin$ as $k$ increases,
let us now generalize our study to a velocity UETC described by
an arbitrary stationary process,
$\EK(\tau_1, \tau_2, k) = \EK(k) \, f(\tau_-, k)$, where $f(\tau_-, k) = \cos (k \cs \tau_-)$ in the sound shell model.
In the general case, the function $\Delta$ in \Eq{GG_function} is
\begin{align}
    \Delta (\delta \tfin, &\,k, p, \tilde p) =
    \int_{\tini^{ }}^{\tfin^{ }} \!\! \frac{\dd \tau^{ }_1}{\tau^{ }_1} 
    \int_{\tini^{ }}^{\tfin^{ }} \!\! \frac{\dd \tau^{ }_2}{\tau^{ }_2} \nonumber \\
    &\, \times
    f (\tau_-, p) f (\tau_-, \tilde p) \,
    \cos (k\tau^{ }_-) \, .
    \label{Delta_stationary_UETC}
\end{align}
Following Ref.~\cite{Caprini:2009fx}, we take the change
of variable $\tau_2 \to \tau_-$,
\begin{align}
    \Delta (\delta \tfin, &\,k, p, \tilde p) =
    \int_{\tini^{ }}^{\tfin^{ }} \!\! \frac{\dd \tau^{ }_1}{\tau^{ }_1} 
    \int_{\tini^{ } - \tau_1}^{\tfin^{ } - \tau_1} \!\! \frac{\dd \tau^{ }_-}{\tau_- + \tau_1} \nonumber \\
    &\, \times
    f (\tau_-, p) f (\tau_-, \tilde p) \,
    \cos (k\tau^{ }_-) \,.
    \label{Delta_general_uetc}
\end{align}

The characteristic linear growth of stationary processes \cite{Kosowsky:2001xp,Gogoberidze:2007an,Hindmarsh:2013xza,Hindmarsh:2019phv,Guo:2020grp}
is found when inverting the order of integration in \Eq{Delta_general_uetc} is allowed \cite{Caprini:2009fx}.
This is justified if the function $f(\tau, k)$ becomes negligibly small in the range $\tau < \tini - \tau_1$
and $\tau > \tfin - \tau_1$ for all $\tau_1 \in (\tini, \tfin)$, such that the integral over $\tau_-$ can be
extended to $\tau_- \in (-\infty, \infty)$ and the limits
of integration do not depend any longer on $\tau_1$ \cite{Caprini:2009fx}.
This condition can be justified, for example, when the UETC decays as
a Gaussian function (e.g., Kraichan decorrelation
\cite{Kraichnan:1965zz}) as we show in \Sec{sec_kraichnan}.
On the other hand, when $f(\tau, k)$ is a periodic function (e.g., the UETC found in the sound shell model) this condition is, in general, unjustified,
unless $f$ becomes sufficiently oscillatory in $\tau_-$. 
This is the case in the $k \tau_- \to \infty$ limit,
where the limits of integration already include several oscillations, so that extending the limits to $\pm \infty$ does not
affect drastically the result of the integral.
This approximation holds in the regime assumed in
Ref.~\cite{Hindmarsh:2019phv}, $k \delta \tfin \to \infty$ (see discussion in \Sec{k3_k9}).
Under this assumption, we find
\begin{align}
    \Delta (\delta \tfin, &\,k, p, \tilde p) =
    \int_{\tini^{ }}^{\tfin^{ }} \!\! \frac{\dd \tau^{ }_1}{\tau^{ }_1} 
    \int_{-\infty}^{\infty} \frac{\dd \tau^{ }_-}{\tau_- + \tau_1} \nonumber \\
    &\, \times
    f (\tau_-, p) f (\tau_-, \tilde p) \,
    \cos (k\tau^{ }_-) \,.
    \label{Delta_stat}
\end{align}
In particular, if one ignores the expansion of the Universe,
the integral over $\tau_1$
directly yields the linear dependence with $\delta \tfin$,
\begin{align}
    \Delta^{\rm flat} & (\delta \tfin, k, p, \tilde p) = \HH_*^2 \delta \tfin  \nonumber \\ \times &\,
    \int_{-\infty}^{\infty} \dd \tau_- 
    f(\tau_-, p) f(\tau_-, \tilde p) \cos (k \tau_-) \ .
    \label{Delta_flat_stat}
\end{align}

\subsection{Sound-shell model UETC}

When we use the UETC found in the sound shell model [see \Eq{eq_EK_Eu_st}],
the solution to \Eq{Delta_flat_stat} is
\begin{align}
    \Delta^{\rm flat}_{mn}  (\delta \tfin, \hat p_{mn}) = &\, \frac{\HH_*^2 \delta \tfin}{4}
    \int_{-\infty}^{\infty} 
    \cos (\hat p_{mn} \tau_-) \dd \tau_- \nonumber \\
    = &\, \frac{\pi}{2}
    \delta \tfin \HH_* \, \delta \bigl(\hat p_{mn}/\HH_*\bigr),
\end{align}
which is equivalent to \Eq{Delta_prime_lim}.
Therefore, we find that, as mentioned above, the assumption
to find \Eq{Delta_flat_stat} and the one used in Ref.~\cite{Hindmarsh:2019phv} to find \Eq{Delta_prime_lim} lead to the same result.

Including the expansion of the Universe, there is still a dependence on $\tau_1$ in the integral over $\tau_-$ in
\Eq{Delta_stat}.
With the change of variables $\{\tau_{1,2}\} \to \{\tau_{\pm}\}$, the term due to the Universe expansion is
$\tau_1 \tau_2 = \tau_+^2 - \fourth \tau_-^2$ [see \Eq{Delta_mn_taupm}].
In Ref.~\cite{Guo:2020grp} the term $\tau_1 \tau_2$ is approximated as $\tau_1 \tau_2 \sim \tau_+^2$ [see their Eq.~(5.22)].
This
is equivalent\footnote{\label{fn_int}
Reference~\cite{Guo:2020grp} uses an integral equivalent to \Eq{Delta_mn_taupm} with an inverted order of integration,
\begin{align}
    \Delta  & (\delta \tfin, k, p, \tilde p) = 
    \nonumber \\ 
    = 2 &\, \int_{0}^{\delta \tfin} \dd \tau_- \int_{\tini + \half \tau_-}^{\tfin - \half \tau_-}
    \frac{f(\tau_-, p) f(\tau_-, \tilde p) \cos (k \tau_-)}{\tau_+^2 - \fourth \tau_-^2} \dd \tau_+ \ ,
    \nonumber
\end{align}
where the limits of integration are shown in \Fig{limits_integral}, and
we have used the change of variable $\tau_- \to - \tau_-$ in the
range $\tau_- \in (-\delta \tfin, 0)$ to find the same integral as the one in the range $(0, \delta \tfin)$.
We find \Eq{Delta_stat_supp} by taking the limits
over $\tau_+$ to $(-\infty, \infty)$ and neglecting
$\tau_-^2$ compared to $4 \tau_+^2$.}
to the omission of the dependence on $\tau_-$ in the term $1/(\tau_- + \tau_1)$ of \Eq{Delta_stat}, yielding
\begin{align}
    \Delta  & (\delta \tfin, k, p, \tilde p) =
    \HH_* \Upsilon(\delta \tfin) \nonumber \\
    &\, \times \int_{-\infty}^{\infty} \dd \tau^{ }_-
    f (\tau_-, p) f (\tau_-, \tilde p) \,
    \cos (k\tau^{ }_-) \, ,
    \label{Delta_stat_supp}
\end{align}
where $\Upsilon$ is the suppression factor defined in Ref.~\cite{Guo:2020grp} and used in recent literature to
account for the expansion of the Universe in the GW 
production from sound waves \cite{Gowling:2022pzb,Gowling:2021gcy,Hindmarsh:2020hop},
\begin{equation}
    \Upsilon (\delta \tfin) = \int_{\tini}^{\tfin}
    \frac{\dd \tau_1}{\HH_* \tau_1^2} = 1 - \frac{1}{\tfin \HH_*}.
    \label{supp_fac}
\end{equation}
This function
reduces to the linear growth $\Upsilon \to \delta \tfin \HH_*$ in the limit $\delta \tfin \HH_* \ll 1$, yielding
\Eq{Delta_flat_stat} in the case of a flat (non-expanding) Universe.
Again, substituting the UETC of \Eq{eq_EK_Eu_st} in \Eq{Delta_stat_supp}, one finds
\begin{equation}
    \Delta_{mn} (\delta \tfin, \hat p_{mn}) = \frac{\pi}{2}
    \Upsilon(\delta \tfin) \, \delta \bigl(\hat p_{mn}/\HH_*\bigr).
    \label{delta_Upsilon}
\end{equation}

The results presented above are justified only in the asymptotic limit $k \delta \tfin \to \infty$, since this is the limit of validity of the
assumptions introduced to invert the order of integration
over $\tau_1$ and $\tau_-$ (or over $\tau_+$ and $\tau_-$). 
In particular, these assumptions imply that the dependence of $\Delta$
on $\delta \tfin$ is encoded solely in the suppression factor $\Upsilon$ [see \Eq{Delta_stat_supp}], which, in the limit of a short GW source, is linear, $\Upsilon\sim\delta \tfin \HH_*$.

The calculation of the integral
over $\tau_1$ and $\tau_2$, performed in \Sec{k_to_0_sec} in the $k \to 0$ limit without any simplifying assumptions,
leads, instead to a dependence with $\tfin$
characterized by $\tilde \Delta$.
This function is given in \Eq{tilde_Delta0} in the $k \to 0$ limit
and even though
it depends on the spectral
shape, it is found to always be
\begin{equation}
    \tilde \Delta_0 (\delta \tfin) \simeq A \ln^2 (\tfin \HH_*),
    \label{lntfin2}
\end{equation}
where $A \in [0.5, 1]$ [see \Fig{Delta_0_ratio} and \Eq{A_fit}].
Moreover,
\Eq{lntfin2} reduces to $(\delta \tfin \HH_*)^2$
when $\delta \tfin$ is short.
Its extension to all $k$ is studied in \Sec{numerical_GW}, where
we find that, when $k \gg 1/\delta \tfin$, the suppression
factor $\Upsilon$ can be found and if, in addition,
the peak is in this regime ($\delta \tfin/R_* \gg 1$),
then it is relevant to describe the GW spectrum around
its peak.

\subsection{Kraichnan decorrelation}
\label{sec_kraichnan}

Let us consider a stationary process, described by a function $f(\tau_-, k)$, that
does not decay fast enough in $\tau_-$ out of the integration
limits in \Eq{Delta_general_uetc}, and does not include many periodic oscillations within the integration limits.
We have argued that, in this case, 
the GW amplitude grows quadratically
with $\delta \tfin$.
To understand this result,
we study the Kraichan decorrelation \cite{Kraichnan:1965zz},
usually applied to the study of turbulence 
\cite{Kosowsky:2001xp,Gogoberidze:2007an,Caprini:2009yp,Niksa:2018ofa,Auclair:2022jod}, where $f$ is a Gaussian function of $\tau_-$,
\begin{equation}
    f(\tau_-, k) = \exp \Bigl(-\half k^2 v_{\rm sw}^2 \tau_-^2\Bigr),
    \label{Kraichnan}
\end{equation}
where $v_{\rm sw} (\tau_1, \tau_2, k)$ is the sweeping velocity \cite{Kraichnan:1965zz}.
We note that this function is
a positive definite kernel only if $v_{\rm sw}$ is a function of 
$\tau_{1,2}$, breaking the stationary assumption \cite{Auclair:2022jod}, 
and otherwise it is not an adequate description of the velocity
field UETC \cite{Caprini:2009yp}.
However, since we want to address the importance of the
aforementioned assumptions for a generic
stationary process qualitatively
in the current work, we use \Eq{Kraichnan} with a time-independent
$v_{\rm sw}$ for simplicity.

\begin{figure}
    \centering
    \includegraphics[width=\columnwidth]{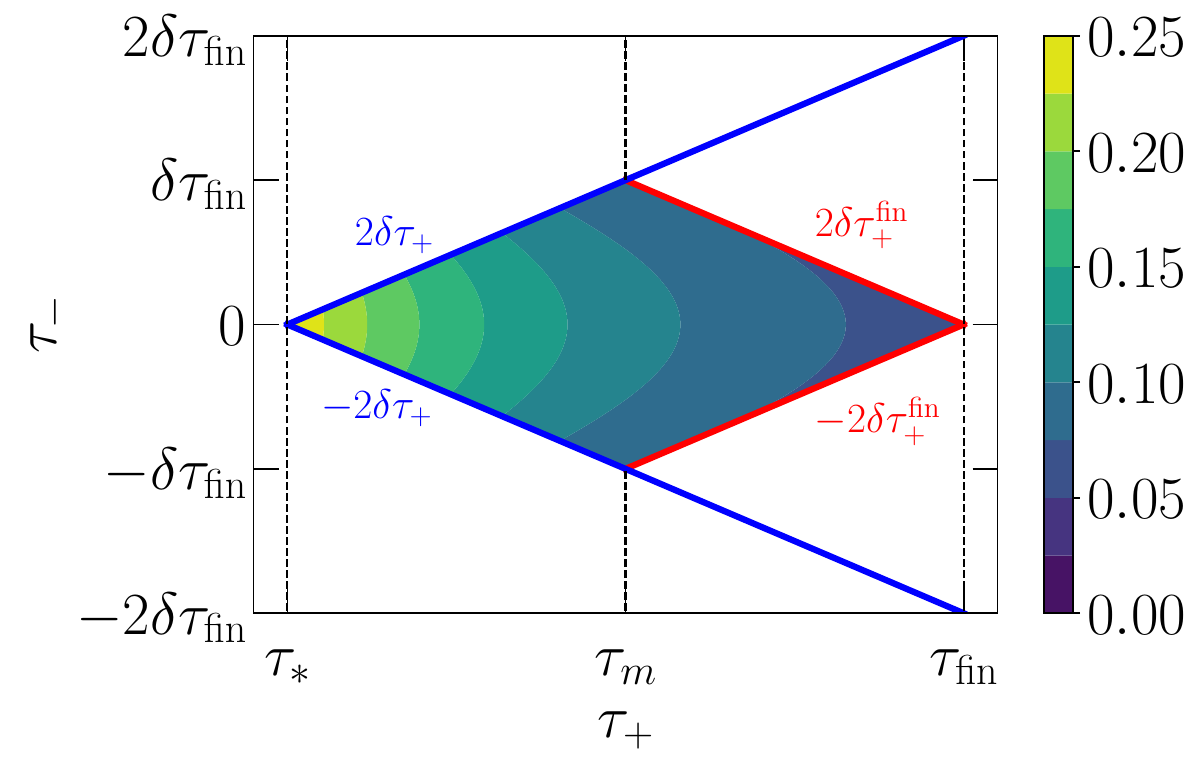}
    \includegraphics[width=\columnwidth]{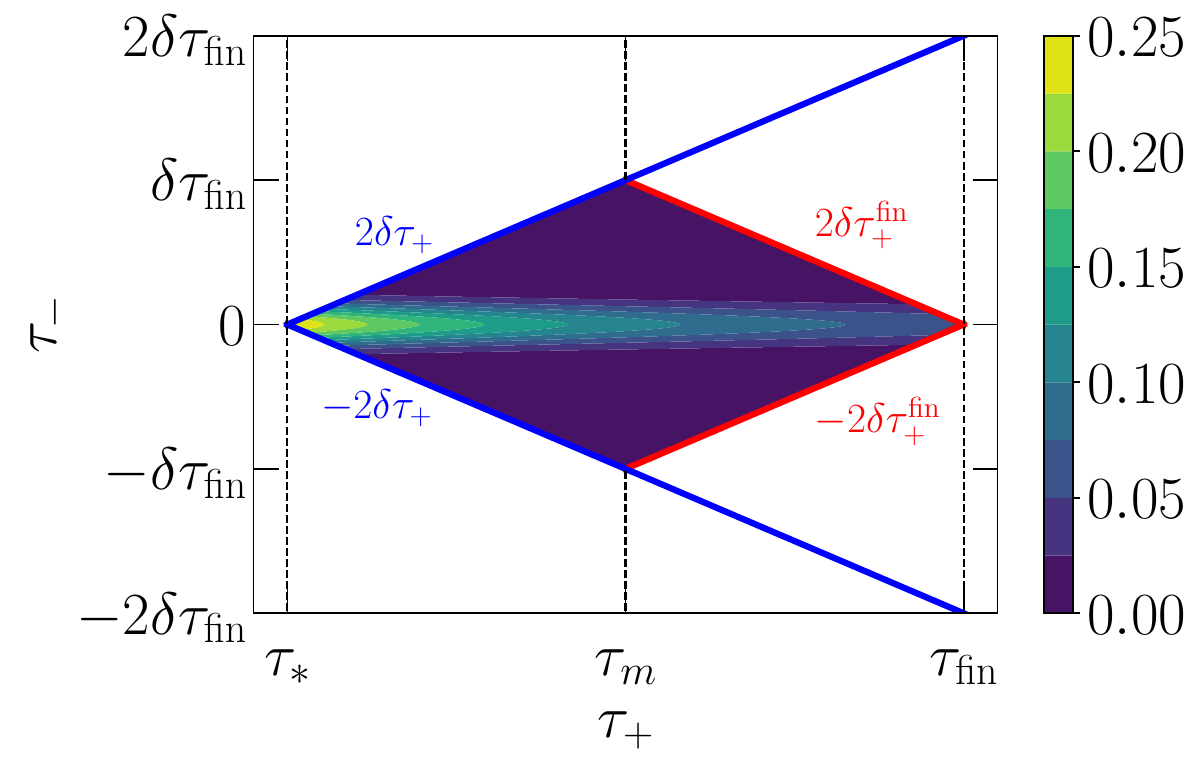}
    \includegraphics[width=\columnwidth]{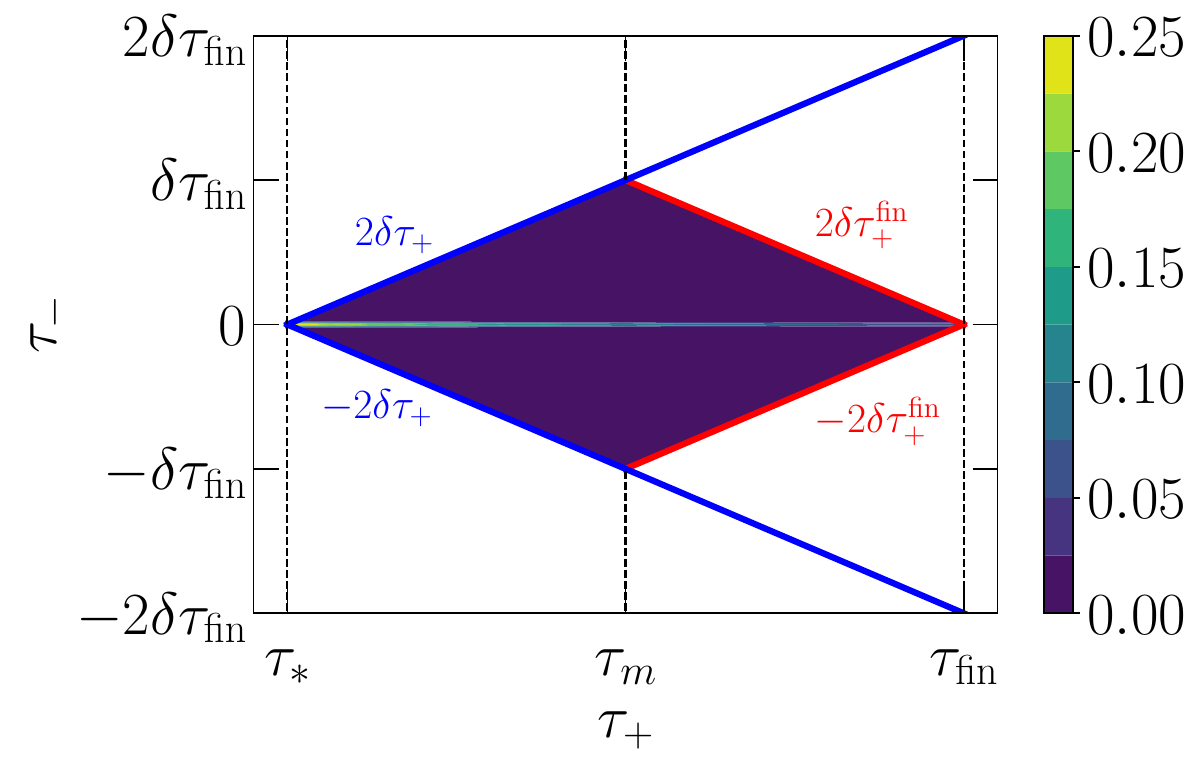}
    \caption{\justifying Integrand leading to the value of $\Delta_0$
    assuming Kraichnan decorrelation for $p^2 v_{\rm sw}^2 = 1$ (upper panel), $10$ (middle), and
    $100$ (lower), in the $k \to 0$ limit.}
    \label{limits_integral2}
\end{figure}

Using this UETC for the velocity field and taking the $k \to 0$ limit (such that $\tilde p \to p$), \Eq{Delta_stationary_UETC} becomes
\begin{align}
    \Delta_0 (\delta \tfin, p) =
    \int_{\tini^{ }}^{\tfin^{ }} \!\! \frac{\dd \tau^{ }_1}{\tau^{ }_1} \int_{\tini^{ }}^{\tfin^{ }} \!\! \frac{\dd \tau^{ }_2}{\tau^{ }_2} e^{-p^2 v_{\rm sw}^2 \tau_-^2} \ .
    \label{Delta0_stat}
\end{align}
The integrand is shown in \Fig{limits_integral2}.

We observe that for large $p^2 v_{\rm sw}^2 \sim {\cal O} (10^2)$,
it is a good approximation to extend the integration limits to $\tau_- \in (-\infty, \infty)$,
while the same is not true at smaller $p^2 v_{\rm sw}^2 \sim {\cal O} (1)$.
In this case we find two limiting cases:
\begin{itemize}
    \item[{\em i)}] if $\delta \tfin \ll 1/(p v_{\rm sw})$, we expand $e^{-p^2 v_{\rm sw}^2 \tau_-^2} \sim 1$, since $\tau_- \in [0, \delta \tfin]$ (see footnote \ref{fn_int}). Then \Eq{Delta0_stat} yields the duration dependence
    found for the UETC of the sound shell model
    in the $k \to 0$ limit: $\ln^2 (\tfin \HH_*)$;
    \item[{\em ii)}] if $\delta \tfin \gg 1/(pv_{\rm sw})$,    the approximation leading to \Eq{Delta_stat_supp} is justified, and we find the suppression factor $\Upsilon$ in the $k \to 0$ limit.
    As discussed above, this regime can also appear in the sound shell model when $k \delta \tfin \gg 1$.
\end{itemize}
Therefore, the resulting dependence of the GW amplitude with $\delta \tfin$ will change for different $v_{\rm sw}$ and it might be a combination of the different modes since one needs to integrate \Eq{tilde_Delta0} over $p$ for the general time-dependence.
In addition, as mentioned above,
$v_{\rm sw}$ is also a function of $\tau_1$ and $\tau_2$, to ensure the positivity of the UETC kernel \cite{Auclair:2022jod}.

\begin{figure}
    \centering
    \includegraphics[width=\columnwidth]{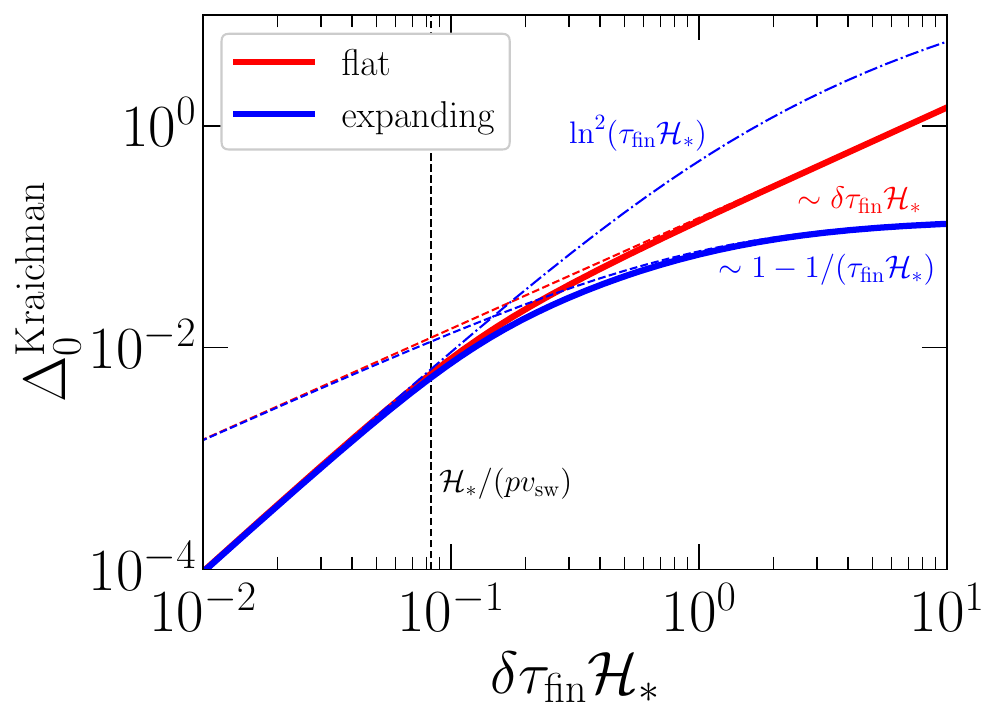}
    \caption{\justifying Dependence of the GW amplitude in the $k \to 0$ limit with the duration of the GW sourcing $\delta \tfin$ for a Kraichnan decorrelation with $p v_{\rm sw} 
    = 12$ for a flat (red) and an expanding (blue) Universe. The two asymptotic limits are separated at $\delta \tfin = 1/(p v_{\rm sw})$, showing the $(\delta \tfin \HH_*)^2$ scaling below this limit, and the suppression factor $\Upsilon$
    above the limit.}
    \label{growth-Kraichnan}
\end{figure}

We recover the previous result analytically when neglecting the expansion of the Universe,\footnote{In this case, one can find an analytical expression for any wave number $k$,
here avoided for the sake of brevity.}
\begin{align}
    \Delta_0^{\rm flat} (\delta \tfin, p)/\HH_*^2 = &\, \frac{\sqrt{\pi}}{p v_{\rm sw}} \delta \tfin \, {\rm Erf} \bigl(p v_{\rm sw}
    \delta \tfin \bigr) \nonumber \\ - &\, \frac{1 - e^{-p^2 v_{\rm sw}^2 \delta \tfin^2}}{p^2 v_{\rm sw}^2} \, ,
\end{align}
where Erf$(x)$ is the error function.
Taking the limits
$\delta \tfin \ll 1/(p v_{\rm sw})$ and $\delta \tfin \gg 1/(p v_{\rm sw})$, we
find the two asymptotic behaviors mentioned above,
\begin{align}
    \Delta_0^{\rm flat} (\delta \tfin p v_{\rm sw} \ll 1)  &\, = (\delta \tfin \HH_*)^2,
    \nonumber \\ \Delta_0^{\rm flat} (\delta \tfin p v_{\rm sw}  \gg 1)  & \, =
    \frac{\sqrt{\pi} \delta \tfin \HH_*}{p v_{\rm sw}/\HH_*}.
\end{align}
Including the effect of the expansion of the Universe
leads to the same short-duration regime, and the limit
at large $\delta \tfin p v_{\rm sw}$ becomes
\begin{equation}
    \Delta (\delta \tfin p v_{\rm sw} \gg 1) = \frac{\sqrt{\pi}}{p v_{\rm sw}/\HH_*} \Upsilon
    (\delta \tfin).
\end{equation}
The two asymptotic limits are shown in \Fig{growth-Kraichnan},
compared to \Eq{Delta0_stat} evaluated numerically.
These results show how we can, in general, find both the
quadratic and linear growth rates, depending on $v_{\rm sw}$,
the specific value
of $k$ (even in the $k \to 0$ limit), and the integrals over $p$ and $\tilde p$ performed to find
the GW spectrum sourced by a stationary process.

\section{GW spectrum from sound waves: results and template}
\label{numerical_GW}

In \Secs{stationary_UETC}{linear_vs_quadratic}, we have studied the GW spectrum in the low-frequency limit $k \to 0$, aiming to understand two
characteristic features: the $k^3$ scaling, and
the amplitude evolution with respect to the duration of the source.
The present section is dedicated to the study of the shape of the GW spectrum at all frequencies.

For a direct comparison of our results for sound waves to those for 
other sources, e.g., decaying vortical turbulence, we adopt a similar normalization as in Ref.~\cite{RoperPol:2022iel} (see also \Sec{sec_norm}).

\subsection{GW spectral shape}

With \Eq{GW_spec_k0_norm} the GW spectrum can be expressed in terms of a normalized spectrum,
$\zeta_{\rm GW}$,
\begin{align}
    \Omega_{\rm GW}(\delta \tfin,&\, R_*, K) = 
    3 \, \bar w^2 \, K^3 \, {\cal T}_{\rm GW} \, {\cal C} \,
    \biggl(\frac{\OmK}{{\cal K}}\biggr)^2\nonumber \\ 
     \times &  \, 
    \, \tilde \Delta_0 (\delta \tfin, R_*) \, \zeta_{\rm GW} ( \delta \tfin,K,R_*) \, .
    \label{OmGW_general1}
\end{align}
In order to describe the spectral
modifications of $\zeta_{\rm GW}$ with respect to $\zeta_\Pi$, we introduce the function $\tilde \Delta \equiv \zeta_{\rm GW}/\zeta_\Pi$.
Then \Eq{OmGW_general1} becomes
\begin{align}
    \Omega_{\rm GW} (\delta & \tfin, R_*,  K) =  3 \, \bar w^2 
    \, K^3 \, {\cal T}_{\rm GW} \, {\cal C} \, \biggl(\frac{\OmK}{{\cal K}}\biggr)^2  \nonumber \\ 
     \times & \, \zeta_{\Pi} (K) \, \tilde \Delta_0 (\delta \tfin, R_*)\, \tilde \Delta (\delta \tfin, R_*, K) \,.
    \label{OmGW_template}
\end{align}
$\tilde \Delta$ generalizes \Eq{tilde_Delta0} to all values of $k$ and $\delta \tfin$,
\begin{align}
    \tilde \Delta (\delta & \tfin, R_*,  K) \nonumber \\ = &\,  \frac{1}{{\cal C} \, \zeta_\Pi (K) \, \tilde \Delta_0 (\delta \tfin, R_*)} \int_0^\infty \dd P P^2 \zetaK (P) 
    \nonumber \\ \times &\,
    \int_{-1}^1 (1 - z^2)^2
    \frac{\zetaK (\tilde P)}{\tilde P^4} \Delta (\delta \tfin, k, p, \tilde p) \dd z \,.
    \label{tilde_Delta_general}
\end{align}
By construction we find that $\tilde \Delta \to 1$, when $\Delta$ does not depend on $p$ nor $k$, i.e., in the short-duration regime, or in the $k \to 0$ limit.

Hence, the parameters that determine the modifications of $\zeta_{\rm GW}$ with respect to $\zeta_\Pi$
are the source duration $\delta \tfin$, and the characteristic scale $R_* = 1/k_*$.

\begin{figure*}
    \centering
    \includegraphics[width=.32\textwidth]{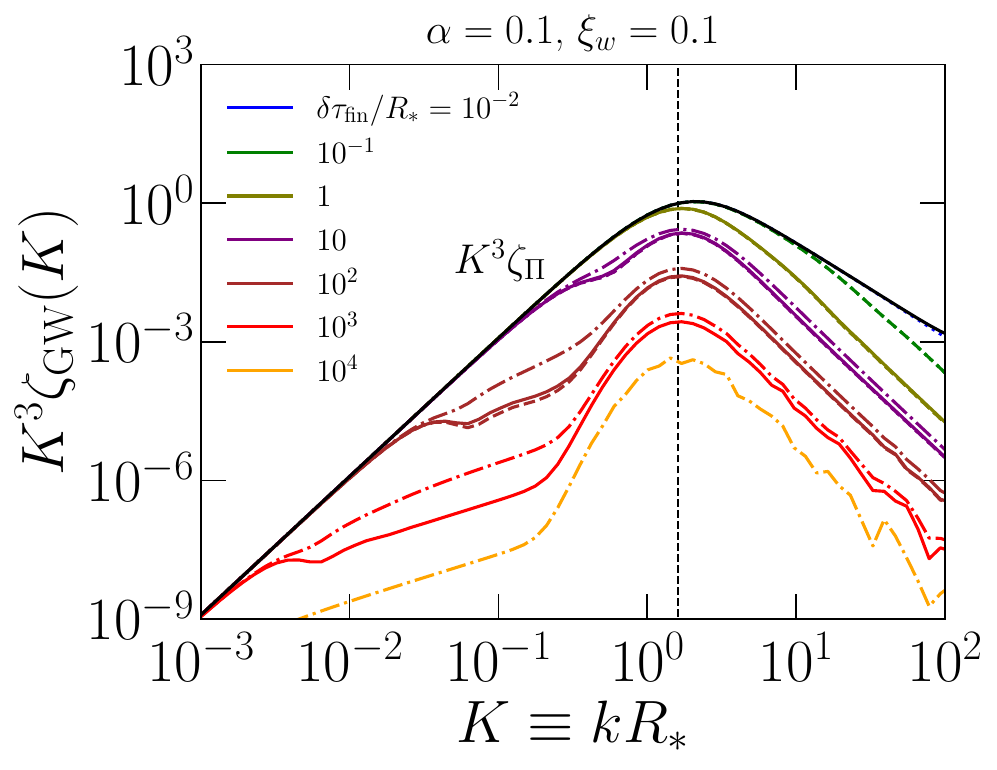}
    \includegraphics[width=.32\textwidth]{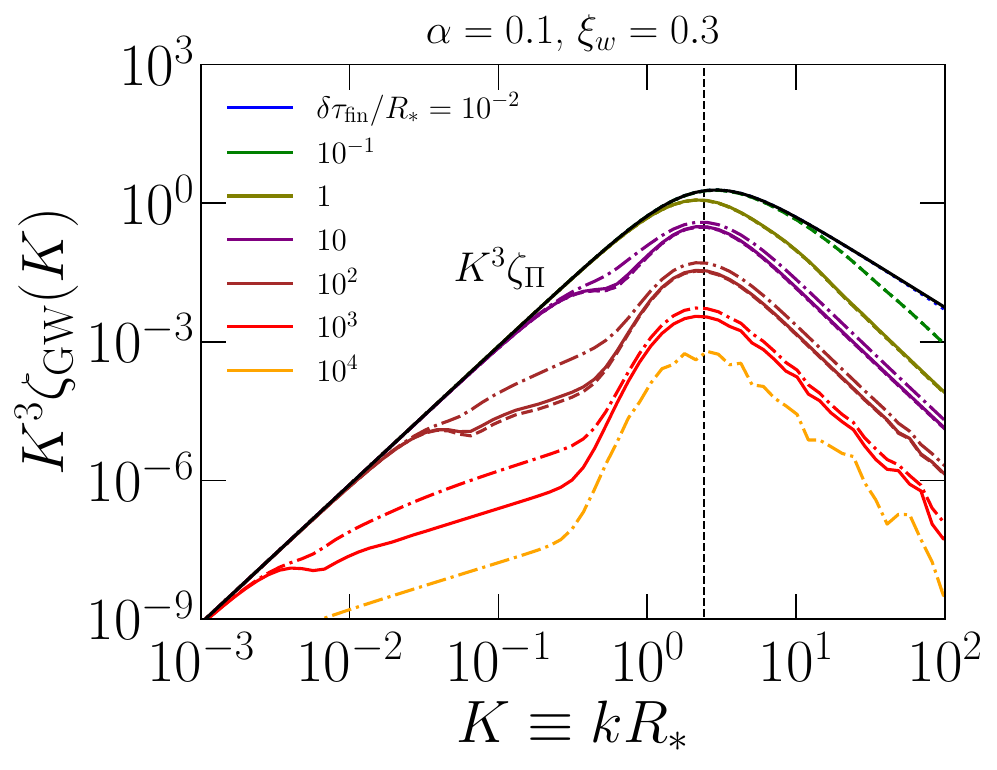}
    \includegraphics[width=.32\textwidth]{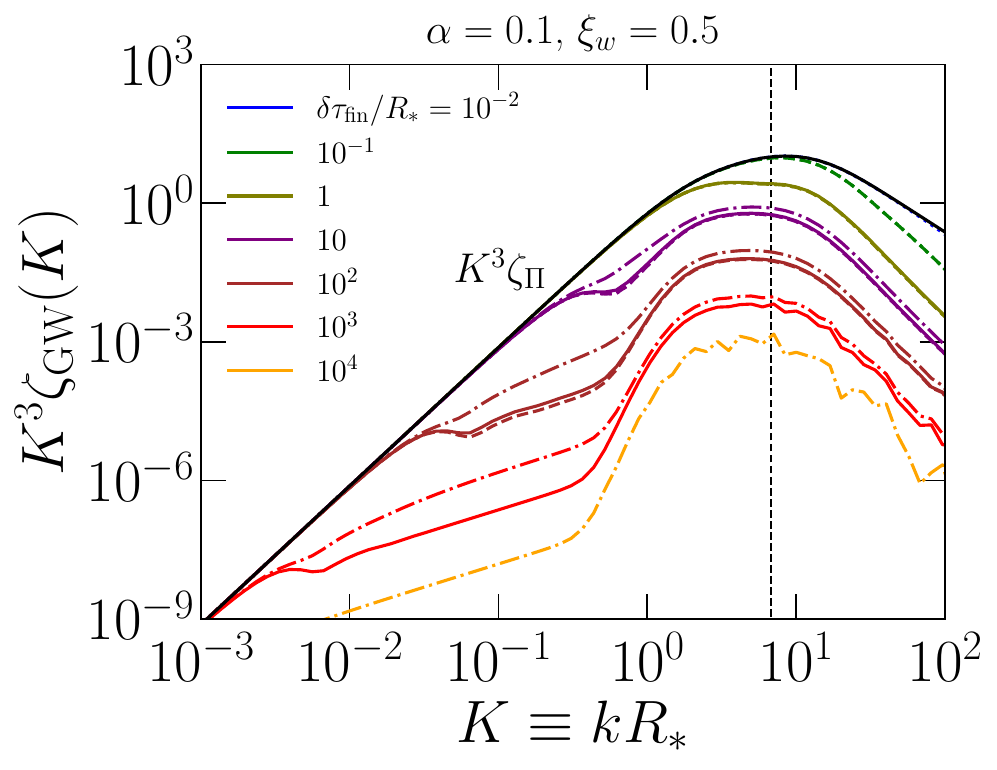}
    \includegraphics[width=.32\textwidth]{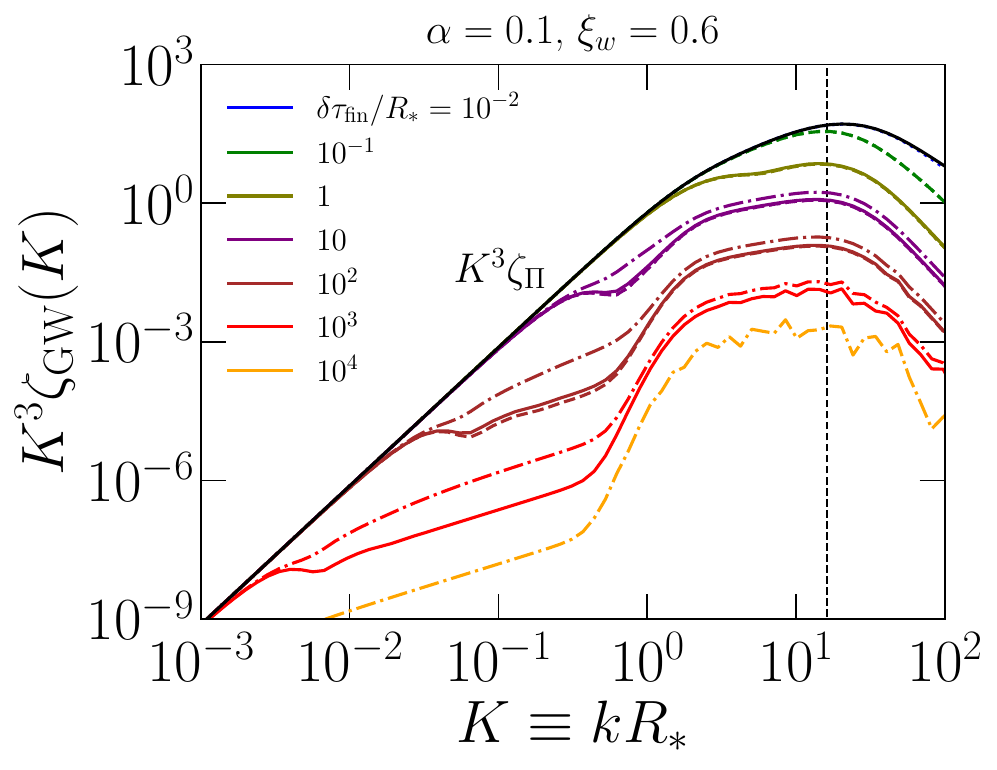}
    \includegraphics[width=.32\textwidth]{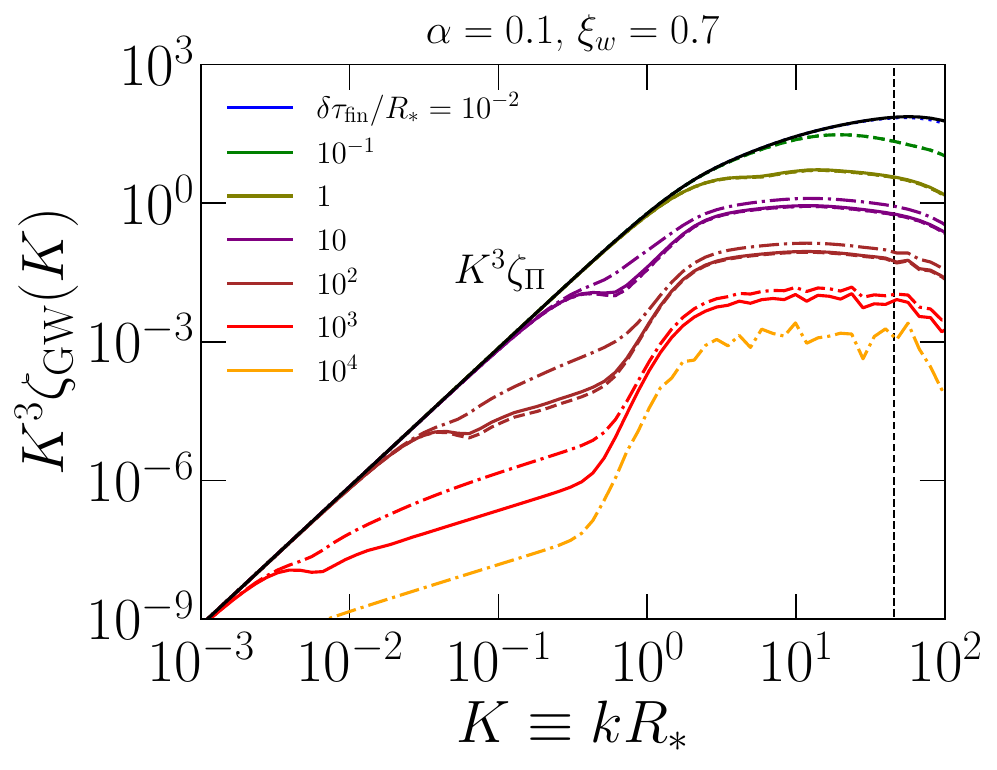}
    \includegraphics[width=.32\textwidth]{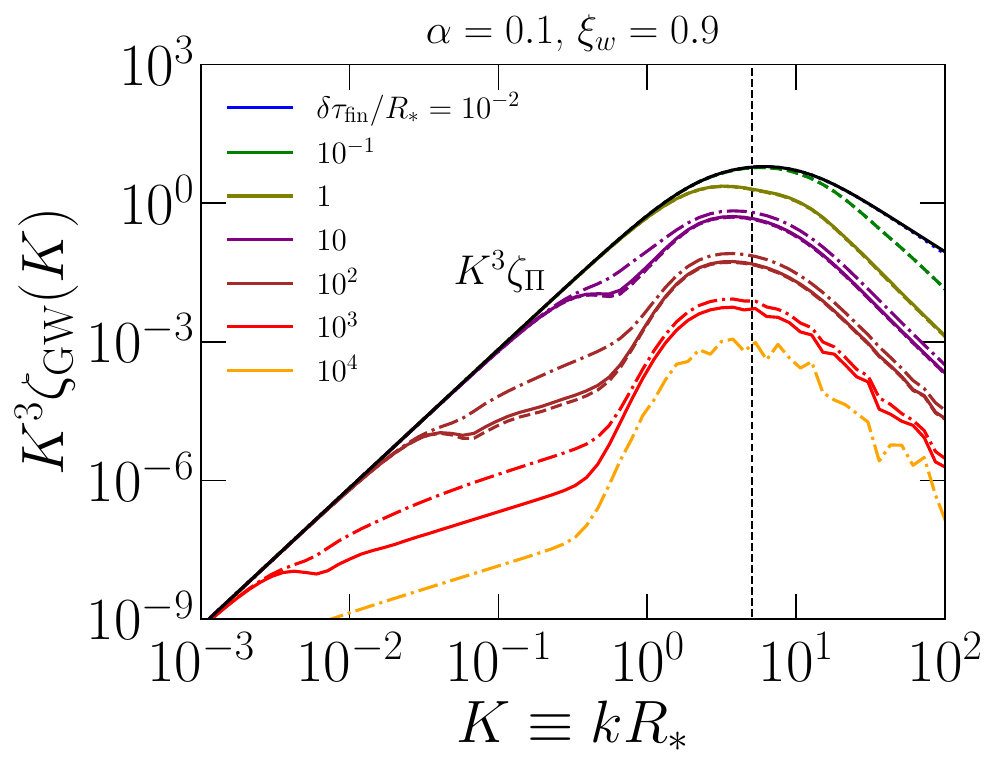}
    \caption{\justifying Normalized GW spectral shape $K^3 \zeta_{\rm GW}$ [see \Eq{OmGW_general1}] for the benchmark phase transitions
    shown in \Fig{kinetic_spectra} in the case of exponential nucleation.
    For comparison we show $K^3 \zeta_\Pi$ (in black), expected in the range $k < 1/\delta \tfin$.
    The modifications with respect to $K^3 \zeta_\Pi$ occur at $k > 1/\delta \tfin$, and different colors correspond to different values of $\delta \tfin/R_*$.
    The exact modifications depend separately on both $R_*$ and $\delta \tfin$, especially when $\delta \tfin/R_* \gg 1$.
    Dotted, dashed, solid, and dash-dotted lines correspond to values $\delta \tfin \HH_*$ of $10^{-2}$, $10^{-1}$, 1, and $10$, respectively.
    The vertical lines indicate the estimated position
    of the GW peak, at $K \approx 0.8 \, K_{\rm GW}$, where $K_{\rm GW}$
    is the position where $K^3 \zeta_\Pi$ is maximum (see values in \Tab{tab:my_label}).
    The second peak of $\zeta_\Pi$ appears also in $\zeta_{\rm GW}$, and is related to the inverse sound shell thickness $1/\Delta R_*\equiv \xi_w/(R_* |\xi_w -c_s|)$ \cite{Hindmarsh:2016lnk}.
    Hence, when $\xi_w$ is closer to $\cs$, the second peak appears
    far from $1/R_*$, yielding a broad plateau around the peak.
    When $\xi_w$ diverges from $\cs$, the second peak becomes closer to the first one at $K \sim 1$, and the plateau disappears.}
    \label{GW_spectra}
\end{figure*}

Depending on how $k$ compares with the inverse source duration $1/\delta\tfin$, the GW spectrum presents different behaviors.

In the regime where
$k \lesssim 1/\delta \tfin$, studied in \Sec{stationary_UETC}, $\tilde \Delta \to 1$.
The dependence of the GW spectrum on the source duration $\delta \tfin$
is then fully encoded in $\tilde \Delta_0 = A \ln^2 (\tfin \HH_*)$, with $A \in [0.5, 1]$ [see \Eq{A_fit}].
The amplitude in this regime does not depend on $R_*$, whose dependence
only appears
through the self-similar $K \equiv k R_*$.
At the same time, the dependence on $K$ survives in $K^3 \zeta_\Pi$, which, as shown in \Fig{zeta_Pi}, follows a broken-power law
that can be fit using \Eq{zeta_Pi_fit}.\footnote{The peak structure in the sound shell model is simple or double, depending on
the specific value of the wall velocity (see \Fig{kinetic_spectra} and \Tab{tab:my_label}).} 
The amplitude of the GW spectrum 
depends on
the specific spectral shape of the kinetic spectrum
via the constants ${\cal K}$ and ${\cal C}$ [see \Eqs{KK}{CC_alp}].
\Tab{tab:my_label} presents values for the
benchmark phase transitions considered here.

At wave numbers $k > 1/\delta \tfin$,
the approximation leading to $\tilde \Delta \sim 1$
is no longer valid, and the function $\tilde \Delta (K)$ depends on both $\delta \tfin$ and $R_*$. 
As a consequence, in this range, the GW spectrum shows a complex dependence on $K$ and $\delta \tfin$
that deviates with respect to the simple $K^3 \zeta_\Pi$ causal growth.
We expect the GW spectrum to transition
from the causal branch at $k \delta \tfin \ll 1$, toward
the spectrum found in Refs.~\cite{Hindmarsh:2019phv,Guo:2020grp} (see \App{HH19}),
which is valid for $k \delta \tfin \gg 1$, as
discussed in \Secs{stationary_UETC}{linear_vs_quadratic}.
This transition among the two asymptotic limits is,
{\em a priori}, unknown and requires a numerical evaluation of \Eq{tilde_Delta_general}.

Numerical examples of the resulting normalized GW spectra,
$K^3 \zeta_{\rm GW}$, are shown in \Fig{GW_spectra} for the benchmark phase transitions of \Fig{kinetic_spectra}, and at different values of $\delta \tfin$
and $R_*$.
We find the predicted $K^3 \zeta_\Pi$ scaling when $k < 1/\delta \tfin$, with the amplitude exactly given
by \Eq{OmGW_template} when setting $\tilde \Delta = 1$. 

\begin{figure}
    \centering
    \includegraphics[width=\columnwidth]{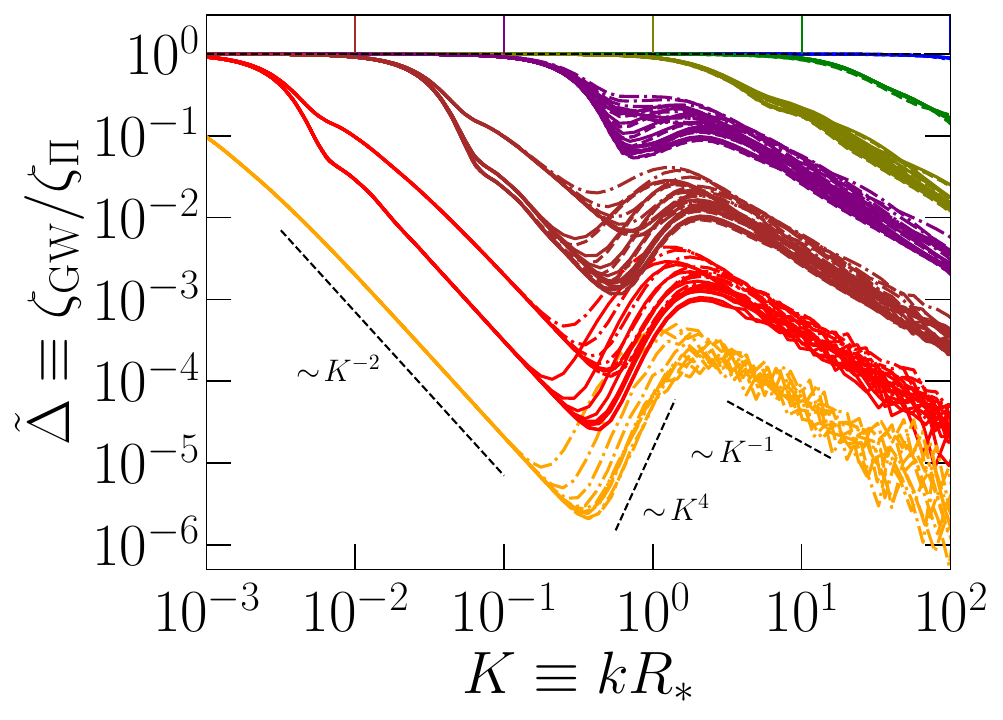}
    \caption{\justifying Ratio $\tilde \Delta \equiv \zeta_{\rm GW}/\zeta_\Pi$ for the benchmark phase transitions and parameters of \Fig{GW_spectra}.
    Line colors and styles are the same
    as those in \Fig{GW_spectra}.}
    \label{normalized_GW_spectra}
\end{figure}

A more complex structure
appears at $k>1/\delta\tfin$, where $\tilde \Delta\equiv \zeta_{\rm GW}/\zeta_\Pi$ plays a major role. 
To underline some generic features, we show $\tilde \Delta$ in
\Fig{normalized_GW_spectra}
at different $\delta \tfin$ and $R_*$.

In the range $1/\delta \tfin \lesssim k  < 1/R_*$,
we find
$\tilde \Delta \sim K^{-2}$, leading to the development of a
linear GW spectrum in $k$. 
A similar transition from a $K^3$ to $K$ slope in the GW spectrum
is also found for vortical (M)HD turbulence \cite{RoperPol:2019wvy,RoperPol:2021xnd,Brandenburg:2021bvg,Brandenburg:2021tmp,Kahniashvili:2020jgm,RoperPol:2022iel,Auclair:2022jod},
and is analytically
described by the constant-in-time approximation \cite{RoperPol:2022iel}.

At larger $k$, a steep growth, $\OmGW \sim K^7$, appears just below the peak of the spectrum. 
This result is close to the $K^9$ growth found in Ref.~\cite{Hindmarsh:2019phv}. 
In fact, in this range, $1/\delta\tfin\ll k\lesssim1/R_*$,
motivating the assumption $k \delta \tfin \to \infty$, required to obtain
the $K^9$ spectrum (see discussion in \Sec{linear_vs_quadratic}).
Note however that, when the source duration becomes a non-negligible
fraction of a Hubble time, $\delta\tfin \HH_* \gtrsim {\cal O} (10^{-1})$, the expansion of the Universe starts playing a significant role. 
In particular, it modifies not only the dependence of the GW spectrum on $\delta\tfin$ but also its spectral shape through $\Delta$ in \Eq{tilde_Delta_general}.

\begin{figure}
    \centering
    \includegraphics[width=\columnwidth]{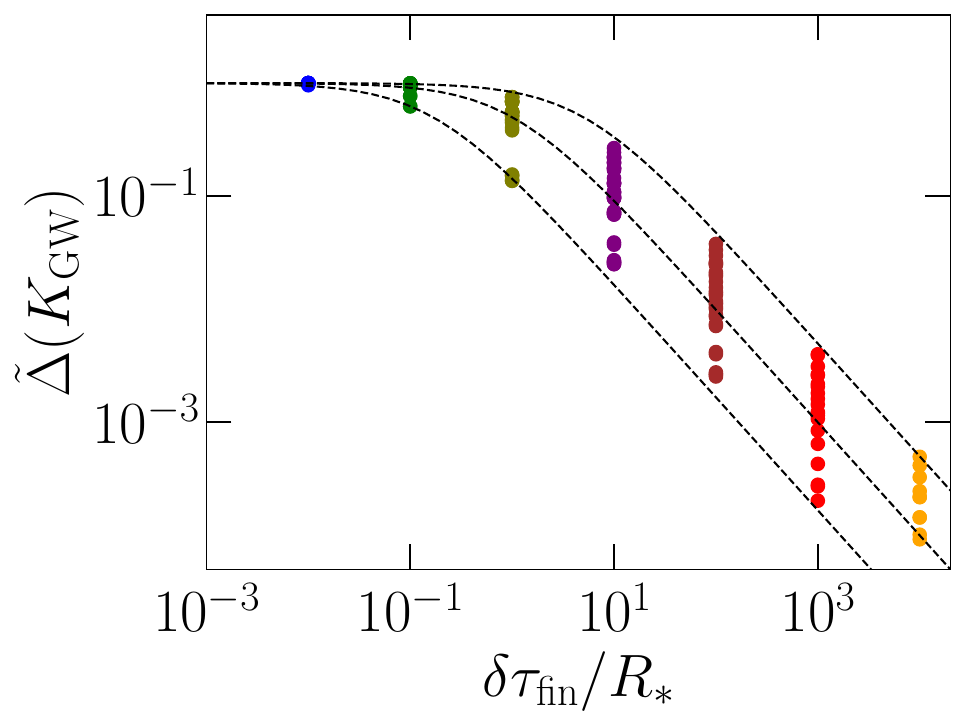}
    \caption{\justifying Dependence of the GW peak amplitude with $\delta \tfin/R_*$, normalized to
    the value in the $\delta \tfin \to 0$ limit.
    Each dot corresponds to a specific line in \Fig{normalized_GW_spectra}.
    The amplitude shows a universal trend with the product $\delta \tfin/R_*$
    that can be approximately fit empirically by the function $\bigl(1 + \delta \tfin/R_*\bigr)^{-1}$, intermediate black dashed line.}
    \label{dependence_peak}
\end{figure}

The peak amplitude of the GW spectrum, which we have previously estimated to be located at $K_{\rm GW}$, where
$K^3 \zeta_\Pi$ is maximum, is modified by $\tilde \Delta$
when the $k \lesssim 1/\delta \tfin$ limit does not hold.
We find that $\tilde \Delta$ modifies the position of the
GW
peak roughly to $K \approx 0.8 \, K_{\rm GW}$ (see \Fig{GW_spectra} and values in \Tab{tab:my_label}).
In addition, $\tilde \Delta$ adds a dependence of the 
GW amplitude on
$\delta \tfin/R_*$, shown in \Fig{dependence_peak}.
This modification at the peak is well approximated by the function $(1 + \delta \tfin/R_*)^{-1}$. 
For the benchmark phase transitions, and 
the values of $\delta \tfin$ and $R_*$ shown in \Fig{normalized_GW_spectra}, we
find that the ratio of the numerical values to the
fit is between 0.2 and 5 (see dashed lines in \Fig{dependence_peak}).

Around the peak, $\tilde \Delta_0\, \tilde \Delta$ 
depends linearly on the 
suppression factor, $\Upsilon$, and $R_* \HH_*$. 
This result agrees with
the one derived in \App{HH19}, following the approximation of Refs.~\cite{Hindmarsh:2019phv,Guo:2020grp}, when $\delta \tfin/R_* \gg 1$, such that the peak $1/R_*$ is within the $k \delta \tfin \gg 1$ regime.
For an accurate prediction of the amplitude at the peak, we thus
take this value into account and multiply it by the value where
the function $K^3 \zeta_\Pi$ is maximal (see \Tab{tab:my_label}).

Finally, at large $K > 1$, we find that the GW spectrum decreases as $1/K$ when compared to $K^3 \zeta_\Pi$. 
Since the latter scales as $K^{-2}$ (see \Fig{zeta_Pi}), the GW spectrum decays as $K^{-3}$ at large values of $k$, which agrees with Refs.~\cite{Hindmarsh:2019phv,Jinno:2020eqg,Jinno:2022mie}.

\begin{figure}
    \centering
    \includegraphics[width=.85\columnwidth]{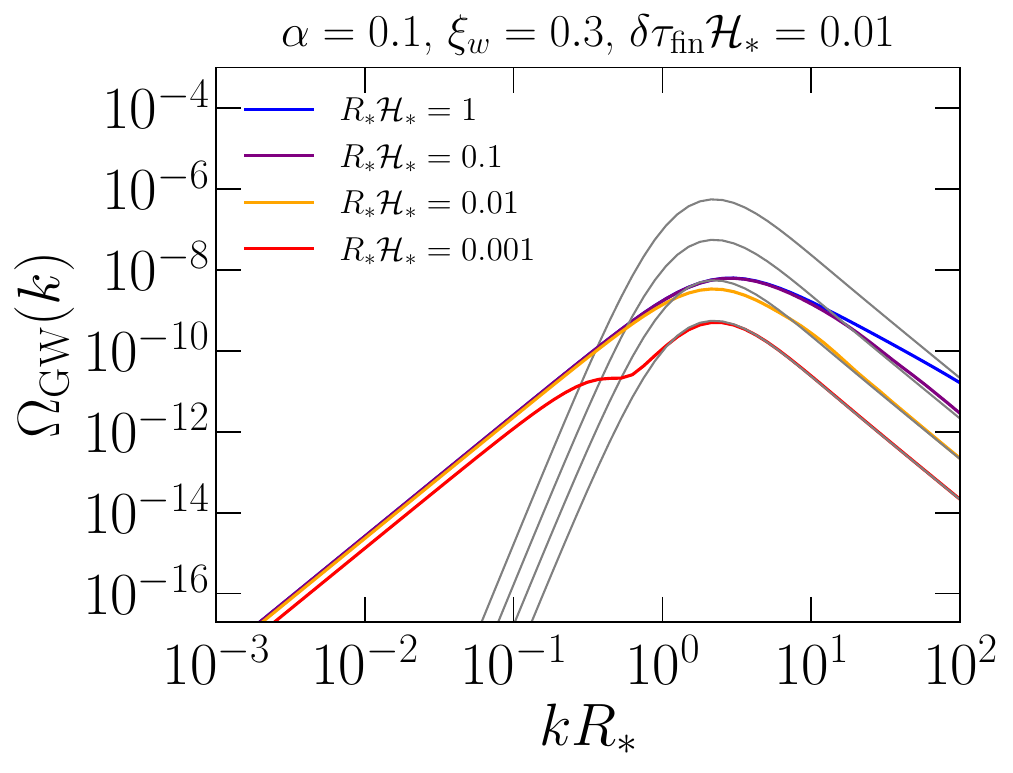}
    \includegraphics[width=.85\columnwidth]{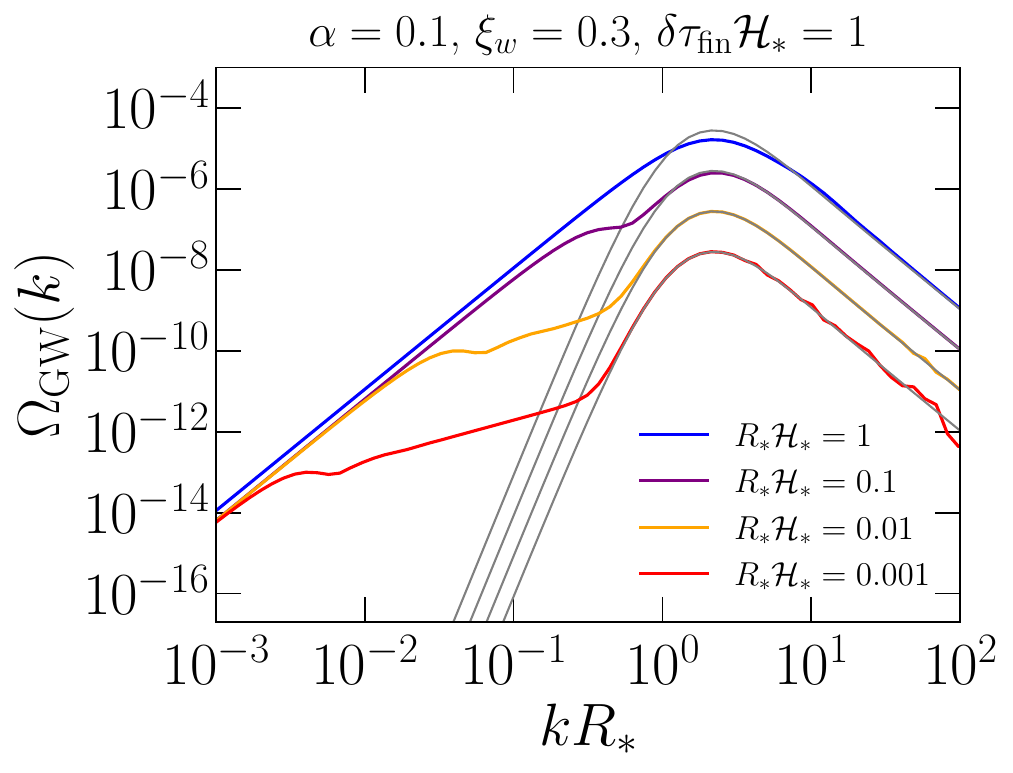}
    \caption{\justifying GW spectrum as a function of $kR_*$ for a benchmark phase transition with $\alpha = 0.1$ and
    $\xi_w = 0.3$, assuming exponential nucleation.
    The results are shown for different values of $R_* \HH_*$, and $\delta \tfin \HH_* = 0.01$ (upper panel)
    and 
    $1$ (lower panel).
    For comparison, the gray lines correspond to the GW spectrum
    using the approximation of Refs.~\cite{Hindmarsh:2019phv,Guo:2020grp} [see \Eq{OmGW_HH19}].
    The values of $\OmGW$ are computed using ${\cal T}_{\rm GW} = 1$ so they should be multiplied by \Eq{eq:evolGW}, choosing the
    specific time of generation, to find the GW spectrum today.}
    \label{GW_comparison}
\end{figure}

To compare the resulting spectral shape of GWs to that of Ref.~\cite{Hindmarsh:2019phv},
where the function $\Delta$ is approximated by a Dirac delta
function, we show in \Fig{GW_comparison} the resulting GW spectra,
obtained for a specific benchmark phase transition with $\alpha = 0.1$
and $\xi_w = 0.3$, for a range of $R_*$ and $\delta \tfin$.
The calculation of the GW spectra under the assumption of
Refs.~\cite{Hindmarsh:2019phv,Guo:2020grp} is given in \App{HH19}.

We show that the GW spectrum found in Ref.~\cite{Hindmarsh:2019phv} is a correct description
for the bump around and above the peak when $\delta \tfin/R_*$ is sufficiently
large (as described above), after taking into account the correction due to the expansion of the Universe \cite{Guo:2020grp}.
The transition toward the GW spectrum in the ``infinite duration'' limit (given in \App{HH19}) is related to the one from the quadratic
to linear growth that we have found in \Sec{sec_kraichnan}, since the
approximation used to extend the limits of integration over $\tau_-$ to $\pm \infty$ in \Eq{Delta_general_uetc} is based on the assumption that $k \delta \tfin \to \infty$.
However, additional linear and cubic regimes appear in $\OmGW$
at frequencies below the peak that were not found in Refs.~\cite{Hindmarsh:2019phv,Guo:2020grp} since the $k \delta \tfin \gg 1$ assumption does not hold in this range of frequencies.
Moreover, when $\delta \tfin/R_*<1$, the peak is in the regime $k_* < 1/\delta \tfin$, so that significant modifications of the GW spectrum
may appear around the peak.

\subsection{Estimation of the source duration}
\label{sec:timenonlin}

Let us now discuss why the variables $\delta \tfin$ and $R_*$ are not completely independent.
The characteristic scale $R_*$ is determined by the mean bubble separation,
which depends on the characteristics of the phase transition via $\beta$ and $\xi_w$ [see relation below \Eq{EK_sw_HH}].

The evaluation of $\delta \tfin$ requires further numerical studies to simulate
the decay of the sound waves, as well as the development
of turbulence.
\begin{figure*}
    \centering
    \includegraphics[width=.66\columnwidth]{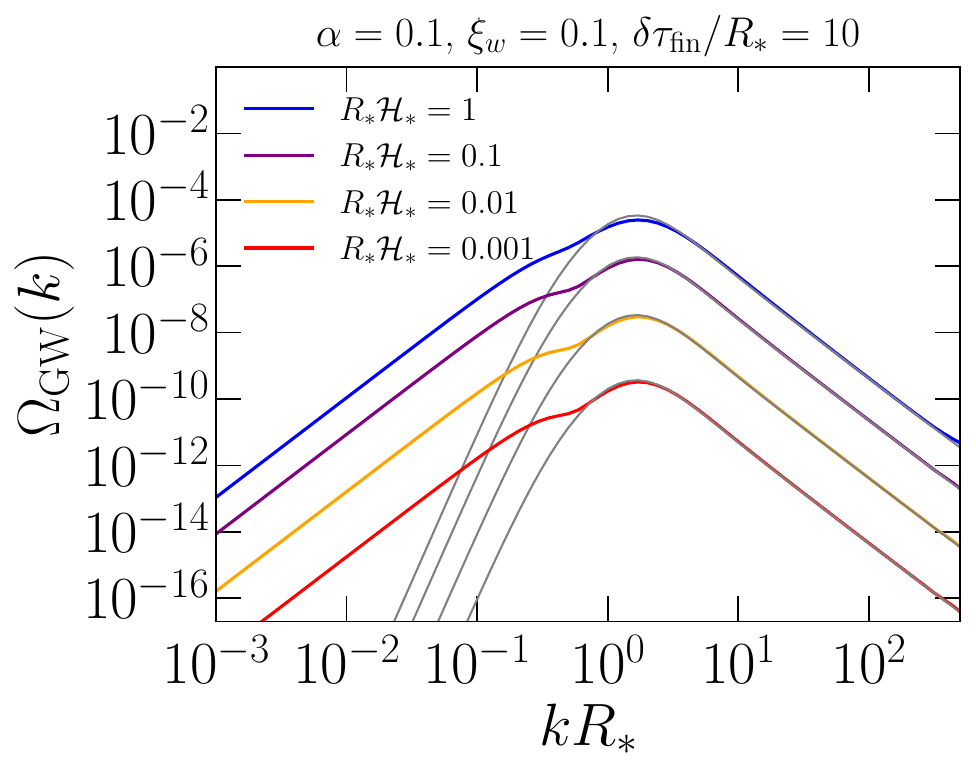}
    \includegraphics[width=.66\columnwidth]{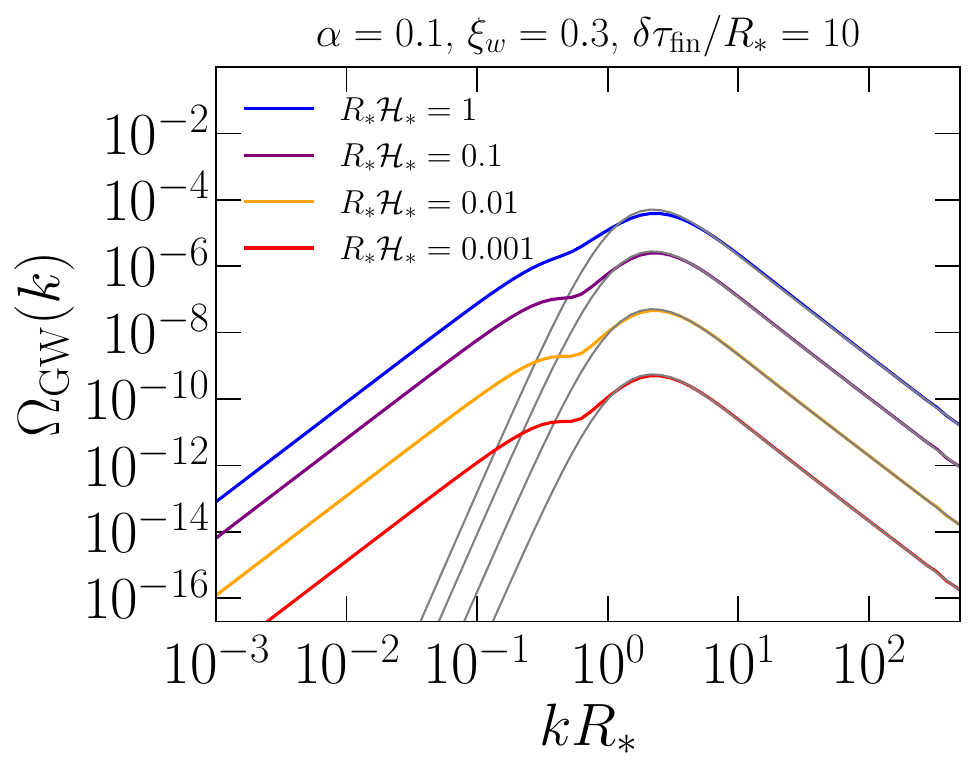}
    \includegraphics[width=.66\columnwidth]{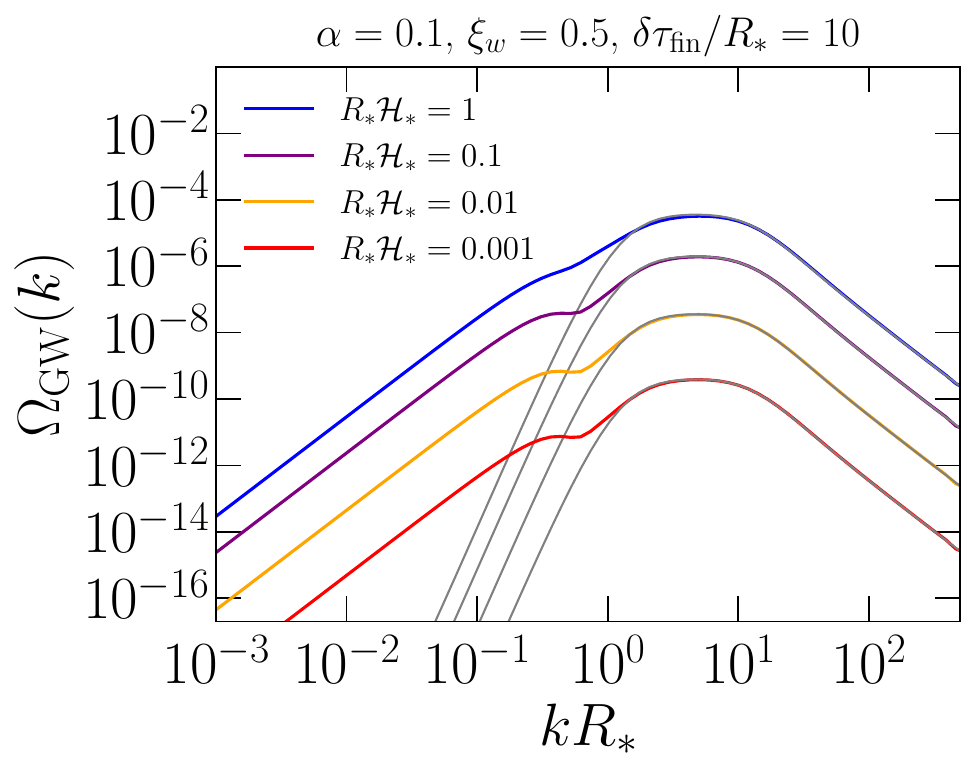}
    \includegraphics[width=.66\columnwidth]{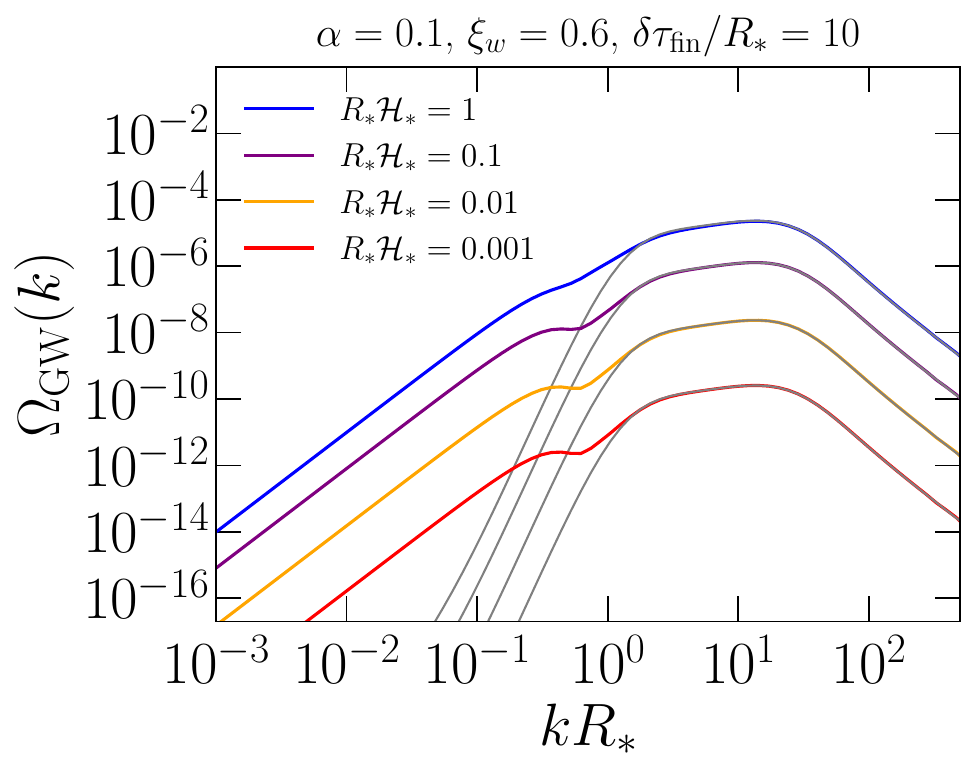}
    \includegraphics[width=.66\columnwidth]{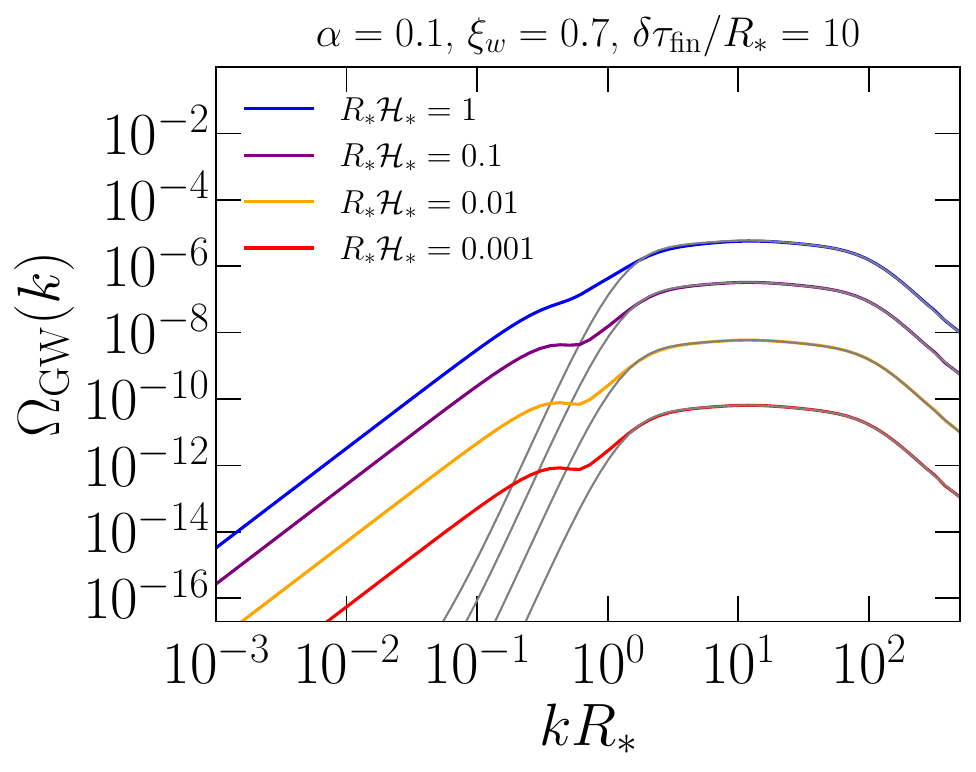}
    \includegraphics[width=.66\columnwidth]{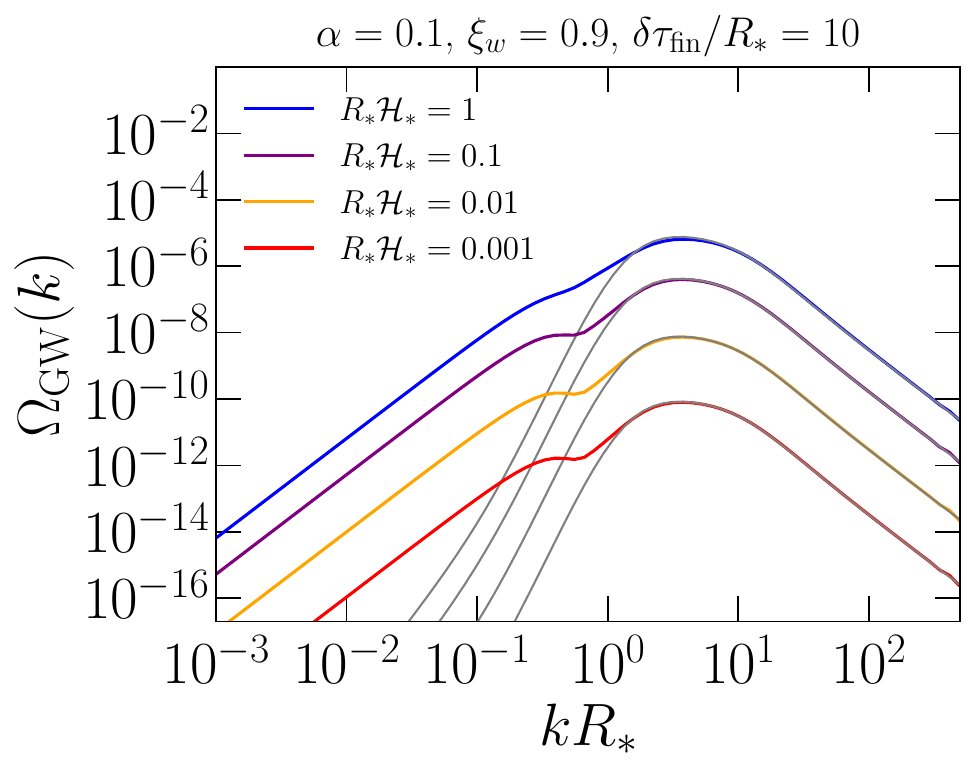}
    \caption{\justifying GW spectrum as a function of $kR_*$ for the benchmark phase transitions shown in \Fig{kinetic_spectra}. The results are shown for different values of $R_* \HH_* = \{0.001, 0.01, 0.1, 1\}$ and taking $\delta \tfin = 10 R_*$,
    corresponding to the time expected to develop non-linearities
    for $\OmK \sim 10^{-2}$.
    For comparison, the gray lines correspond to the GW spectrum
    using the approximation of Refs.~\cite{Hindmarsh:2019phv,Guo:2020grp} [see \Eq{OmGW_HH19}].
    \label{GW_comparison2}
    The values of $\OmGW$ are computed using ${\cal T}_{\rm GW} = 1$ so they should be multiplied by \Eq{eq:evolGW}, choosing the
    specific time of generation, to find the GW spectrum today.}
\end{figure*}
A first estimation of $\delta \tfin$ is the eddy turnover
time,
i.e., the time that it takes the plasma to develop
non-linearities, $\delta \tau_{\rm nl} \sim R_*/\sqrt{\OmK}$ \cite{Caprini:2019egz}, and it directly depends on $R_*$.
Setting $\delta \tfin \sim \delta \tau_{\rm nl}$ and
$\OmK \sim 10^{-2}$ for the benchmark phase transitions with $\alpha = 0.1$ (see \Tab{tab:my_label}),
we find $\delta \tfin/R_* \sim 10$.
For this estimate, the condition $\delta \tfin/R_* \gg 1$ is
valid, and the prescription of Refs.~\cite{Hindmarsh:2019phv,Guo:2020grp} gives a correct estimate of
the amplitude around the peak.
However, it fails at frequencies below the peak, as expected.

We show in \Fig{GW_comparison2} the GW spectrum found in the current work and compare
it to the one given by \Eq{OmGW_HH19}, based on the assumptions of Refs.~\cite{Hindmarsh:2019phv,Guo:2020grp},
when we set $\delta \tfin \sim 10 R_*$.
We find that, in this case, the suppression factor $\Upsilon$ is justified
to describe the growth rate with $\tfin$ at the peak.
At frequencies below the peak, we find, in this case, that the
linear growth with $k$ is almost completely absent and the causality
tail, proportional to $k^3$, appears close to the peak, similar
to the results of numerical simulations \cite{Jinno:2022mie}
and other analytical estimates \cite{Cai:2023guc}.
However, for the exact dependence with $\delta \tfin$ of
the full spectral shape, we need to use the prescription 
developed in the current work.
In particular, we find that the causality tail grows proportional
to $\ln^2 (\tfin \HH_*)$.

\subsection{Present-time spectral amplitude}
\label{sec5B}

The present-time GW energy density spectrum today is directly
found using \Eq{OmGW_template} with the transfer function ${\cal T}_{\rm GW}$
given in \Eq{eq:evolGW} taking into account the value of $g_*$ at the time
of GW generation.
We note that the numerical values in \Figs{GW_comparison}{GW_comparison2} are computed using
${\cal T}_{\rm GW} = 1$, so those need to be multiplied by
the corresponding value of ${\cal T}_{\rm GW}$ to produce
the GW spectrum at present time.

Frequencies can be obtained from $k$ using the dispersion relation
of GWs, $2 \pi f = k$, and redshifting the mean-size of the bubbles $R_*$ to the present day,
\begin{align}
    R_0^{-1}  = &\, \frac{H_*}{R_* \HH_*} \frac{a_*}{a_0} = \frac{H_*}{R_* \HH_*} \frac{T_0}{T_*} \biggl(\frac{g_0}{g_*}\biggr)^{1\over 3}
    \nonumber \\
     \simeq &\,  \frac{1.65 \times 10^{-5} \, \Hz}{R_* \HH_*}
     \frac{T_*}{100 \, {\rm GeV}}
     \biggl(\frac{g_*}{100}\biggr)^{1\over 6},
    \label{HHstar}
\end{align}
where we have used $g_0 = 3.91$ and $T_0 = 2.725$ K \cite{Kolb:1990vq,Fixsen:2009ug}.

\section{Conclusions}
\label{conclusions}

We have studied the GW production from sound waves in a first-order phase transition during radiation domination.
Sound waves are expected to be the dominant contribution to the
SGWB, unless the bubbles
run away, the phase transition is supercooled,\footnote{In this case
bubble collisions may represent the dominant contribution to the GW signal \cite{Caprini:2015zlo}.} or the efficiency in
generating turbulence from bubble collisions is large.

We adopt the framework of the sound shell model to estimate the UETC of the
velocity field \cite{Hindmarsh:2016lnk}.
For the single-bubble velocity and energy density profiles,
we follow the description of
Ref.~\cite{Espinosa:2010hh} and present the details of 
our calculation in an accompanying paper \cite{kin_sp_SSM}.
The sound shell model predicted a $k^9$ growth of the spectrum at small frequencies $k$, and a linear dependence on the source duration $\delta \tfin$ in Ref.~\cite{Hindmarsh:2019phv} that can be generalized to the suppression factor $\Upsilon = 1 - 1/(1+\delta\tfin\HH_*)$ when including the effect of the expansion of the Universe \cite{Guo:2020grp}. With this work, we have found
that their prescription holds only in the regime $k \gg 1/\delta \tfin$.
We have addressed this issue and generalized their results to all frequencies.

Our results show that at small frequencies $k\to 0$, the GW spectrum presents a causal tail,
proportional to $k^3$. 
The amplitude of this tail
has a universal dependence on the physical parameters that describe the source.
In particular, it is independent of $R_*$, and it grows with the duration of the source as $\ln^2 (1 + \delta\tfin \HH_*)$,
which yields a quadratic dependence when the source duration
is short.

Around $k \gtrsim 1/\delta \tfin$, an intermediate linear spectrum, $\OmGW \sim k$, may appear, extending until 
a steep slope just below the peak takes over,
which leads to the formation of a bump around the peak.
When we estimate the duration of the GW sourcing as the time scale for the production of non-linearities in the 
plasma, we find that, for the benchmark phase transitions considered in this work with $\alpha = 0.1$, $\delta \tfin/R_* \sim 10$.
In this case, the linear regime in $\OmGW$ is almost absent,
and the GW spectrum soon develops the causal $k^3$ tail at frequencies below the peak.
When $\delta \tfin/R_*$ becomes larger, the intermediate linear
regime extends between the peak and the causal tail.
This bump is a characteristic sign of a GW spectrum sourced by sound
waves, since this distinctive feature does not appear in the
GW spectrum sourced by vortical turbulence \cite{RoperPol:2019wvy,Brandenburg:2021tmp,Kahniashvili:2020jgm,Brandenburg:2021bvg,RoperPol:2021xnd,RoperPol:2022iel,Auclair:2022jod}.
A similar bump was previously found numerically for acoustic
turbulence in Ref.~\cite{RoperPol:2019wvy} and confirmed in
Ref.~\cite{Sharma:2023mao}.
As long as the source duration is sufficiently large, $\delta \tfin/R_* \gg 1$, we find that the amplitude around the peak is
well described by the approach of Refs.~\cite{Hindmarsh:2019phv,Guo:2020grp}.

Our results reconcile the predictions of the sound shell model
with the numerical simulations of Ref.~\cite{Jinno:2022mie}, where a 
cubic dependence of the GW spectrum at low $k$ is also found.
Furthermore, they are in agreement with the findings of Ref.~\cite{Sharma:2023mao}, where numerical simulations are also performed, supporting the theoretical results of the sound shell model. 

We have presented a theoretical description of the origin
of the linear and quadratic growth with $\delta \tfin$ that can
appear when GWs are sourced by a general stationary process as,
in the sound shell model, by a stationary UETC of the velocity field given by \Eq{eq_EK_Eu_st}.

The resulting GW spectrum has been presented in a semi-analytical
framework by separating  each of the
different contributions that can affect its final spectral shape
and amplitude.
Understanding each of the different contributions
separately is important to test the validity of each
of the underlying assumptions in future work.
This framework allows for direct extensions of our results to include
different models or assumptions.

We present the detailed calculation of the anisotropic stresses
of the velocity field, following the sound shell model, in
an accompanying paper \cite{kin_sp_SSM}.
We have also addressed the issue of causality that motivated
the choice of initial conditions for sound waves in Ref.~\cite{Hindmarsh:2019phv}, but we defer a detailed discussion of this issue to Ref.~\cite{kin_sp_SSM}.

Our work has consequences on the interpretation of current observations
of pulsar timing arrays under the assumption that the QCD phase
transition is of first order.
There are several analyses in the literature that have used the $k^9$ spectrum and the inclusion of a $k^3$ tail could lead to
significantly different constraints on the phase transition
parameters.
This is especially important if one considers the smallest
frequency bins reported by the PTA collaborations, which are
below the characteristic frequency of the QCD phase transition
where the signal is expected to be dominated by the $k^3$
tail or by the intermediate linear growth, $k$.
Even at frequencies right below the peak, we expect
the $k^9$ behavior to be shallower.
Especially with the improvement of the PTA data in this range of
frequencies expected in the next years, the study of the
GW spectrum from sound waves with the presented modifications
will become completely relevant.

Similarly, our model has implications for current
estimations of the phase transition parameters that
can be probed by LISA when one considers a first-order electroweak
phase transition, since several analyses are currently using
the $k^9$ model for the GW signal.

At larger frequencies, our model can be used to test the potential
observability of higher-energy phase transitions with next-generation ground-based detectors, like Einstein Telescope or Cosmic Explorer, and to put constraints on the current and forthcoming observing runs
by the LIGO--Virgo--KAGRA collaborations, especially in view
of the advent of improvements in their sensitivities.\\

Data availability --- The calculations and routines to compute the
radial fluid profiles and the resulting spectra of
the velocity field, the anisotropic stresses, and the GWs presented in this work will be publicly available on GitHub \cite{GH},
alongside those used in the accompanying paper \cite{kin_sp_SSM}.

\begin{acknowledgements}
    We are grateful to Jorinde van de Vis and Mikko Laine for useful discussions, and to Ramkishor Sharma, Jani Dahl, Axel Brandenburg, and Mark Hindmarsh for sharing their draft \cite{Sharma:2023mao}.
   ARP is supported by the Swiss National Science Foundation (SNSF Ambizione grant \href{https://data.snf.ch/grants/grant/208807}{182044}). 
    CC and SP are supported by the Swiss National Science Foundation (SNSF Project Funding grant \href{https://data.snf.ch/grants/grant/212125}{212125}).
   SP is supported by the Swiss National Science Foundation under grant 
   \href{https://data.snf.ch/grants/grant/188712}{188712}.
   ARP and SP acknowledge the hospitality of CERN, where
   part of this work has taken place.
\end{acknowledgements}

\bigskip

\appendix
\addcontentsline{toc}{section}{Appendix}

\section{Full time evolution of the GW spectrum}
\label{app_GW_tdep}

In this section, we compute the time evolution of the GW
spectrum while the source is active, according to the sound shell
model.
The GW spectrum is usually averaged over oscillations
in time, as we are interested in its present time observable,
i.e., at very late times $\tau_0 \gg \tfin$.
However, if the GW spectrum is compared with the results from
simulations to, for example,
test the validity of the sound shell model, it is required to compute its exact time evolution while the source is active at $\tau<\tfin$. The average over
oscillations is then not well motivated and it could lead to
wrong results.
We note that one has to pay particular attention to this aspect
when using Weinberg's formula as, for example, in Refs.~\cite{Jinno:2020eqg,Jinno:2022mie}, since this approach
already assumes that the GWs have reached their free propagation 
regime at all $k$, which can potentially lead to wrong results
in the IR tail of the GW spectrum,
and it does not allow to study their time evolution.

We start with the GW spectrum, given by \Eq{OmGW_full},
and use the UETC of the anisotropic stresses of \Eq{eq_EPi}
with the stationary assumption for the velocity field UETC,
see \Eq{eq_EK_Eu_st}.
We then find an expression analogous to that of \Eq{OmGW_GG} but
in this case, the function $\Delta$ is a time-dependent expression
given as
\begin{align}
    \Delta  (\tau, &\,k, p, \tilde p) \equiv 2 \int_{\tau_*}^{\tau} \frac{\dd \tau_1}{\tau_1} \int_{\tau_*}^{\tau} \frac{\dd \tau_2}{\tau_2} \cos (\cs p \tau_-) 
    \nonumber \\
    \times &\, \cos (\cs \tilde p \tau_-)
    \cos k (\tau - \tau_1) \cos k (\tau - \tau_2).
\end{align}
We can express the product of $\cos$ as
\begin{equation}
    \cos (\cs p \tau_-) \cos (\cs \tilde p \tau_-) = \frac{1}{2}
    \sum_{m = \pm 1} \cos (\hat p_m \tau_-),
\end{equation}
with $\hat p_m = \cs (p + m \tilde p)$.
Then, using $\cos k (\tau - \tau_i) = \cos k \tau \cos k \tau_i + \sin k \tau \sin k \tau_i$ for $i = 1, 2$, one gets,
\begin{widetext}
\begin{align}
    \Delta   (\tau, k, p, \tilde p) 
    = \sum_{m = \pm 1}
    \Biggl[ \biggl(
    \int_{\tini}^{\tau} \frac{\dd \tau_1}{\tau_1} \bigl[
    \cos k \tau \cos k \tau_1 + &\, 
    \sin k \tau \sin k \tau_1] 
    \cos (\hat p_m \tau_1)
    \biggr)^2 \nonumber \\ +  \biggl(
    \int_{\tini}^{\tau} \frac{\dd \tau_1}{\tau_1} \bigl[&\, 
    \cos k \tau \cos k \tau_1 + 
    \sin k \tau \sin k \tau_1] 
    \sin (\hat p_m \tau_1) \biggr)^2 \Biggr] \nonumber \\ 
    = \frac{1}{4} \sum_{m, n = \pm 1} \biggl[
    \Delta \Ci^2 (\tau, \hat p_{mn}) +  \Delta \Si^2 (&\,\tau,  \hat p_{mn})  \nonumber\\
    + \cos 2k\tau \Bigl( \Delta\Ci (\tau,\hat p_{mn})&\Delta\Ci (\tau,\hat p_{m,-n}) +\Delta\Si (\tau,\hat p_{mn})\Delta\Si (\tau,\hat p_{m,-n}) \Bigr) \biggr] \,,
    \label{final_Delta_tau}
\end{align}
\end{widetext}
where the functions $\Delta \Ci_{mn}$ and $\Delta \Si_{mn}$
have been defined in \Eqs{Ci_Delta}{Si_Delta}.
We note that if one uses \Eq{GG_kp_stat} substituting $\tfin \to \tau$,
\Eq{final_Delta_tau} is not recovered, since the latter presents
an additional term that is relevant during the phase of GW
production.
Hence, when comparing to numerical simulations, one should use
\Eq{final_Delta_tau} to study the validity of the stationary
assumption for the UETC found in the sound shell model.

\section{GW spectrum in the infinite duration approximation}
\label{HH19}

In this section, we take the approximation of $\Delta$ as a Dirac
delta function [see \Eq{delta_Upsilon}] that has been used in
Refs.~\cite{Hindmarsh:2019phv,Guo:2020grp} to find the GW spectrum
from sound waves in the sound-shell model approximation.
We have shown in \Secs{stationary_UETC}{linear_vs_quadratic} that this assumption is not valid in the $k \to 0$ limit and have
presented the resulting GW spectrum in \Sec{numerical_GW}, so we compare here
what are the differences in the resulting spectral shape.

The GW spectrum, which we denote as HH19 (for Hindmarsh and Hijazi 2019 \cite{Hindmarsh:2019phv}),
is found substituting \Eq{delta_Upsilon} into \Eq{OmGW_GG},
\begin{align}
     \OmGW^{\rm HH19}(K) =  &\, \frac{3 \pi}{2}   K^2 \, 
     \Upsilon(\tfin) \, \frac{\HH_* R_*}{\cs} \, \bar w^2  
     \, {\cal T}_{\rm GW}
     \, \biggl(\frac{\OmK}{{\cal K}}\biggr)^2   \nonumber \\  \times \int_0^\infty P  &\, \zetaK (P) \dd P
    \int_{|P-K|}^{P+K} (1 - z^2)^2  \frac{\dd \tilde P}{\tilde P^3} \nonumber \\  \times \,  \zetaK (\tilde P&\,) \, 
    \delta(P + \tilde P - K/\cs)\ .
    \label{OMM}
\end{align}

Under this assumption,
one can perform the integral in \Eq{OMM}
over $\tilde P$ by substituting $\tilde P = K/\cs - P$ when
$|K-P| \leq K/\cs - P \leq K + P$, which yields the condition $P \in [P_-, P_+]$ being $P_{\pm} = \half K (1 \pm \cs)/\cs$,
and
\begin{equation}
	z = \frac{1}{\cs} - \frac{K(1 - \cs^2)}{2P\cs^2}.
\end{equation}
Then the GW spectrum becomes
\begin{align}
    \OmGW^{\rm HH19}(K) = &\,  \frac{3 \pi}{2}  K^2 \, \Upsilon(\tfin) \, \frac{\HH_* R_*}{\cs} \, \bar w^2 
    \, {\cal T}_{\rm GW}
    \, \biggl(\frac{\OmK}{{\cal K}}\biggr)^2   \nonumber \\  \times &\, \int_{P_-}^{P_+} P   \zetaK (P)
    (1 - z^2)^2 \nonumber \\ &\,  \times  \frac{\zetaK (K/\cs - P)}{(K/\cs - P)^3}
    \dd P.
    \label{OmGW_HH19}
\end{align}
The resulting GW spectrum is shown in \Figs{GW_comparison}{GW_comparison2}, compared with
the full calculation.
We find that \Eq{OmGW_HH19} provides a good approximation when
$k \gg 1/\delta \tfin$.

\bibliography{library}

\begin{thebibliography}{97}%
\makeatletter
\providecommand \@ifxundefined [1]{%
 \@ifx{#1\undefined}
}%
\providecommand \@ifnum [1]{%
 \ifnum #1\expandafter \@firstoftwo
 \else \expandafter \@secondoftwo
 \fi
}%
\providecommand \@ifx [1]{%
 \ifx #1\expandafter \@firstoftwo
 \else \expandafter \@secondoftwo
 \fi
}%
\providecommand \natexlab [1]{#1}%
\providecommand \enquote  [1]{``#1''}%
\providecommand \bibnamefont  [1]{#1}%
\providecommand \bibfnamefont [1]{#1}%
\providecommand \citenamefont [1]{#1}%
\providecommand \href@noop [0]{\@secondoftwo}%
\providecommand \href [0]{\begingroup \@sanitize@url \@href}%
\providecommand \@href[1]{\@@startlink{#1}\@@href}%
\providecommand \@@href[1]{\endgroup#1\@@endlink}%
\providecommand \@sanitize@url [0]{\catcode `\\12\catcode `\$12\catcode
  `\&12\catcode `\#12\catcode `\^12\catcode `\_12\catcode `\%12\relax}%
\providecommand \@@startlink[1]{}%
\providecommand \@@endlink[0]{}%
\providecommand \url  [0]{\begingroup\@sanitize@url \@url }%
\providecommand \@url [1]{\endgroup\@href {#1}{\urlprefix }}%
\providecommand \urlprefix  [0]{URL }%
\providecommand \Eprint [0]{\href }%
\providecommand \doibase [0]{http://dx.doi.org/}%
\providecommand \selectlanguage [0]{\@gobble}%
\providecommand \bibinfo  [0]{\@secondoftwo}%
\providecommand \bibfield  [0]{\@secondoftwo}%
\providecommand \translation [1]{[#1]}%
\providecommand \BibitemOpen [0]{}%
\providecommand \bibitemStop [0]{}%
\providecommand \bibitemNoStop [0]{.\EOS\space}%
\providecommand \EOS [0]{\spacefactor3000\relax}%
\providecommand \BibitemShut  [1]{\csname bibitem#1\endcsname}%
\let\auto@bib@innerbib\@empty
\bibitem [{\citenamefont {Kirzhnits}\ and\ \citenamefont
  {Linde}(1976)}]{Kirzhnits:1976ts}%
  \BibitemOpen
  \bibfield  {author} {\bibinfo {author} {\bibfnamefont {D.~A.}\ \bibnamefont
  {Kirzhnits}}\ and\ \bibinfo {author} {\bibfnamefont {Andrei~D.}\ \bibnamefont
  {Linde}},\ }\bibfield  {title} {\enquote {\bibinfo {title} {{Symmetry
  behavior in gauge theories}},}\ }\href {\doibase
  10.1016/0003-4916(76)90279-7} {\bibfield  {journal} {\bibinfo  {journal}
  {Ann. Phys.}\ }\textbf {\bibinfo {volume} {101}},\ \bibinfo {pages} {195}
  (\bibinfo {year} {1976})}\BibitemShut {NoStop}%
\bibitem [{\citenamefont {Coleman}(1977)}]{Coleman:1977py}%
  \BibitemOpen
  \bibfield  {author} {\bibinfo {author} {\bibfnamefont {Sidney~R.}\
  \bibnamefont {Coleman}},\ }\bibfield  {title} {\enquote {\bibinfo {title}
  {{The fate of the false vacuum. 1. Semiclassical theory}},}\ }\href {\doibase
  10.1103/PhysRevD.16.1248} {\bibfield  {journal} {\bibinfo  {journal} {Phys.
  Rev. D}\ }\textbf {\bibinfo {volume} {15}},\ \bibinfo {pages} {2929}
  (\bibinfo {year} {1977})},\ \bibinfo {note} {[Erratum: Phys.~Rev.~D {\bf 16},
  1248 (1977)]}\BibitemShut {NoStop}%
\bibitem [{\citenamefont {Linde}(1983)}]{Linde:1981zj}%
  \BibitemOpen
  \bibfield  {author} {\bibinfo {author} {\bibfnamefont {Andrei~D.}\
  \bibnamefont {Linde}},\ }\bibfield  {title} {\enquote {\bibinfo {title}
  {{Decay of the false vacuum at finite temperature}},}\ }\href {\doibase
  10.1016/0550-3213(83)90072-X} {\bibfield  {journal} {\bibinfo  {journal}
  {Nucl. Phys. B}\ }\textbf {\bibinfo {volume} {216}},\ \bibinfo {pages} {421}
  (\bibinfo {year} {1983})},\ \bibinfo {note} {[Erratum: Nucl.~Phys.~B {\bf
  223}, 544 (1983)]}\BibitemShut {NoStop}%
\bibitem [{\citenamefont {Kajantie}\ \emph {et~al.}(1997)\citenamefont
  {Kajantie}, \citenamefont {Laine}, \citenamefont {Rummukainen},\ and\
  \citenamefont {Shaposhnikov}}]{Kajantie:1996qd}%
  \BibitemOpen
  \bibfield  {author} {\bibinfo {author} {\bibfnamefont {K.}~\bibnamefont
  {Kajantie}}, \bibinfo {author} {\bibfnamefont {M.}~\bibnamefont {Laine}},
  \bibinfo {author} {\bibfnamefont {K.}~\bibnamefont {Rummukainen}}, \ and\
  \bibinfo {author} {\bibfnamefont {Mikhail~E.}\ \bibnamefont {Shaposhnikov}},\
  }\bibfield  {title} {\enquote {\bibinfo {title} {{A nonperturbative analysis
  of the finite T phase transition in SU(2) x U(1) electroweak theory}},}\
  }\href {\doibase 10.1016/S0550-3213(97)00164-8} {\bibfield  {journal}
  {\bibinfo  {journal} {Nucl. Phys. B}\ }\textbf {\bibinfo {volume} {493}},\
  \bibinfo {pages} {413} (\bibinfo {year} {1997})},\ \Eprint
  {http://arxiv.org/abs/hep-lat/9612006} {arXiv:hep-lat/9612006} \BibitemShut
  {NoStop}%
\bibitem [{\citenamefont {Stephanov}(2006)}]{Stephanov:2006dn}%
  \BibitemOpen
  \bibfield  {author} {\bibinfo {author} {\bibfnamefont {M.~A.}\ \bibnamefont
  {Stephanov}},\ }\bibfield  {title} {\enquote {\bibinfo {title} {{QCD critical
  point and complex chemical potential singularities}},}\ }\href {\doibase
  10.1103/PhysRevD.73.094508} {\bibfield  {journal} {\bibinfo  {journal} {Phys.
  Rev. D}\ }\textbf {\bibinfo {volume} {73}},\ \bibinfo {pages} {094508}
  (\bibinfo {year} {2006})},\ \Eprint {http://arxiv.org/abs/hep-lat/0603014}
  {arXiv:hep-lat/0603014} \BibitemShut {NoStop}%
\bibitem [{\citenamefont {Caprini}\ \emph {et~al.}(2020)\citenamefont {Caprini}
  \emph {et~al.}}]{Caprini:2019egz}%
  \BibitemOpen
  \bibfield  {author} {\bibinfo {author} {\bibfnamefont {Chiara}\ \bibnamefont
  {Caprini}} \emph {et~al.},\ }\bibfield  {title} {\enquote {\bibinfo {title}
  {{Detecting gravitational waves from cosmological phase transitions with
  LISA: An update}},}\ }\href {\doibase 10.1088/1475-7516/2020/03/024}
  {\bibfield  {journal} {\bibinfo  {journal} {J. Cosmol. Astropart. Phys.}\
  }\textbf {\bibinfo {volume} {03}},\ \bibinfo {pages} {024} (\bibinfo {year}
  {2020})},\ \Eprint {http://arxiv.org/abs/1910.13125} {arXiv:1910.13125
  [astro-ph.CO]} \BibitemShut {NoStop}%
\bibitem [{\citenamefont {Schwarz}\ and\ \citenamefont
  {Stuke}(2009)}]{Schwarz:2009ii}%
  \BibitemOpen
  \bibfield  {author} {\bibinfo {author} {\bibfnamefont {Dominik~J.}\
  \bibnamefont {Schwarz}}\ and\ \bibinfo {author} {\bibfnamefont {Maik}\
  \bibnamefont {Stuke}},\ }\bibfield  {title} {\enquote {\bibinfo {title}
  {{Lepton asymmetry and the cosmic QCD transition}},}\ }\href {\doibase
  10.1088/1475-7516/2009/11/025} {\bibfield  {journal} {\bibinfo  {journal} {J.
  Cosmol. Astropart. Phys.}\ }\textbf {\bibinfo {volume} {11}},\ \bibinfo
  {pages} {025} (\bibinfo {year} {2009})},\ \bibinfo {note} {[Erratum: J.
  Cosmol. Astropart. Phys. {\bf 10}, E01 (2010)]},\ \Eprint
  {http://arxiv.org/abs/0906.3434} {arXiv:0906.3434 [hep-ph]} \BibitemShut
  {NoStop}%
\bibitem [{\citenamefont {Wygas}\ \emph {et~al.}(2018)\citenamefont {Wygas},
  \citenamefont {Oldengott}, \citenamefont {B\"odeker},\ and\ \citenamefont
  {Schwarz}}]{Wygas:2018otj}%
  \BibitemOpen
  \bibfield  {author} {\bibinfo {author} {\bibfnamefont {Mandy~M.}\
  \bibnamefont {Wygas}}, \bibinfo {author} {\bibfnamefont {Isabel~M.}\
  \bibnamefont {Oldengott}}, \bibinfo {author} {\bibfnamefont {Dietrich}\
  \bibnamefont {B\"odeker}}, \ and\ \bibinfo {author} {\bibfnamefont
  {Dominik~J.}\ \bibnamefont {Schwarz}},\ }\bibfield  {title} {\enquote
  {\bibinfo {title} {{Cosmic QCD epoch at nonvanishing lepton asymmetry}},}\
  }\href {\doibase 10.1103/PhysRevLett.121.201302} {\bibfield  {journal}
  {\bibinfo  {journal} {Phys. Rev. Lett.}\ }\textbf {\bibinfo {volume} {121}},\
  \bibinfo {pages} {201302} (\bibinfo {year} {2018})},\ \Eprint
  {http://arxiv.org/abs/1807.10815} {arXiv:1807.10815 [hep-ph]} \BibitemShut
  {NoStop}%
\bibitem [{\citenamefont {Middeldorf-Wygas}\ \emph {et~al.}(2022)\citenamefont
  {Middeldorf-Wygas}, \citenamefont {Oldengott}, \citenamefont {B\"odeker},\
  and\ \citenamefont {Schwarz}}]{Middeldorf-Wygas:2020glx}%
  \BibitemOpen
  \bibfield  {author} {\bibinfo {author} {\bibfnamefont {Mandy~M.}\
  \bibnamefont {Middeldorf-Wygas}}, \bibinfo {author} {\bibfnamefont
  {Isabel~M.}\ \bibnamefont {Oldengott}}, \bibinfo {author} {\bibfnamefont
  {Dietrich}\ \bibnamefont {B\"odeker}}, \ and\ \bibinfo {author}
  {\bibfnamefont {Dominik~J.}\ \bibnamefont {Schwarz}},\ }\bibfield  {title}
  {\enquote {\bibinfo {title} {{Cosmic QCD transition for large lepton flavor
  asymmetries}},}\ }\href {\doibase 10.1103/PhysRevD.105.123533} {\bibfield
  {journal} {\bibinfo  {journal} {Phys. Rev. D}\ }\textbf {\bibinfo {volume}
  {105}},\ \bibinfo {pages} {123533} (\bibinfo {year} {2022})},\ \Eprint
  {http://arxiv.org/abs/2009.00036} {arXiv:2009.00036 [hep-ph]} \BibitemShut
  {NoStop}%
\bibitem [{\citenamefont {Vovchenko}\ \emph {et~al.}(2021)\citenamefont
  {Vovchenko}, \citenamefont {Brandt}, \citenamefont {Cuteri}, \citenamefont
  {Endr\H{o}di}, \citenamefont {Hajkarim},\ and\ \citenamefont
  {Schaffner-Bielich}}]{Vovchenko:2020crk}%
  \BibitemOpen
  \bibfield  {author} {\bibinfo {author} {\bibfnamefont {Volodymyr}\
  \bibnamefont {Vovchenko}}, \bibinfo {author} {\bibfnamefont {Bastian~B.}\
  \bibnamefont {Brandt}}, \bibinfo {author} {\bibfnamefont {Francesca}\
  \bibnamefont {Cuteri}}, \bibinfo {author} {\bibfnamefont {Gergely}\
  \bibnamefont {Endr\H{o}di}}, \bibinfo {author} {\bibfnamefont {Fazlollah}\
  \bibnamefont {Hajkarim}}, \ and\ \bibinfo {author} {\bibfnamefont {J\"urgen}\
  \bibnamefont {Schaffner-Bielich}},\ }\bibfield  {title} {\enquote {\bibinfo
  {title} {{Pion condensation in the early universe at nonvanishing lepton
  flavor asymmetry and its gravitational wave signatures}},}\ }\href {\doibase
  10.1103/PhysRevLett.126.012701} {\bibfield  {journal} {\bibinfo  {journal}
  {Phys. Rev. Lett.}\ }\textbf {\bibinfo {volume} {126}},\ \bibinfo {pages}
  {012701} (\bibinfo {year} {2021})},\ \Eprint
  {http://arxiv.org/abs/2009.02309} {arXiv:2009.02309 [hep-ph]} \BibitemShut
  {NoStop}%
\bibitem [{\citenamefont {Cao}(2023)}]{Cao:2022fow}%
  \BibitemOpen
  \bibfield  {author} {\bibinfo {author} {\bibfnamefont {Gaoqing}\ \bibnamefont
  {Cao}},\ }\bibfield  {title} {\enquote {\bibinfo {title} {{First-order QCD
  transition in a primordial magnetic field}},}\ }\href {\doibase
  10.1103/PhysRevD.107.014021} {\bibfield  {journal} {\bibinfo  {journal}
  {Phys. Rev. D}\ }\textbf {\bibinfo {volume} {107}},\ \bibinfo {pages}
  {014021} (\bibinfo {year} {2023})},\ \Eprint
  {http://arxiv.org/abs/2210.09794} {arXiv:2210.09794 [nucl-th]} \BibitemShut
  {NoStop}%
\bibitem [{\citenamefont {Steinhardt}(1982)}]{Steinhardt:1981ct}%
  \BibitemOpen
  \bibfield  {author} {\bibinfo {author} {\bibfnamefont {Paul~Joseph}\
  \bibnamefont {Steinhardt}},\ }\bibfield  {title} {\enquote {\bibinfo {title}
  {{Relativistic detonation waves and bubble growth in false vacuum decay}},}\
  }\href {\doibase 10.1103/PhysRevD.25.2074} {\bibfield  {journal} {\bibinfo
  {journal} {Phys. Rev. D}\ }\textbf {\bibinfo {volume} {25}},\ \bibinfo
  {pages} {2074} (\bibinfo {year} {1982})}\BibitemShut {NoStop}%
\bibitem [{\citenamefont {Enqvist}\ \emph {et~al.}(1992)\citenamefont
  {Enqvist}, \citenamefont {Ignatius}, \citenamefont {Kajantie},\ and\
  \citenamefont {Rummukainen}}]{Enqvist:1991xw}%
  \BibitemOpen
  \bibfield  {author} {\bibinfo {author} {\bibfnamefont {K.}~\bibnamefont
  {Enqvist}}, \bibinfo {author} {\bibfnamefont {J.}~\bibnamefont {Ignatius}},
  \bibinfo {author} {\bibfnamefont {K.}~\bibnamefont {Kajantie}}, \ and\
  \bibinfo {author} {\bibfnamefont {K.}~\bibnamefont {Rummukainen}},\
  }\bibfield  {title} {\enquote {\bibinfo {title} {{Nucleation and bubble
  growth in a first order cosmological electroweak phase transition}},}\ }\href
  {\doibase 10.1103/PhysRevD.45.3415} {\bibfield  {journal} {\bibinfo
  {journal} {Phys. Rev. D}\ }\textbf {\bibinfo {volume} {45}},\ \bibinfo
  {pages} {3415} (\bibinfo {year} {1992})}\BibitemShut {NoStop}%
\bibitem [{\citenamefont {Ignatius}\ \emph {et~al.}(1994)\citenamefont
  {Ignatius}, \citenamefont {Kajantie}, \citenamefont {Kurki-Suonio},\ and\
  \citenamefont {Laine}}]{Ignatius:1993qn}%
  \BibitemOpen
  \bibfield  {author} {\bibinfo {author} {\bibfnamefont {J.}~\bibnamefont
  {Ignatius}}, \bibinfo {author} {\bibfnamefont {K.}~\bibnamefont {Kajantie}},
  \bibinfo {author} {\bibfnamefont {H.}~\bibnamefont {Kurki-Suonio}}, \ and\
  \bibinfo {author} {\bibfnamefont {M.}~\bibnamefont {Laine}},\ }\bibfield
  {title} {\enquote {\bibinfo {title} {{The growth of bubbles in cosmological
  phase transitions}},}\ }\href {\doibase 10.1103/PhysRevD.49.3854} {\bibfield
  {journal} {\bibinfo  {journal} {Phys. Rev. D}\ }\textbf {\bibinfo {volume}
  {49}},\ \bibinfo {pages} {3854} (\bibinfo {year} {1994})},\ \Eprint
  {http://arxiv.org/abs/astro-ph/9309059} {arXiv:astro-ph/9309059} \BibitemShut
  {NoStop}%
\bibitem [{\citenamefont {Witten}(1984)}]{Witten:1984rs}%
  \BibitemOpen
  \bibfield  {author} {\bibinfo {author} {\bibfnamefont {Edward}\ \bibnamefont
  {Witten}},\ }\bibfield  {title} {\enquote {\bibinfo {title} {{Cosmic
  separation of phases}},}\ }\href {\doibase 10.1103/PhysRevD.30.272}
  {\bibfield  {journal} {\bibinfo  {journal} {Phys. Rev. D}\ }\textbf {\bibinfo
  {volume} {30}},\ \bibinfo {pages} {272} (\bibinfo {year} {1984})}\BibitemShut
  {NoStop}%
\bibitem [{\citenamefont {Hogan}(1986)}]{Hogan:1986qda}%
  \BibitemOpen
  \bibfield  {author} {\bibinfo {author} {\bibfnamefont {C.~J.}\ \bibnamefont
  {Hogan}},\ }\bibfield  {title} {\enquote {\bibinfo {title} {{Gravitational
  radiation from cosmological phase transitions}},}\ }\href@noop {} {\bibfield
  {journal} {\bibinfo  {journal} {Mon. Not. R. Astron. Soc.}\ }\textbf
  {\bibinfo {volume} {218}},\ \bibinfo {pages} {629} (\bibinfo {year}
  {1986})}\BibitemShut {NoStop}%
\bibitem [{\citenamefont {Kosowsky}\ \emph
  {et~al.}(1992{\natexlab{a}})\citenamefont {Kosowsky}, \citenamefont
  {Turner},\ and\ \citenamefont {Watkins}}]{Kosowsky:1991ua}%
  \BibitemOpen
  \bibfield  {author} {\bibinfo {author} {\bibfnamefont {Arthur}\ \bibnamefont
  {Kosowsky}}, \bibinfo {author} {\bibfnamefont {Michael~S.}\ \bibnamefont
  {Turner}}, \ and\ \bibinfo {author} {\bibfnamefont {Richard}\ \bibnamefont
  {Watkins}},\ }\bibfield  {title} {\enquote {\bibinfo {title} {{Gravitational
  radiation from colliding vacuum bubbles}},}\ }\href {\doibase
  10.1103/PhysRevD.45.4514} {\bibfield  {journal} {\bibinfo  {journal} {Phys.
  Rev. D}\ }\textbf {\bibinfo {volume} {45}},\ \bibinfo {pages} {4514}
  (\bibinfo {year} {1992}{\natexlab{a}})}\BibitemShut {NoStop}%
\bibitem [{\citenamefont {Kosowsky}\ \emph
  {et~al.}(1992{\natexlab{b}})\citenamefont {Kosowsky}, \citenamefont
  {Turner},\ and\ \citenamefont {Watkins}}]{Kosowsky:1992rz}%
  \BibitemOpen
  \bibfield  {author} {\bibinfo {author} {\bibfnamefont {Arthur}\ \bibnamefont
  {Kosowsky}}, \bibinfo {author} {\bibfnamefont {Michael~S.}\ \bibnamefont
  {Turner}}, \ and\ \bibinfo {author} {\bibfnamefont {Richard}\ \bibnamefont
  {Watkins}},\ }\bibfield  {title} {\enquote {\bibinfo {title} {{Gravitational
  waves from first order cosmological phase transitions}},}\ }\href {\doibase
  10.1103/PhysRevLett.69.2026} {\bibfield  {journal} {\bibinfo  {journal}
  {Phys. Rev. Lett.}\ }\textbf {\bibinfo {volume} {69}},\ \bibinfo {pages}
  {2026--2029} (\bibinfo {year} {1992}{\natexlab{b}})}\BibitemShut {NoStop}%
\bibitem [{\citenamefont {Kosowsky}\ and\ \citenamefont
  {Turner}(1993)}]{Kosowsky:1992vn}%
  \BibitemOpen
  \bibfield  {author} {\bibinfo {author} {\bibfnamefont {Arthur}\ \bibnamefont
  {Kosowsky}}\ and\ \bibinfo {author} {\bibfnamefont {Michael~S.}\ \bibnamefont
  {Turner}},\ }\bibfield  {title} {\enquote {\bibinfo {title} {{Gravitational
  radiation from colliding vacuum bubbles: Envelope approximation to many
  bubble collisions}},}\ }\href {\doibase 10.1103/PhysRevD.47.4372} {\bibfield
  {journal} {\bibinfo  {journal} {Phys. Rev. D}\ }\textbf {\bibinfo {volume}
  {47}},\ \bibinfo {pages} {4372} (\bibinfo {year} {1993})},\ \Eprint
  {http://arxiv.org/abs/astro-ph/9211004} {arXiv:astro-ph/9211004} \BibitemShut
  {NoStop}%
\bibitem [{\citenamefont {Caprini}\ \emph {et~al.}(2008)\citenamefont
  {Caprini}, \citenamefont {Durrer},\ and\ \citenamefont
  {Servant}}]{Caprini:2007xq}%
  \BibitemOpen
  \bibfield  {author} {\bibinfo {author} {\bibfnamefont {Chiara}\ \bibnamefont
  {Caprini}}, \bibinfo {author} {\bibfnamefont {Ruth}\ \bibnamefont {Durrer}},
  \ and\ \bibinfo {author} {\bibfnamefont {Geraldine}\ \bibnamefont
  {Servant}},\ }\bibfield  {title} {\enquote {\bibinfo {title} {{Gravitational
  wave generation from bubble collisions in first-order phase transitions: An
  analytic approach}},}\ }\href {\doibase 10.1103/PhysRevD.77.124015}
  {\bibfield  {journal} {\bibinfo  {journal} {Phys. Rev. D}\ }\textbf {\bibinfo
  {volume} {77}},\ \bibinfo {pages} {124015} (\bibinfo {year} {2008})},\
  \Eprint {http://arxiv.org/abs/0711.2593} {arXiv:0711.2593 [astro-ph]}
  \BibitemShut {NoStop}%
\bibitem [{\citenamefont {Huber}\ and\ \citenamefont
  {Konstandin}(2008{\natexlab{a}})}]{Huber:2008hg}%
  \BibitemOpen
  \bibfield  {author} {\bibinfo {author} {\bibfnamefont {Stephan~J.}\
  \bibnamefont {Huber}}\ and\ \bibinfo {author} {\bibfnamefont {Thomas}\
  \bibnamefont {Konstandin}},\ }\bibfield  {title} {\enquote {\bibinfo {title}
  {{Gravitational wave production by collisions: More bubbles}},}\ }\href
  {\doibase 10.1088/1475-7516/2008/09/022} {\bibfield  {journal} {\bibinfo
  {journal} {J. Cosmol. Astropart. Phys.}\ }\textbf {\bibinfo {volume} {09}},\
  \bibinfo {pages} {022} (\bibinfo {year} {2008}{\natexlab{a}})},\ \Eprint
  {http://arxiv.org/abs/0806.1828} {arXiv:0806.1828 [hep-ph]} \BibitemShut
  {NoStop}%
\bibitem [{\citenamefont {Jinno}\ and\ \citenamefont
  {Takimoto}(2019)}]{Jinno:2017fby}%
  \BibitemOpen
  \bibfield  {author} {\bibinfo {author} {\bibfnamefont {Ryusuke}\ \bibnamefont
  {Jinno}}\ and\ \bibinfo {author} {\bibfnamefont {Masahiro}\ \bibnamefont
  {Takimoto}},\ }\bibfield  {title} {\enquote {\bibinfo {title} {{Gravitational
  waves from bubble dynamics: Beyond the envelope}},}\ }\href {\doibase
  10.1088/1475-7516/2019/01/060} {\bibfield  {journal} {\bibinfo  {journal} {J.
  Cosmol. Astropart. Phys.}\ }\textbf {\bibinfo {volume} {01}},\ \bibinfo
  {pages} {060} (\bibinfo {year} {2019})},\ \Eprint
  {http://arxiv.org/abs/1707.03111} {arXiv:1707.03111 [hep-ph]} \BibitemShut
  {NoStop}%
\bibitem [{\citenamefont {Cutting}\ \emph {et~al.}(2018)\citenamefont
  {Cutting}, \citenamefont {Hindmarsh},\ and\ \citenamefont
  {Weir}}]{Cutting:2018tjt}%
  \BibitemOpen
  \bibfield  {author} {\bibinfo {author} {\bibfnamefont {Daniel}\ \bibnamefont
  {Cutting}}, \bibinfo {author} {\bibfnamefont {Mark}\ \bibnamefont
  {Hindmarsh}}, \ and\ \bibinfo {author} {\bibfnamefont {David~J.}\
  \bibnamefont {Weir}},\ }\bibfield  {title} {\enquote {\bibinfo {title}
  {{Gravitational waves from vacuum first-order phase transitions: From the
  envelope to the lattice}},}\ }\href {\doibase 10.1103/PhysRevD.97.123513}
  {\bibfield  {journal} {\bibinfo  {journal} {Phys. Rev. D}\ }\textbf {\bibinfo
  {volume} {97}},\ \bibinfo {pages} {123513} (\bibinfo {year} {2018})},\
  \Eprint {http://arxiv.org/abs/1802.05712} {arXiv:1802.05712 [astro-ph.CO]}
  \BibitemShut {NoStop}%
\bibitem [{\citenamefont {Hindmarsh}\ \emph {et~al.}(2014)\citenamefont
  {Hindmarsh}, \citenamefont {Huber}, \citenamefont {Rummukainen},\ and\
  \citenamefont {Weir}}]{Hindmarsh:2013xza}%
  \BibitemOpen
  \bibfield  {author} {\bibinfo {author} {\bibfnamefont {Mark}\ \bibnamefont
  {Hindmarsh}}, \bibinfo {author} {\bibfnamefont {Stephan~J.}\ \bibnamefont
  {Huber}}, \bibinfo {author} {\bibfnamefont {Kari}\ \bibnamefont
  {Rummukainen}}, \ and\ \bibinfo {author} {\bibfnamefont {David~J.}\
  \bibnamefont {Weir}},\ }\bibfield  {title} {\enquote {\bibinfo {title}
  {{Gravitational waves from the sound of a first order phase transition}},}\
  }\href {\doibase 10.1103/PhysRevLett.112.041301} {\bibfield  {journal}
  {\bibinfo  {journal} {Phys. Rev. Lett.}\ }\textbf {\bibinfo {volume} {112}},\
  \bibinfo {pages} {041301} (\bibinfo {year} {2014})},\ \Eprint
  {http://arxiv.org/abs/1304.2433} {arXiv:1304.2433 [hep-ph]} \BibitemShut
  {NoStop}%
\bibitem [{\citenamefont {Hindmarsh}\ \emph {et~al.}(2015)\citenamefont
  {Hindmarsh}, \citenamefont {Huber}, \citenamefont {Rummukainen},\ and\
  \citenamefont {Weir}}]{Hindmarsh:2015qta}%
  \BibitemOpen
  \bibfield  {author} {\bibinfo {author} {\bibfnamefont {Mark}\ \bibnamefont
  {Hindmarsh}}, \bibinfo {author} {\bibfnamefont {Stephan~J.}\ \bibnamefont
  {Huber}}, \bibinfo {author} {\bibfnamefont {Kari}\ \bibnamefont
  {Rummukainen}}, \ and\ \bibinfo {author} {\bibfnamefont {David~J.}\
  \bibnamefont {Weir}},\ }\bibfield  {title} {\enquote {\bibinfo {title}
  {{Numerical simulations of acoustically generated gravitational waves at a
  first order phase transition}},}\ }\href {\doibase
  10.1103/PhysRevD.92.123009} {\bibfield  {journal} {\bibinfo  {journal} {Phys.
  Rev. D}\ }\textbf {\bibinfo {volume} {92}},\ \bibinfo {pages} {123009}
  (\bibinfo {year} {2015})},\ \Eprint {http://arxiv.org/abs/1504.03291}
  {arXiv:1504.03291 [astro-ph.CO]} \BibitemShut {NoStop}%
\bibitem [{\citenamefont {Hindmarsh}(2018)}]{Hindmarsh:2016lnk}%
  \BibitemOpen
  \bibfield  {author} {\bibinfo {author} {\bibfnamefont {Mark}\ \bibnamefont
  {Hindmarsh}},\ }\bibfield  {title} {\enquote {\bibinfo {title} {{Sound shell
  model for acoustic gravitational wave production at a first-order phase
  transition in the early Universe}},}\ }\href {\doibase
  10.1103/PhysRevLett.120.071301} {\bibfield  {journal} {\bibinfo  {journal}
  {Phys. Rev. Lett.}\ }\textbf {\bibinfo {volume} {120}},\ \bibinfo {pages}
  {071301} (\bibinfo {year} {2018})},\ \Eprint
  {http://arxiv.org/abs/1608.04735} {arXiv:1608.04735 [astro-ph.CO]}
  \BibitemShut {NoStop}%
\bibitem [{\citenamefont {Hindmarsh}\ \emph {et~al.}(2017)\citenamefont
  {Hindmarsh}, \citenamefont {Huber}, \citenamefont {Rummukainen},\ and\
  \citenamefont {Weir}}]{Hindmarsh:2017gnf}%
  \BibitemOpen
  \bibfield  {author} {\bibinfo {author} {\bibfnamefont {Mark}\ \bibnamefont
  {Hindmarsh}}, \bibinfo {author} {\bibfnamefont {Stephan~J.}\ \bibnamefont
  {Huber}}, \bibinfo {author} {\bibfnamefont {Kari}\ \bibnamefont
  {Rummukainen}}, \ and\ \bibinfo {author} {\bibfnamefont {David~J.}\
  \bibnamefont {Weir}},\ }\bibfield  {title} {\enquote {\bibinfo {title}
  {{Shape of the acoustic gravitational wave power spectrum from a first order
  phase transition}},}\ }\href {\doibase 10.1103/PhysRevD.96.103520} {\bibfield
   {journal} {\bibinfo  {journal} {Phys. Rev. D}\ }\textbf {\bibinfo {volume}
  {96}},\ \bibinfo {pages} {103520} (\bibinfo {year} {2017})},\ \bibinfo {note}
  {[Erratum: Phys.~Rev.~D {\bf 101}, 089902 (2020)]},\ \Eprint
  {http://arxiv.org/abs/1704.05871} {arXiv:1704.05871 [astro-ph.CO]}
  \BibitemShut {NoStop}%
\bibitem [{\citenamefont {Hindmarsh}\ and\ \citenamefont
  {Hijazi}(2019)}]{Hindmarsh:2019phv}%
  \BibitemOpen
  \bibfield  {author} {\bibinfo {author} {\bibfnamefont {Mark}\ \bibnamefont
  {Hindmarsh}}\ and\ \bibinfo {author} {\bibfnamefont {Mulham}\ \bibnamefont
  {Hijazi}},\ }\bibfield  {title} {\enquote {\bibinfo {title} {{Gravitational
  waves from first order cosmological phase transitions in the sound shell
  model}},}\ }\href {\doibase 10.1088/1475-7516/2019/12/062} {\bibfield
  {journal} {\bibinfo  {journal} {J. Cosmol. Astropart. Phys.}\ }\textbf
  {\bibinfo {volume} {12}},\ \bibinfo {pages} {062} (\bibinfo {year} {2019})},\
  \Eprint {http://arxiv.org/abs/1909.10040} {arXiv:1909.10040 [astro-ph.CO]}
  \BibitemShut {NoStop}%
\bibitem [{\citenamefont {Jinno}\ \emph {et~al.}(2021)\citenamefont {Jinno},
  \citenamefont {Konstandin},\ and\ \citenamefont {Rubira}}]{Jinno:2020eqg}%
  \BibitemOpen
  \bibfield  {author} {\bibinfo {author} {\bibfnamefont {Ryusuke}\ \bibnamefont
  {Jinno}}, \bibinfo {author} {\bibfnamefont {Thomas}\ \bibnamefont
  {Konstandin}}, \ and\ \bibinfo {author} {\bibfnamefont {Henrique}\
  \bibnamefont {Rubira}},\ }\bibfield  {title} {\enquote {\bibinfo {title} {{A
  hybrid simulation of gravitational wave production in first-order phase
  transitions}},}\ }\href {\doibase 10.1088/1475-7516/2021/04/014} {\bibfield
  {journal} {\bibinfo  {journal} {J. Cosmol. Astropart. Phys.}\ }\textbf
  {\bibinfo {volume} {04}},\ \bibinfo {pages} {014} (\bibinfo {year} {2021})},\
  \Eprint {http://arxiv.org/abs/2010.00971} {arXiv:2010.00971 [astro-ph.CO]}
  \BibitemShut {NoStop}%
\bibitem [{\citenamefont {Jinno}\ \emph {et~al.}(2023)\citenamefont {Jinno},
  \citenamefont {Konstandin}, \citenamefont {Rubira},\ and\ \citenamefont
  {Stomberg}}]{Jinno:2022mie}%
  \BibitemOpen
  \bibfield  {author} {\bibinfo {author} {\bibfnamefont {Ryusuke}\ \bibnamefont
  {Jinno}}, \bibinfo {author} {\bibfnamefont {Thomas}\ \bibnamefont
  {Konstandin}}, \bibinfo {author} {\bibfnamefont {Henrique}\ \bibnamefont
  {Rubira}}, \ and\ \bibinfo {author} {\bibfnamefont {Isak}\ \bibnamefont
  {Stomberg}},\ }\bibfield  {title} {\enquote {\bibinfo {title} {{Higgsless
  simulations of cosmological phase transitions and gravitational waves}},}\
  }\href {\doibase 10.1088/1475-7516/2023/02/011} {\bibfield  {journal}
  {\bibinfo  {journal} {J. Cosmol. Astropart. Phys.}\ }\textbf {\bibinfo
  {volume} {02}},\ \bibinfo {pages} {011} (\bibinfo {year} {2023})},\ \Eprint
  {http://arxiv.org/abs/2209.04369} {arXiv:2209.04369 [astro-ph.CO]}
  \BibitemShut {NoStop}%
\bibitem [{\citenamefont {Kosowsky}\ \emph {et~al.}(2002)\citenamefont
  {Kosowsky}, \citenamefont {Mack},\ and\ \citenamefont
  {Kahniashvili}}]{Kosowsky:2001xp}%
  \BibitemOpen
  \bibfield  {author} {\bibinfo {author} {\bibfnamefont {Arthur}\ \bibnamefont
  {Kosowsky}}, \bibinfo {author} {\bibfnamefont {Andrew}\ \bibnamefont {Mack}},
  \ and\ \bibinfo {author} {\bibfnamefont {Tinatin}\ \bibnamefont
  {Kahniashvili}},\ }\bibfield  {title} {\enquote {\bibinfo {title}
  {{Gravitational radiation from cosmological turbulence}},}\ }\href {\doibase
  10.1103/PhysRevD.66.024030} {\bibfield  {journal} {\bibinfo  {journal} {Phys.
  Rev. D}\ }\textbf {\bibinfo {volume} {66}},\ \bibinfo {pages} {024030}
  (\bibinfo {year} {2002})},\ \Eprint {http://arxiv.org/abs/astro-ph/0111483}
  {arXiv:astro-ph/0111483} \BibitemShut {NoStop}%
\bibitem [{\citenamefont {Gogoberidze}\ \emph {et~al.}(2007)\citenamefont
  {Gogoberidze}, \citenamefont {Kahniashvili},\ and\ \citenamefont
  {Kosowsky}}]{Gogoberidze:2007an}%
  \BibitemOpen
  \bibfield  {author} {\bibinfo {author} {\bibfnamefont {Grigol}\ \bibnamefont
  {Gogoberidze}}, \bibinfo {author} {\bibfnamefont {Tina}\ \bibnamefont
  {Kahniashvili}}, \ and\ \bibinfo {author} {\bibfnamefont {Arthur}\
  \bibnamefont {Kosowsky}},\ }\bibfield  {title} {\enquote {\bibinfo {title}
  {{The spectrum of gravitational radiation from primordial turbulence}},}\
  }\href {\doibase 10.1103/PhysRevD.76.083002} {\bibfield  {journal} {\bibinfo
  {journal} {Phys. Rev. D}\ }\textbf {\bibinfo {volume} {76}},\ \bibinfo
  {pages} {083002} (\bibinfo {year} {2007})},\ \Eprint
  {http://arxiv.org/abs/0705.1733} {arXiv:0705.1733 [astro-ph]} \BibitemShut
  {NoStop}%
\bibitem [{\citenamefont {Caprini}\ \emph
  {et~al.}(2009{\natexlab{a}})\citenamefont {Caprini}, \citenamefont {Durrer},
  \citenamefont {Konstandin},\ and\ \citenamefont {Servant}}]{Caprini:2009fx}%
  \BibitemOpen
  \bibfield  {author} {\bibinfo {author} {\bibfnamefont {Chiara}\ \bibnamefont
  {Caprini}}, \bibinfo {author} {\bibfnamefont {Ruth}\ \bibnamefont {Durrer}},
  \bibinfo {author} {\bibfnamefont {Thomas}\ \bibnamefont {Konstandin}}, \ and\
  \bibinfo {author} {\bibfnamefont {Geraldine}\ \bibnamefont {Servant}},\
  }\bibfield  {title} {\enquote {\bibinfo {title} {{General properties of the
  gravitational wave spectrum from phase transitions}},}\ }\href {\doibase
  10.1103/PhysRevD.79.083519} {\bibfield  {journal} {\bibinfo  {journal} {Phys.
  Rev. D}\ }\textbf {\bibinfo {volume} {79}},\ \bibinfo {pages} {083519}
  (\bibinfo {year} {2009}{\natexlab{a}})},\ \Eprint
  {http://arxiv.org/abs/0901.1661} {arXiv:0901.1661 [astro-ph.CO]} \BibitemShut
  {NoStop}%
\bibitem [{\citenamefont {Caprini}\ \emph
  {et~al.}(2009{\natexlab{b}})\citenamefont {Caprini}, \citenamefont {Durrer},\
  and\ \citenamefont {Servant}}]{Caprini:2009yp}%
  \BibitemOpen
  \bibfield  {author} {\bibinfo {author} {\bibfnamefont {Chiara}\ \bibnamefont
  {Caprini}}, \bibinfo {author} {\bibfnamefont {Ruth}\ \bibnamefont {Durrer}},
  \ and\ \bibinfo {author} {\bibfnamefont {Geraldine}\ \bibnamefont
  {Servant}},\ }\bibfield  {title} {\enquote {\bibinfo {title} {{The stochastic
  gravitational wave background from turbulence and magnetic fields generated
  by a first-order phase transition}},}\ }\href {\doibase
  10.1088/1475-7516/2009/12/024} {\bibfield  {journal} {\bibinfo  {journal} {J.
  Cosmol. Astropart. Phys.}\ }\textbf {\bibinfo {volume} {12}},\ \bibinfo
  {pages} {024} (\bibinfo {year} {2009}{\natexlab{b}})},\ \Eprint
  {http://arxiv.org/abs/0909.0622} {arXiv:0909.0622 [astro-ph.CO]} \BibitemShut
  {NoStop}%
\bibitem [{\citenamefont {Niksa}\ \emph {et~al.}(2018)\citenamefont {Niksa},
  \citenamefont {Schlederer},\ and\ \citenamefont {Sigl}}]{Niksa:2018ofa}%
  \BibitemOpen
  \bibfield  {author} {\bibinfo {author} {\bibfnamefont {Peter}\ \bibnamefont
  {Niksa}}, \bibinfo {author} {\bibfnamefont {Martin}\ \bibnamefont
  {Schlederer}}, \ and\ \bibinfo {author} {\bibfnamefont {G\"unter}\
  \bibnamefont {Sigl}},\ }\bibfield  {title} {\enquote {\bibinfo {title}
  {{Gravitational waves produced by compressible MHD turbulence from
  cosmological phase transitions}},}\ }\href {\doibase
  10.1088/1361-6382/aac89c} {\bibfield  {journal} {\bibinfo  {journal}
  {Classical Quantum Gravity}\ }\textbf {\bibinfo {volume} {35}},\ \bibinfo
  {pages} {144001} (\bibinfo {year} {2018})},\ \Eprint
  {http://arxiv.org/abs/1803.02271} {arXiv:1803.02271 [astro-ph.CO]}
  \BibitemShut {NoStop}%
\bibitem [{\citenamefont {Roper~Pol}\ \emph
  {et~al.}(2020{\natexlab{a}})\citenamefont {Roper~Pol}, \citenamefont
  {Brandenburg}, \citenamefont {Kahniashvili}, \citenamefont {Kosowsky},\ and\
  \citenamefont {Mandal}}]{RoperPol:2018sap}%
  \BibitemOpen
  \bibfield  {author} {\bibinfo {author} {\bibfnamefont {Alberto}\ \bibnamefont
  {Roper~Pol}}, \bibinfo {author} {\bibfnamefont {Axel}\ \bibnamefont
  {Brandenburg}}, \bibinfo {author} {\bibfnamefont {Tina}\ \bibnamefont
  {Kahniashvili}}, \bibinfo {author} {\bibfnamefont {Arthur}\ \bibnamefont
  {Kosowsky}}, \ and\ \bibinfo {author} {\bibfnamefont {Sayan}\ \bibnamefont
  {Mandal}},\ }\bibfield  {title} {\enquote {\bibinfo {title} {{The timestep
  constraint in solving the gravitational wave equations sourced by
  hydromagnetic turbulence}},}\ }\href {\doibase 10.1080/03091929.2019.1653460}
  {\bibfield  {journal} {\bibinfo  {journal} {Geophys. Astrophys. Fluid Dyn.}\
  }\textbf {\bibinfo {volume} {114}},\ \bibinfo {pages} {130} (\bibinfo {year}
  {2020}{\natexlab{a}})},\ \Eprint {http://arxiv.org/abs/1807.05479}
  {arXiv:1807.05479 [physics.flu-dyn]} \BibitemShut {NoStop}%
\bibitem [{\citenamefont {Roper~Pol}\ \emph
  {et~al.}(2020{\natexlab{b}})\citenamefont {Roper~Pol}, \citenamefont
  {Mandal}, \citenamefont {Brandenburg}, \citenamefont {Kahniashvili},\ and\
  \citenamefont {Kosowsky}}]{RoperPol:2019wvy}%
  \BibitemOpen
  \bibfield  {author} {\bibinfo {author} {\bibfnamefont {Alberto}\ \bibnamefont
  {Roper~Pol}}, \bibinfo {author} {\bibfnamefont {Sayan}\ \bibnamefont
  {Mandal}}, \bibinfo {author} {\bibfnamefont {Axel}\ \bibnamefont
  {Brandenburg}}, \bibinfo {author} {\bibfnamefont {Tina}\ \bibnamefont
  {Kahniashvili}}, \ and\ \bibinfo {author} {\bibfnamefont {Arthur}\
  \bibnamefont {Kosowsky}},\ }\bibfield  {title} {\enquote {\bibinfo {title}
  {{Numerical simulations of gravitational waves from early-universe
  turbulence}},}\ }\href {\doibase 10.1103/PhysRevD.102.083512} {\bibfield
  {journal} {\bibinfo  {journal} {Phys. Rev. D}\ }\textbf {\bibinfo {volume}
  {102}},\ \bibinfo {pages} {083512} (\bibinfo {year} {2020}{\natexlab{b}})},\
  \Eprint {http://arxiv.org/abs/1903.08585} {arXiv:1903.08585 [astro-ph.CO]}
  \BibitemShut {NoStop}%
\bibitem [{\citenamefont {Kahniashvili}\ \emph {et~al.}(2021)\citenamefont
  {Kahniashvili}, \citenamefont {Brandenburg}, \citenamefont {Gogoberidze},
  \citenamefont {Mandal},\ and\ \citenamefont
  {Roper~Pol}}]{Kahniashvili:2020jgm}%
  \BibitemOpen
  \bibfield  {author} {\bibinfo {author} {\bibfnamefont {Tina}\ \bibnamefont
  {Kahniashvili}}, \bibinfo {author} {\bibfnamefont {Axel}\ \bibnamefont
  {Brandenburg}}, \bibinfo {author} {\bibfnamefont {Grigol}\ \bibnamefont
  {Gogoberidze}}, \bibinfo {author} {\bibfnamefont {Sayan}\ \bibnamefont
  {Mandal}}, \ and\ \bibinfo {author} {\bibfnamefont {Alberto}\ \bibnamefont
  {Roper~Pol}},\ }\bibfield  {title} {\enquote {\bibinfo {title} {{Circular
  polarization of gravitational waves from early-Universe helical
  turbulence}},}\ }\href {\doibase 10.1103/PhysRevResearch.3.013193} {\bibfield
   {journal} {\bibinfo  {journal} {Phys. Rev. Res.}\ }\textbf {\bibinfo
  {volume} {3}},\ \bibinfo {pages} {013193} (\bibinfo {year} {2021})},\ \Eprint
  {http://arxiv.org/abs/2011.05556} {arXiv:2011.05556 [astro-ph.CO]}
  \BibitemShut {NoStop}%
\bibitem [{\citenamefont {Brandenburg}\ \emph
  {et~al.}(2021{\natexlab{a}})\citenamefont {Brandenburg}, \citenamefont
  {Clarke}, \citenamefont {He},\ and\ \citenamefont
  {Kahniashvili}}]{Brandenburg:2021tmp}%
  \BibitemOpen
  \bibfield  {author} {\bibinfo {author} {\bibfnamefont {Axel}\ \bibnamefont
  {Brandenburg}}, \bibinfo {author} {\bibfnamefont {Emma}\ \bibnamefont
  {Clarke}}, \bibinfo {author} {\bibfnamefont {Yutong}\ \bibnamefont {He}}, \
  and\ \bibinfo {author} {\bibfnamefont {Tina}\ \bibnamefont {Kahniashvili}},\
  }\bibfield  {title} {\enquote {\bibinfo {title} {{Can we observe the QCD
  phase transition-generated gravitational waves through pulsar timing
  arrays?}}}\ }\href {\doibase 10.1103/PhysRevD.104.043513} {\bibfield
  {journal} {\bibinfo  {journal} {Phys. Rev. D}\ }\textbf {\bibinfo {volume}
  {104}},\ \bibinfo {pages} {043513} (\bibinfo {year} {2021}{\natexlab{a}})},\
  \Eprint {http://arxiv.org/abs/2102.12428} {arXiv:2102.12428 [astro-ph.CO]}
  \BibitemShut {NoStop}%
\bibitem [{\citenamefont {Brandenburg}\ \emph
  {et~al.}(2021{\natexlab{b}})\citenamefont {Brandenburg}, \citenamefont
  {Gogoberidze}, \citenamefont {Kahniashvili}, \citenamefont {Mandal},
  \citenamefont {Roper~Pol},\ and\ \citenamefont
  {Shenoy}}]{Brandenburg:2021bvg}%
  \BibitemOpen
  \bibfield  {author} {\bibinfo {author} {\bibfnamefont {Axel}\ \bibnamefont
  {Brandenburg}}, \bibinfo {author} {\bibfnamefont {Grigol}\ \bibnamefont
  {Gogoberidze}}, \bibinfo {author} {\bibfnamefont {Tina}\ \bibnamefont
  {Kahniashvili}}, \bibinfo {author} {\bibfnamefont {Sayan}\ \bibnamefont
  {Mandal}}, \bibinfo {author} {\bibfnamefont {Alberto}\ \bibnamefont
  {Roper~Pol}}, \ and\ \bibinfo {author} {\bibfnamefont {Nakul}\ \bibnamefont
  {Shenoy}},\ }\bibfield  {title} {\enquote {\bibinfo {title} {{The scalar,
  vector, and tensor modes in gravitational wave turbulence simulations}},}\
  }\href {\doibase 10.1088/1361-6382/ac011c} {\bibfield  {journal} {\bibinfo
  {journal} {Classical Quantum Gravity}\ }\textbf {\bibinfo {volume} {38}},\
  \bibinfo {pages} {145002} (\bibinfo {year} {2021}{\natexlab{b}})},\ \Eprint
  {http://arxiv.org/abs/2103.01140} {arXiv:2103.01140 [gr-qc]} \BibitemShut
  {NoStop}%
\bibitem [{\citenamefont {Roper~Pol}\ \emph
  {et~al.}(2022{\natexlab{a}})\citenamefont {Roper~Pol}, \citenamefont
  {Mandal}, \citenamefont {Brandenburg},\ and\ \citenamefont
  {Kahniashvili}}]{RoperPol:2021xnd}%
  \BibitemOpen
  \bibfield  {author} {\bibinfo {author} {\bibfnamefont {Alberto}\ \bibnamefont
  {Roper~Pol}}, \bibinfo {author} {\bibfnamefont {Sayan}\ \bibnamefont
  {Mandal}}, \bibinfo {author} {\bibfnamefont {Axel}\ \bibnamefont
  {Brandenburg}}, \ and\ \bibinfo {author} {\bibfnamefont {Tina}\ \bibnamefont
  {Kahniashvili}},\ }\bibfield  {title} {\enquote {\bibinfo {title}
  {{Polarization of gravitational waves from helical MHD turbulent sources}},}\
  }\href {\doibase 10.1088/1475-7516/2022/04/019} {\bibfield  {journal}
  {\bibinfo  {journal} {J. Cosmol. Astropart. Phys.}\ }\textbf {\bibinfo
  {volume} {04}},\ \bibinfo {pages} {019} (\bibinfo {year}
  {2022}{\natexlab{a}})},\ \Eprint {http://arxiv.org/abs/2107.05356}
  {arXiv:2107.05356 [gr-qc]} \BibitemShut {NoStop}%
\bibitem [{\citenamefont {Roper~Pol}\ \emph
  {et~al.}(2022{\natexlab{b}})\citenamefont {Roper~Pol}, \citenamefont
  {Caprini}, \citenamefont {Neronov},\ and\ \citenamefont
  {Semikoz}}]{RoperPol:2022iel}%
  \BibitemOpen
  \bibfield  {author} {\bibinfo {author} {\bibfnamefont {Alberto}\ \bibnamefont
  {Roper~Pol}}, \bibinfo {author} {\bibfnamefont {Chiara}\ \bibnamefont
  {Caprini}}, \bibinfo {author} {\bibfnamefont {Andrii}\ \bibnamefont
  {Neronov}}, \ and\ \bibinfo {author} {\bibfnamefont {Dmitri}\ \bibnamefont
  {Semikoz}},\ }\bibfield  {title} {\enquote {\bibinfo {title} {{Gravitational
  wave signal from primordial magnetic fields in the Pulsar Timing Array
  frequency band}},}\ }\href {\doibase 10.1103/PhysRevD.105.123502} {\bibfield
  {journal} {\bibinfo  {journal} {Phys. Rev. D}\ }\textbf {\bibinfo {volume}
  {105}},\ \bibinfo {pages} {123502} (\bibinfo {year} {2022}{\natexlab{b}})},\
  \Eprint {http://arxiv.org/abs/2201.05630} {arXiv:2201.05630 [astro-ph.CO]}
  \BibitemShut {NoStop}%
\bibitem [{\citenamefont {Auclair}\ \emph {et~al.}(2022)\citenamefont
  {Auclair}, \citenamefont {Caprini}, \citenamefont {Cutting}, \citenamefont
  {Hindmarsh}, \citenamefont {Rummukainen}, \citenamefont {Steer},\ and\
  \citenamefont {Weir}}]{Auclair:2022jod}%
  \BibitemOpen
  \bibfield  {author} {\bibinfo {author} {\bibfnamefont {Pierre}\ \bibnamefont
  {Auclair}}, \bibinfo {author} {\bibfnamefont {Chiara}\ \bibnamefont
  {Caprini}}, \bibinfo {author} {\bibfnamefont {Daniel}\ \bibnamefont
  {Cutting}}, \bibinfo {author} {\bibfnamefont {Mark}\ \bibnamefont
  {Hindmarsh}}, \bibinfo {author} {\bibfnamefont {Kari}\ \bibnamefont
  {Rummukainen}}, \bibinfo {author} {\bibfnamefont {Dani\`ele~A.}\ \bibnamefont
  {Steer}}, \ and\ \bibinfo {author} {\bibfnamefont {David~J.}\ \bibnamefont
  {Weir}},\ }\bibfield  {title} {\enquote {\bibinfo {title} {{Generation of
  gravitational waves from freely decaying turbulence}},}\ }\href {\doibase
  10.1088/1475-7516/2022/09/029} {\bibfield  {journal} {\bibinfo  {journal} {J.
  Cosmol. Astropart. Phys.}\ }\textbf {\bibinfo {volume} {09}},\ \bibinfo
  {pages} {029} (\bibinfo {year} {2022})},\ \Eprint
  {http://arxiv.org/abs/2205.02588} {arXiv:2205.02588 [astro-ph.CO]}
  \BibitemShut {NoStop}%
\bibitem [{\citenamefont {Sharma}\ and\ \citenamefont
  {Brandenburg}(2022)}]{Sharma:2022ysf}%
  \BibitemOpen
  \bibfield  {author} {\bibinfo {author} {\bibfnamefont {Ramkishor}\
  \bibnamefont {Sharma}}\ and\ \bibinfo {author} {\bibfnamefont {Axel}\
  \bibnamefont {Brandenburg}},\ }\bibfield  {title} {\enquote {\bibinfo {title}
  {{Low frequency tail of gravitational wave spectra from hydromagnetic
  turbulence}},}\ }\href {\doibase 10.1103/PhysRevD.106.103536} {\bibfield
  {journal} {\bibinfo  {journal} {Phys. Rev. D}\ }\textbf {\bibinfo {volume}
  {106}},\ \bibinfo {pages} {103536} (\bibinfo {year} {2022})},\ \Eprint
  {http://arxiv.org/abs/2206.00055} {arXiv:2206.00055 [astro-ph.CO]}
  \BibitemShut {NoStop}%
\bibitem [{\citenamefont {Caprini}\ and\ \citenamefont
  {Figueroa}(2018)}]{Caprini:2018mtu}%
  \BibitemOpen
  \bibfield  {author} {\bibinfo {author} {\bibfnamefont {Chiara}\ \bibnamefont
  {Caprini}}\ and\ \bibinfo {author} {\bibfnamefont {Daniel~G.}\ \bibnamefont
  {Figueroa}},\ }\bibfield  {title} {\enquote {\bibinfo {title} {{Cosmological
  backgrounds of gravitational waves}},}\ }\href {\doibase
  10.1088/1361-6382/aac608} {\bibfield  {journal} {\bibinfo  {journal}
  {Classical Quantum Gravity}\ }\textbf {\bibinfo {volume} {35}},\ \bibinfo
  {pages} {163001} (\bibinfo {year} {2018})},\ \Eprint
  {http://arxiv.org/abs/1801.04268} {arXiv:1801.04268 [astro-ph.CO]}
  \BibitemShut {NoStop}%
\bibitem [{\citenamefont {Espinosa}\ \emph {et~al.}(2010)\citenamefont
  {Espinosa}, \citenamefont {Konstandin}, \citenamefont {No},\ and\
  \citenamefont {Servant}}]{Espinosa:2010hh}%
  \BibitemOpen
  \bibfield  {author} {\bibinfo {author} {\bibfnamefont {Jose~R.}\ \bibnamefont
  {Espinosa}}, \bibinfo {author} {\bibfnamefont {Thomas}\ \bibnamefont
  {Konstandin}}, \bibinfo {author} {\bibfnamefont {Jose~M.}\ \bibnamefont
  {No}}, \ and\ \bibinfo {author} {\bibfnamefont {Geraldine}\ \bibnamefont
  {Servant}},\ }\bibfield  {title} {\enquote {\bibinfo {title} {{Energy budget
  of cosmological first-order phase transitions}},}\ }\href {\doibase
  10.1088/1475-7516/2010/06/028} {\bibfield  {journal} {\bibinfo  {journal} {J.
  Cosmol. Astropart. Phys.}\ }\textbf {\bibinfo {volume} {06}},\ \bibinfo
  {pages} {028} (\bibinfo {year} {2010})},\ \Eprint
  {http://arxiv.org/abs/1004.4187} {arXiv:1004.4187 [hep-ph]} \BibitemShut
  {NoStop}%
\bibitem [{\citenamefont {Bodeker}\ and\ \citenamefont
  {Moore}(2009)}]{Bodeker:2009qy}%
  \BibitemOpen
  \bibfield  {author} {\bibinfo {author} {\bibfnamefont {Dietrich}\
  \bibnamefont {Bodeker}}\ and\ \bibinfo {author} {\bibfnamefont {Guy~D.}\
  \bibnamefont {Moore}},\ }\bibfield  {title} {\enquote {\bibinfo {title} {{Can
  electroweak bubble walls run away?}}}\ }\href {\doibase
  10.1088/1475-7516/2009/05/009} {\bibfield  {journal} {\bibinfo  {journal} {J.
  Cosmol. Astropart. Phys.}\ }\textbf {\bibinfo {volume} {05}},\ \bibinfo
  {pages} {009} (\bibinfo {year} {2009})},\ \Eprint
  {http://arxiv.org/abs/0903.4099} {arXiv:0903.4099 [hep-ph]} \BibitemShut
  {NoStop}%
\bibitem [{\citenamefont {Bodeker}\ and\ \citenamefont
  {Moore}(2017)}]{Bodeker:2017cim}%
  \BibitemOpen
  \bibfield  {author} {\bibinfo {author} {\bibfnamefont {Dietrich}\
  \bibnamefont {Bodeker}}\ and\ \bibinfo {author} {\bibfnamefont {Guy~D.}\
  \bibnamefont {Moore}},\ }\bibfield  {title} {\enquote {\bibinfo {title}
  {{Electroweak bubble wall speed limit}},}\ }\href {\doibase
  10.1088/1475-7516/2017/05/025} {\bibfield  {journal} {\bibinfo  {journal} {J.
  Cosmol. Astropart. Phys.}\ }\textbf {\bibinfo {volume} {05}},\ \bibinfo
  {pages} {025} (\bibinfo {year} {2017})},\ \Eprint
  {http://arxiv.org/abs/1703.08215} {arXiv:1703.08215 [hep-ph]} \BibitemShut
  {NoStop}%
\bibitem [{\citenamefont {Caprini}\ \emph {et~al.}(2016)\citenamefont {Caprini}
  \emph {et~al.}}]{Caprini:2015zlo}%
  \BibitemOpen
  \bibfield  {author} {\bibinfo {author} {\bibfnamefont {Chiara}\ \bibnamefont
  {Caprini}} \emph {et~al.},\ }\bibfield  {title} {\enquote {\bibinfo {title}
  {{Science with the space-based interferometer eLISA. II: Gravitational waves
  from cosmological phase transitions}},}\ }\href {\doibase
  10.1088/1475-7516/2016/04/001} {\bibfield  {journal} {\bibinfo  {journal} {J.
  Cosmol. Astropart. Phys.}\ }\textbf {\bibinfo {volume} {04}},\ \bibinfo
  {pages} {001} (\bibinfo {year} {2016})},\ \Eprint
  {http://arxiv.org/abs/1512.06239} {arXiv:1512.06239 [astro-ph.CO]}
  \BibitemShut {NoStop}%
\bibitem [{\citenamefont {von Harling}\ and\ \citenamefont
  {Servant}(2018)}]{vonHarling:2017yew}%
  \BibitemOpen
  \bibfield  {author} {\bibinfo {author} {\bibfnamefont {Benedict}\
  \bibnamefont {von Harling}}\ and\ \bibinfo {author} {\bibfnamefont
  {Geraldine}\ \bibnamefont {Servant}},\ }\bibfield  {title} {\enquote
  {\bibinfo {title} {{QCD-induced electroweak phase transition}},}\ }\href
  {\doibase 10.1007/JHEP01(2018)159} {\bibfield  {journal} {\bibinfo  {journal}
  {J. High Energy Phys.}\ }\textbf {\bibinfo {volume} {01}},\ \bibinfo {pages}
  {159} (\bibinfo {year} {2018})},\ \Eprint {http://arxiv.org/abs/1711.11554}
  {arXiv:1711.11554 [hep-ph]} \BibitemShut {NoStop}%
\bibitem [{\citenamefont {Kobakhidze}\ \emph {et~al.}(2017)\citenamefont
  {Kobakhidze}, \citenamefont {Lagger}, \citenamefont {Manning},\ and\
  \citenamefont {Yue}}]{Kobakhidze:2017mru}%
  \BibitemOpen
  \bibfield  {author} {\bibinfo {author} {\bibfnamefont {Archil}\ \bibnamefont
  {Kobakhidze}}, \bibinfo {author} {\bibfnamefont {Cyril}\ \bibnamefont
  {Lagger}}, \bibinfo {author} {\bibfnamefont {Adrian}\ \bibnamefont
  {Manning}}, \ and\ \bibinfo {author} {\bibfnamefont {Jason}\ \bibnamefont
  {Yue}},\ }\bibfield  {title} {\enquote {\bibinfo {title} {{Gravitational
  waves from a supercooled electroweak phase transition and their detection
  with pulsar timing arrays}},}\ }\href {\doibase
  10.1140/epjc/s10052-017-5132-y} {\bibfield  {journal} {\bibinfo  {journal}
  {Eur. Phys. J. C}\ }\textbf {\bibinfo {volume} {77}},\ \bibinfo {pages} {570}
  (\bibinfo {year} {2017})},\ \Eprint {http://arxiv.org/abs/1703.06552}
  {arXiv:1703.06552 [hep-ph]} \BibitemShut {NoStop}%
\bibitem [{\citenamefont {Kosowsky}(1996)}]{Kosowsky:1994cy}%
  \BibitemOpen
  \bibfield  {author} {\bibinfo {author} {\bibfnamefont {Arthur}\ \bibnamefont
  {Kosowsky}},\ }\bibfield  {title} {\enquote {\bibinfo {title} {{Cosmic
  microwave background polarization}},}\ }\href {\doibase
  10.1006/aphy.1996.0020} {\bibfield  {journal} {\bibinfo  {journal} {Ann.
  Phys.}\ }\textbf {\bibinfo {volume} {246}},\ \bibinfo {pages} {49} (\bibinfo
  {year} {1996})},\ \Eprint {http://arxiv.org/abs/astro-ph/9501045}
  {arXiv:astro-ph/9501045} \BibitemShut {NoStop}%
\bibitem [{\citenamefont {Quashnock}\ \emph {et~al.}(1989)\citenamefont
  {Quashnock}, \citenamefont {Loeb},\ and\ \citenamefont
  {Spergel}}]{Quashnock:1988vs}%
  \BibitemOpen
  \bibfield  {author} {\bibinfo {author} {\bibfnamefont {Jean~M.}\ \bibnamefont
  {Quashnock}}, \bibinfo {author} {\bibfnamefont {Abraham}\ \bibnamefont
  {Loeb}}, \ and\ \bibinfo {author} {\bibfnamefont {David~N.}\ \bibnamefont
  {Spergel}},\ }\bibfield  {title} {\enquote {\bibinfo {title} {{Magnetic field
  generation during the cosmological QCD phase transition}},}\ }\href {\doibase
  10.1086/185528} {\bibfield  {journal} {\bibinfo  {journal} {Astrophys. J.
  Lett.}\ }\textbf {\bibinfo {volume} {344}},\ \bibinfo {pages} {L49} (\bibinfo
  {year} {1989})}\BibitemShut {NoStop}%
\bibitem [{\citenamefont {Brandenburg}\ \emph {et~al.}(1996)\citenamefont
  {Brandenburg}, \citenamefont {Enqvist},\ and\ \citenamefont
  {Olesen}}]{Brandenburg:1996fc}%
  \BibitemOpen
  \bibfield  {author} {\bibinfo {author} {\bibfnamefont {Axel}\ \bibnamefont
  {Brandenburg}}, \bibinfo {author} {\bibfnamefont {Kari}\ \bibnamefont
  {Enqvist}}, \ and\ \bibinfo {author} {\bibfnamefont {Poul}\ \bibnamefont
  {Olesen}},\ }\bibfield  {title} {\enquote {\bibinfo {title} {{Large scale
  magnetic fields from hydromagnetic turbulence in the very early universe}},}\
  }\href {\doibase 10.1103/PhysRevD.54.1291} {\bibfield  {journal} {\bibinfo
  {journal} {Phys. Rev. D}\ }\textbf {\bibinfo {volume} {54}},\ \bibinfo
  {pages} {1291} (\bibinfo {year} {1996})},\ \Eprint
  {http://arxiv.org/abs/astro-ph/9602031} {arXiv:astro-ph/9602031} \BibitemShut
  {NoStop}%
\bibitem [{\citenamefont {Ahonen}\ and\ \citenamefont
  {Enqvist}(1996)}]{Ahonen:1996nq}%
  \BibitemOpen
  \bibfield  {author} {\bibinfo {author} {\bibfnamefont {Jarkko}\ \bibnamefont
  {Ahonen}}\ and\ \bibinfo {author} {\bibfnamefont {Kari}\ \bibnamefont
  {Enqvist}},\ }\bibfield  {title} {\enquote {\bibinfo {title} {{Electrical
  conductivity in the early universe}},}\ }\href {\doibase
  10.1016/0370-2693(96)00633-8} {\bibfield  {journal} {\bibinfo  {journal}
  {Phys. Lett. B}\ }\textbf {\bibinfo {volume} {382}},\ \bibinfo {pages} {40}
  (\bibinfo {year} {1996})},\ \Eprint {http://arxiv.org/abs/hep-ph/9602357}
  {arXiv:hep-ph/9602357} \BibitemShut {NoStop}%
\bibitem [{\citenamefont {Arnold}\ \emph {et~al.}(2000)\citenamefont {Arnold},
  \citenamefont {Moore},\ and\ \citenamefont {Yaffe}}]{Arnold:2000dr}%
  \BibitemOpen
  \bibfield  {author} {\bibinfo {author} {\bibfnamefont {Peter~Brockway}\
  \bibnamefont {Arnold}}, \bibinfo {author} {\bibfnamefont {Guy~D.}\
  \bibnamefont {Moore}}, \ and\ \bibinfo {author} {\bibfnamefont {Laurence~G.}\
  \bibnamefont {Yaffe}},\ }\bibfield  {title} {\enquote {\bibinfo {title}
  {{Transport coefficients in high temperature gauge theories. 1. Leading log
  results}},}\ }\href {\doibase 10.1088/1126-6708/2000/11/001} {\bibfield
  {journal} {\bibinfo  {journal} {J. High Energy Phys.}\ }\textbf {\bibinfo
  {volume} {11}},\ \bibinfo {pages} {001} (\bibinfo {year} {2000})},\ \Eprint
  {http://arxiv.org/abs/hep-ph/0010177} {arXiv:hep-ph/0010177} \BibitemShut
  {NoStop}%
\bibitem [{\citenamefont {Cutting}\ \emph {et~al.}(2020)\citenamefont
  {Cutting}, \citenamefont {Hindmarsh},\ and\ \citenamefont
  {Weir}}]{Cutting:2019zws}%
  \BibitemOpen
  \bibfield  {author} {\bibinfo {author} {\bibfnamefont {Daniel}\ \bibnamefont
  {Cutting}}, \bibinfo {author} {\bibfnamefont {Mark}\ \bibnamefont
  {Hindmarsh}}, \ and\ \bibinfo {author} {\bibfnamefont {David~J.}\
  \bibnamefont {Weir}},\ }\bibfield  {title} {\enquote {\bibinfo {title}
  {{Vorticity, kinetic energy, and suppressed gravitational wave production in
  strong first order phase transitions}},}\ }\href {\doibase
  10.1103/PhysRevLett.125.021302} {\bibfield  {journal} {\bibinfo  {journal}
  {Phys. Rev. Lett.}\ }\textbf {\bibinfo {volume} {125}},\ \bibinfo {pages}
  {021302} (\bibinfo {year} {2020})},\ \Eprint
  {http://arxiv.org/abs/1906.00480} {arXiv:1906.00480 [hep-ph]} \BibitemShut
  {NoStop}%
\bibitem [{\citenamefont {Roper~Pol}\ \emph {et~al.}()\citenamefont
  {Roper~Pol}, \citenamefont {Neronov}, \citenamefont {Caprini}, \citenamefont
  {Boyer},\ and\ \citenamefont {Semikoz}}]{RoperPol:2023bqa}%
  \BibitemOpen
  \bibfield  {author} {\bibinfo {author} {\bibfnamefont {A.}~\bibnamefont
  {Roper~Pol}}, \bibinfo {author} {\bibfnamefont {A.}~\bibnamefont {Neronov}},
  \bibinfo {author} {\bibfnamefont {C.}~\bibnamefont {Caprini}}, \bibinfo
  {author} {\bibfnamefont {T.}~\bibnamefont {Boyer}}, \ and\ \bibinfo {author}
  {\bibfnamefont {D.}~\bibnamefont {Semikoz}},\ }\bibfield  {title} {\enquote
  {\bibinfo {title} {{LISA and $\gamma$-ray telescopes as multi-messenger
  probes of a first-order cosmological phase transition}},}\ }\href@noop {} {\
  }\Eprint {http://arxiv.org/abs/2307.10744} {arXiv:2307.10744 [astro-ph.CO]}
  \BibitemShut {NoStop}%
\bibitem [{\citenamefont {Guo}\ \emph {et~al.}(2021)\citenamefont {Guo},
  \citenamefont {Sinha}, \citenamefont {Vagie},\ and\ \citenamefont
  {White}}]{Guo:2020grp}%
  \BibitemOpen
  \bibfield  {author} {\bibinfo {author} {\bibfnamefont {Huai-Ke}\ \bibnamefont
  {Guo}}, \bibinfo {author} {\bibfnamefont {Kuver}\ \bibnamefont {Sinha}},
  \bibinfo {author} {\bibfnamefont {Daniel}\ \bibnamefont {Vagie}}, \ and\
  \bibinfo {author} {\bibfnamefont {Graham}\ \bibnamefont {White}},\ }\bibfield
   {title} {\enquote {\bibinfo {title} {{Phase transitions in an expanding
  universe: Stochastic gravitational waves in standard and non-standard
  histories}},}\ }\href {\doibase 10.1088/1475-7516/2021/01/001} {\bibfield
  {journal} {\bibinfo  {journal} {J. Cosmol. Astropart. Phys.}\ }\textbf
  {\bibinfo {volume} {01}},\ \bibinfo {pages} {001} (\bibinfo {year} {2021})},\
  \Eprint {http://arxiv.org/abs/2007.08537} {arXiv:2007.08537 [hep-ph]}
  \BibitemShut {NoStop}%
\bibitem [{\citenamefont {Cai}\ \emph {et~al.}(2023)\citenamefont {Cai},
  \citenamefont {Wang},\ and\ \citenamefont {Yuwen}}]{Cai:2023guc}%
  \BibitemOpen
  \bibfield  {author} {\bibinfo {author} {\bibfnamefont {Rong-Gen}\
  \bibnamefont {Cai}}, \bibinfo {author} {\bibfnamefont {Shao-Jiang}\
  \bibnamefont {Wang}}, \ and\ \bibinfo {author} {\bibfnamefont {Zi-Yan}\
  \bibnamefont {Yuwen}},\ }\bibfield  {title} {\enquote {\bibinfo {title}
  {{Hydrodynamic sound shell model}},}\ }\href {\doibase
  10.1103/PhysRevD.108.L021502} {\bibfield  {journal} {\bibinfo  {journal}
  {Phys. Rev. D}\ }\textbf {\bibinfo {volume} {108}},\ \bibinfo {pages}
  {L021502} (\bibinfo {year} {2023})},\ \Eprint
  {http://arxiv.org/abs/2305.00074} {arXiv:2305.00074 [gr-qc]} \BibitemShut
  {NoStop}%
\bibitem [{\citenamefont {Antoniadis}\ \emph
  {et~al.}(2023{\natexlab{a}})\citenamefont {Antoniadis} \emph
  {et~al.}}]{EPTA:2023fyk}%
  \BibitemOpen
  \bibfield  {author} {\bibinfo {author} {\bibfnamefont {J.}~\bibnamefont
  {Antoniadis}} \emph {et~al.} (\bibinfo {collaboration} {EPTA-InPTA
  Collaborations}),\ }\bibfield  {title} {\enquote {\bibinfo {title} {{The
  second data release from the European Pulsar Timing Array III. Search for
  gravitational wave signals}},}\ }\href {\doibase 10.1051/0004-6361/202346844}
  {\bibfield  {journal} {\bibinfo  {journal} {Astron. Astrophys.}\ }\textbf
  {\bibinfo {volume} {678}},\ \bibinfo {pages} {A50} (\bibinfo {year}
  {2023}{\natexlab{a}})},\ \Eprint {http://arxiv.org/abs/2306.16214}
  {arXiv:2306.16214 [astro-ph.HE]} \BibitemShut {NoStop}%
\bibitem [{\citenamefont {Antoniadis}\ \emph
  {et~al.}(2023{\natexlab{b}})\citenamefont {Antoniadis} \emph
  {et~al.}}]{EPTA:2023xxk}%
  \BibitemOpen
  \bibfield  {author} {\bibinfo {author} {\bibfnamefont {J.}~\bibnamefont
  {Antoniadis}} \emph {et~al.} (\bibinfo {collaboration} {EPTA-InPTA
  Collaborations}),\ }\bibfield  {title} {\enquote {\bibinfo {title} {{The
  second data release from the European Pulsar Timing Array: V. Implications
  for massive black holes, dark matter and the early Universe}},}\ }\href@noop
  {} {\  (\bibinfo {year} {2023}{\natexlab{b}})},\ \Eprint
  {http://arxiv.org/abs/2306.16227} {arXiv:2306.16227 [astro-ph.CO]}
  \BibitemShut {NoStop}%
\bibitem [{\citenamefont {Agazie}\ \emph {et~al.}(2023)\citenamefont {Agazie}
  \emph {et~al.}}]{NANOGrav:2023gor}%
  \BibitemOpen
  \bibfield  {author} {\bibinfo {author} {\bibfnamefont {Gabriella}\
  \bibnamefont {Agazie}} \emph {et~al.} (\bibinfo {collaboration} {NANOGrav
  Collaboration}),\ }\bibfield  {title} {\enquote {\bibinfo {title} {{The
  NANOGrav 15 yr data set: Evidence for a gravitational-wave background}},}\
  }\href {\doibase 10.3847/2041-8213/acdac6} {\bibfield  {journal} {\bibinfo
  {journal} {Astrophys. J. Lett.}\ }\textbf {\bibinfo {volume} {951}},\
  \bibinfo {pages} {L8} (\bibinfo {year} {2023})},\ \Eprint
  {http://arxiv.org/abs/2306.16213} {arXiv:2306.16213 [astro-ph.HE]}
  \BibitemShut {NoStop}%
\bibitem [{\citenamefont {Afzal}\ \emph {et~al.}(2023)\citenamefont {Afzal}
  \emph {et~al.}}]{NANOGrav:2023hvm}%
  \BibitemOpen
  \bibfield  {author} {\bibinfo {author} {\bibfnamefont {Adeela}\ \bibnamefont
  {Afzal}} \emph {et~al.} (\bibinfo {collaboration} {NANOGrav Collaboration}),\
  }\bibfield  {title} {\enquote {\bibinfo {title} {{The NANOGrav 15 yr data
  set: Search for signals from new physics}},}\ }\href {\doibase
  10.3847/2041-8213/acdc91} {\bibfield  {journal} {\bibinfo  {journal}
  {Astrophys. J. Lett.}\ }\textbf {\bibinfo {volume} {951}},\ \bibinfo {pages}
  {L11} (\bibinfo {year} {2023})},\ \Eprint {http://arxiv.org/abs/2306.16219}
  {arXiv:2306.16219 [astro-ph.HE]} \BibitemShut {NoStop}%
\bibitem [{\citenamefont {Reardon}\ \emph {et~al.}(2023)\citenamefont {Reardon}
  \emph {et~al.}}]{Reardon:2023gzh}%
  \BibitemOpen
  \bibfield  {author} {\bibinfo {author} {\bibfnamefont {Daniel~J.}\
  \bibnamefont {Reardon}} \emph {et~al.},\ }\bibfield  {title} {\enquote
  {\bibinfo {title} {{Search for an isotropic gravitational-wave background
  with the Parkes Pulsar Timing Array}},}\ }\href {\doibase
  10.3847/2041-8213/acdd02} {\bibfield  {journal} {\bibinfo  {journal}
  {Astrophys. J. Lett.}\ }\textbf {\bibinfo {volume} {951}},\ \bibinfo {pages}
  {L6} (\bibinfo {year} {2023})},\ \Eprint {http://arxiv.org/abs/2306.16215}
  {arXiv:2306.16215 [astro-ph.HE]} \BibitemShut {NoStop}%
\bibitem [{\citenamefont {Xu}\ \emph {et~al.}(2023)\citenamefont {Xu} \emph
  {et~al.}}]{Xu:2023wog}%
  \BibitemOpen
  \bibfield  {author} {\bibinfo {author} {\bibfnamefont {Heng}\ \bibnamefont
  {Xu}} \emph {et~al.},\ }\bibfield  {title} {\enquote {\bibinfo {title}
  {{Searching for the nano-Hertz stochastic gravitational wave background with
  the Chinese Pulsar Timing Array data release I}},}\ }\href {\doibase
  10.1088/1674-4527/acdfa5} {\bibfield  {journal} {\bibinfo  {journal} {Res.
  Astron. Astrophys.}\ }\textbf {\bibinfo {volume} {23}},\ \bibinfo {pages}
  {075024} (\bibinfo {year} {2023})},\ \Eprint
  {http://arxiv.org/abs/2306.16216} {arXiv:2306.16216 [astro-ph.HE]}
  \BibitemShut {NoStop}%
\bibitem [{\citenamefont {Madge}\ \emph {et~al.}(2023)\citenamefont {Madge},
  \citenamefont {Morgante}, \citenamefont {Puchades-Ib\'a\~nez}, \citenamefont
  {Ramberg}, \citenamefont {Ratzinger}, \citenamefont {Schenk},\ and\
  \citenamefont {Schwaller}}]{Madge:2023dxc}%
  \BibitemOpen
  \bibfield  {author} {\bibinfo {author} {\bibfnamefont {Eric}\ \bibnamefont
  {Madge}}, \bibinfo {author} {\bibfnamefont {Enrico}\ \bibnamefont
  {Morgante}}, \bibinfo {author} {\bibfnamefont {Cristina}\ \bibnamefont
  {Puchades-Ib\'a\~nez}}, \bibinfo {author} {\bibfnamefont {Nicklas}\
  \bibnamefont {Ramberg}}, \bibinfo {author} {\bibfnamefont {Wolfram}\
  \bibnamefont {Ratzinger}}, \bibinfo {author} {\bibfnamefont {Sebastian}\
  \bibnamefont {Schenk}}, \ and\ \bibinfo {author} {\bibfnamefont {Pedro}\
  \bibnamefont {Schwaller}},\ }\bibfield  {title} {\enquote {\bibinfo {title}
  {{Primordial gravitational waves in the nano-Hertz regime and PTA data
  \textemdash{} towards solving the GW inverse problem}},}\ }\href {\doibase
  10.1007/JHEP10(2023)171} {\bibfield  {journal} {\bibinfo  {journal} {J. High
  Energy Phys.}\ }\textbf {\bibinfo {volume} {10}},\ \bibinfo {pages} {171}
  (\bibinfo {year} {2023})},\ \Eprint {http://arxiv.org/abs/2306.14856}
  {arXiv:2306.14856 [hep-ph]} \BibitemShut {NoStop}%
\bibitem [{\citenamefont {Bringmann}\ \emph {et~al.}(2023)\citenamefont
  {Bringmann}, \citenamefont {Depta}, \citenamefont {Konstandin}, \citenamefont
  {Schmidt-Hoberg},\ and\ \citenamefont {Tasillo}}]{Bringmann:2023opz}%
  \BibitemOpen
  \bibfield  {author} {\bibinfo {author} {\bibfnamefont {Torsten}\ \bibnamefont
  {Bringmann}}, \bibinfo {author} {\bibfnamefont {Paul~Frederik}\ \bibnamefont
  {Depta}}, \bibinfo {author} {\bibfnamefont {Thomas}\ \bibnamefont
  {Konstandin}}, \bibinfo {author} {\bibfnamefont {Kai}\ \bibnamefont
  {Schmidt-Hoberg}}, \ and\ \bibinfo {author} {\bibfnamefont {Carlo}\
  \bibnamefont {Tasillo}},\ }\bibfield  {title} {\enquote {\bibinfo {title}
  {{Does NANOGrav observe a dark sector phase transition?}}}\ }\href {\doibase
  10.1088/1475-7516/2023/11/053} {\bibfield  {journal} {\bibinfo  {journal} {J.
  Cosmol. Astropart. Phys}\ }\textbf {\bibinfo {volume} {11}},\ \bibinfo
  {pages} {053} (\bibinfo {year} {2023})},\ \Eprint
  {http://arxiv.org/abs/2306.09411} {arXiv:2306.09411 [astro-ph.CO]}
  \BibitemShut {NoStop}%
\bibitem [{\citenamefont {Zu}\ \emph {et~al.}()\citenamefont {Zu},
  \citenamefont {Zhang}, \citenamefont {Li}, \citenamefont {Gu}, \citenamefont
  {Tsai},\ and\ \citenamefont {Fan}}]{Zu:2023olm}%
  \BibitemOpen
  \bibfield  {author} {\bibinfo {author} {\bibfnamefont {Lei}\ \bibnamefont
  {Zu}}, \bibinfo {author} {\bibfnamefont {Chi}\ \bibnamefont {Zhang}},
  \bibinfo {author} {\bibfnamefont {Yao-Yu}\ \bibnamefont {Li}}, \bibinfo
  {author} {\bibfnamefont {Yu-Chao}\ \bibnamefont {Gu}}, \bibinfo {author}
  {\bibfnamefont {Yue-Lin~Sming}\ \bibnamefont {Tsai}}, \ and\ \bibinfo
  {author} {\bibfnamefont {Yi-Zhong}\ \bibnamefont {Fan}},\ }\bibfield  {title}
  {\enquote {\bibinfo {title} {{Mirror QCD phase transition as the origin of
  the nanohertz stochastic gravitational-wave background}},}\ }\href@noop {} {\
  }\Eprint {http://arxiv.org/abs/2306.16769} {arXiv:2306.16769 [astro-ph.HE]}
  \BibitemShut {NoStop}%
\bibitem [{\citenamefont {Addazi}\ \emph {et~al.}(2024)\citenamefont {Addazi},
  \citenamefont {Cai}, \citenamefont {Marciano},\ and\ \citenamefont
  {Visinelli}}]{Addazi:2023jvg}%
  \BibitemOpen
  \bibfield  {author} {\bibinfo {author} {\bibfnamefont {Andrea}\ \bibnamefont
  {Addazi}}, \bibinfo {author} {\bibfnamefont {Yi-Fu}\ \bibnamefont {Cai}},
  \bibinfo {author} {\bibfnamefont {Antonino}\ \bibnamefont {Marciano}}, \ and\
  \bibinfo {author} {\bibfnamefont {Luca}\ \bibnamefont {Visinelli}},\
  }\bibfield  {title} {\enquote {\bibinfo {title} {{Have pulsar timing array
  methods detected a cosmological phase transition?}}}\ }\href {\doibase
  10.1103/PhysRevD.109.015028} {\bibfield  {journal} {\bibinfo  {journal}
  {Phys. Rev. D}\ }\textbf {\bibinfo {volume} {109}},\ \bibinfo {pages}
  {015028} (\bibinfo {year} {2024})},\ \Eprint
  {http://arxiv.org/abs/2306.17205} {arXiv:2306.17205 [astro-ph.CO]}
  \BibitemShut {NoStop}%
\bibitem [{\citenamefont {Han}\ \emph {et~al.}()\citenamefont {Han},
  \citenamefont {Xie}, \citenamefont {Yang},\ and\ \citenamefont
  {Zhang}}]{Han:2023olf}%
  \BibitemOpen
  \bibfield  {author} {\bibinfo {author} {\bibfnamefont {Chengcheng}\
  \bibnamefont {Han}}, \bibinfo {author} {\bibfnamefont {Ke-Pan}\ \bibnamefont
  {Xie}}, \bibinfo {author} {\bibfnamefont {Jin~Min}\ \bibnamefont {Yang}}, \
  and\ \bibinfo {author} {\bibfnamefont {Mengchao}\ \bibnamefont {Zhang}},\
  }\bibfield  {title} {\enquote {\bibinfo {title} {{Self-interacting dark
  matter implied by nano-Hertz gravitational waves}},}\ }\href@noop {} {\
  }\Eprint {http://arxiv.org/abs/2306.16966} {arXiv:2306.16966 [hep-ph]}
  \BibitemShut {NoStop}%
\bibitem [{\citenamefont {Megias}\ \emph {et~al.}(2023)\citenamefont {Megias},
  \citenamefont {Nardini},\ and\ \citenamefont {Quiros}}]{Megias:2023kiy}%
  \BibitemOpen
  \bibfield  {author} {\bibinfo {author} {\bibfnamefont {Eugenio}\ \bibnamefont
  {Megias}}, \bibinfo {author} {\bibfnamefont {Germano}\ \bibnamefont
  {Nardini}}, \ and\ \bibinfo {author} {\bibfnamefont {Mariano}\ \bibnamefont
  {Quiros}},\ }\bibfield  {title} {\enquote {\bibinfo {title} {{Pulsar timing
  array stochastic background from light Kaluza-Klein resonances}},}\ }\href
  {\doibase 10.1103/PhysRevD.108.095017} {\bibfield  {journal} {\bibinfo
  {journal} {Phys. Rev. D}\ }\textbf {\bibinfo {volume} {108}},\ \bibinfo
  {pages} {095017} (\bibinfo {year} {2023})},\ \Eprint
  {http://arxiv.org/abs/2306.17071} {arXiv:2306.17071 [hep-ph]} \BibitemShut
  {NoStop}%
\bibitem [{\citenamefont {Li}\ and\ \citenamefont {Xie}(2023)}]{Li:2023bxy}%
  \BibitemOpen
  \bibfield  {author} {\bibinfo {author} {\bibfnamefont {Shao-Ping}\
  \bibnamefont {Li}}\ and\ \bibinfo {author} {\bibfnamefont {Ke-Pan}\
  \bibnamefont {Xie}},\ }\bibfield  {title} {\enquote {\bibinfo {title}
  {{Collider test of nano-Hertz gravitational waves from pulsar timing
  arrays}},}\ }\href {\doibase 10.1103/PhysRevD.108.055018} {\bibfield
  {journal} {\bibinfo  {journal} {Phys. Rev. D}\ }\textbf {\bibinfo {volume}
  {108}},\ \bibinfo {pages} {055018} (\bibinfo {year} {2023})},\ \Eprint
  {http://arxiv.org/abs/2307.01086} {arXiv:2307.01086 [hep-ph]} \BibitemShut
  {NoStop}%
\bibitem [{\citenamefont {Di~Bari}\ and\ \citenamefont
  {Rahat}()}]{DiBari:2023upq}%
  \BibitemOpen
  \bibfield  {author} {\bibinfo {author} {\bibfnamefont {Pasquale}\
  \bibnamefont {Di~Bari}}\ and\ \bibinfo {author} {\bibfnamefont
  {Moinul~Hossain}\ \bibnamefont {Rahat}},\ }\bibfield  {title} {\enquote
  {\bibinfo {title} {{The split majoron model confronts the NANOGrav
  signal}},}\ }\href@noop {} {\ }\Eprint {http://arxiv.org/abs/2307.03184}
  {arXiv:2307.03184 [hep-ph]} \BibitemShut {NoStop}%
\bibitem [{\citenamefont {Bai}\ \emph {et~al.}(2023)\citenamefont {Bai},
  \citenamefont {Chen},\ and\ \citenamefont {Korwar}}]{Bai:2023cqj}%
  \BibitemOpen
  \bibfield  {author} {\bibinfo {author} {\bibfnamefont {Yang}\ \bibnamefont
  {Bai}}, \bibinfo {author} {\bibfnamefont {Ting-Kuo}\ \bibnamefont {Chen}}, \
  and\ \bibinfo {author} {\bibfnamefont {Mrunal}\ \bibnamefont {Korwar}},\
  }\bibfield  {title} {\enquote {\bibinfo {title} {{QCD-collapsed domain walls:
  QCD phase transition and gravitational wave spectroscopy}},}\ }\href
  {\doibase 10.1007/JHEP12(2023)194} {\bibfield  {journal} {\bibinfo  {journal}
  {J. High Energy Phys.}\ }\textbf {\bibinfo {volume} {12}},\ \bibinfo {pages}
  {194} (\bibinfo {year} {2023})},\ \Eprint {http://arxiv.org/abs/2306.17160}
  {arXiv:2306.17160 [hep-ph]} \BibitemShut {NoStop}%
\bibitem [{\citenamefont {Ghosh}\ \emph {et~al.}()\citenamefont {Ghosh},
  \citenamefont {Ghoshal}, \citenamefont {Guo}, \citenamefont {Hajkarim},
  \citenamefont {King}, \citenamefont {Sinha}, \citenamefont {Wang},\ and\
  \citenamefont {White}}]{Ghosh:2023aum}%
  \BibitemOpen
  \bibfield  {author} {\bibinfo {author} {\bibfnamefont {Tathagata}\
  \bibnamefont {Ghosh}}, \bibinfo {author} {\bibfnamefont {Anish}\ \bibnamefont
  {Ghoshal}}, \bibinfo {author} {\bibfnamefont {Huai-Ke}\ \bibnamefont {Guo}},
  \bibinfo {author} {\bibfnamefont {Fazlollah}\ \bibnamefont {Hajkarim}},
  \bibinfo {author} {\bibfnamefont {Stephen~F.}\ \bibnamefont {King}}, \bibinfo
  {author} {\bibfnamefont {Kuver}\ \bibnamefont {Sinha}}, \bibinfo {author}
  {\bibfnamefont {Xin}\ \bibnamefont {Wang}}, \ and\ \bibinfo {author}
  {\bibfnamefont {Graham}\ \bibnamefont {White}},\ }\bibfield  {title}
  {\enquote {\bibinfo {title} {{Did we hear the sound of the Universe boiling?
  Analysis using the full fluid velocity profiles and NANOGrav 15-year
  data}},}\ }\href@noop {} {\ }\Eprint {http://arxiv.org/abs/2307.02259}
  {arXiv:2307.02259 [astro-ph.HE]} \BibitemShut {NoStop}%
\bibitem [{\citenamefont {Figueroa}\ \emph {et~al.}()\citenamefont {Figueroa},
  \citenamefont {Pieroni}, \citenamefont {Ricciardone},\ and\ \citenamefont
  {Simakachorn}}]{Figueroa:2023zhu}%
  \BibitemOpen
  \bibfield  {author} {\bibinfo {author} {\bibfnamefont {Daniel~G.}\
  \bibnamefont {Figueroa}}, \bibinfo {author} {\bibfnamefont {Mauro}\
  \bibnamefont {Pieroni}}, \bibinfo {author} {\bibfnamefont {Angelo}\
  \bibnamefont {Ricciardone}}, \ and\ \bibinfo {author} {\bibfnamefont {Peera}\
  \bibnamefont {Simakachorn}},\ }\bibfield  {title} {\enquote {\bibinfo {title}
  {{Cosmological background interpretation of pulsar timing array data}},}\
  }\href@noop {} {\ }\Eprint {http://arxiv.org/abs/2307.02399}
  {arXiv:2307.02399 [astro-ph.CO]} \BibitemShut {NoStop}%
\bibitem [{\citenamefont {Amaro-Seoane}\ \emph {et~al.}()\citenamefont
  {Amaro-Seoane} \emph {et~al.}}]{LISA:2017pwj}%
  \BibitemOpen
  \bibfield  {author} {\bibinfo {author} {\bibfnamefont {Pau}\ \bibnamefont
  {Amaro-Seoane}} \emph {et~al.} (\bibinfo {collaboration} {LISA
  Collaboration}),\ }\bibfield  {title} {\enquote {\bibinfo {title} {{Laser
  Interferometer Space Antenna}},}\ }\href@noop {} {\ }\Eprint
  {http://arxiv.org/abs/1702.00786} {arXiv:1702.00786 [astro-ph.IM]}
  \BibitemShut {NoStop}%
\bibitem [{\citenamefont {Caprini}\ \emph {et~al.}(2019)\citenamefont
  {Caprini}, \citenamefont {Figueroa}, \citenamefont {Flauger}, \citenamefont
  {Nardini}, \citenamefont {Peloso}, \citenamefont {Pieroni}, \citenamefont
  {Ricciardone},\ and\ \citenamefont {Tasinato}}]{Caprini:2019pxz}%
  \BibitemOpen
  \bibfield  {author} {\bibinfo {author} {\bibfnamefont {Chiara}\ \bibnamefont
  {Caprini}}, \bibinfo {author} {\bibfnamefont {Daniel~G.}\ \bibnamefont
  {Figueroa}}, \bibinfo {author} {\bibfnamefont {Raphael}\ \bibnamefont
  {Flauger}}, \bibinfo {author} {\bibfnamefont {Germano}\ \bibnamefont
  {Nardini}}, \bibinfo {author} {\bibfnamefont {Marco}\ \bibnamefont {Peloso}},
  \bibinfo {author} {\bibfnamefont {Mauro}\ \bibnamefont {Pieroni}}, \bibinfo
  {author} {\bibfnamefont {Angelo}\ \bibnamefont {Ricciardone}}, \ and\
  \bibinfo {author} {\bibfnamefont {Gianmassimo}\ \bibnamefont {Tasinato}},\
  }\bibfield  {title} {\enquote {\bibinfo {title} {{Reconstructing the spectral
  shape of a stochastic gravitational wave background with LISA}},}\ }\href
  {\doibase 10.1088/1475-7516/2019/11/017} {\bibfield  {journal} {\bibinfo
  {journal} {J. Cosmol. Astropart. Phys.}\ }\textbf {\bibinfo {volume} {11}},\
  \bibinfo {pages} {017} (\bibinfo {year} {2019})},\ \Eprint
  {http://arxiv.org/abs/1906.09244} {arXiv:1906.09244 [astro-ph.CO]}
  \BibitemShut {NoStop}%
\bibitem [{\citenamefont {Gowling}\ and\ \citenamefont
  {Hindmarsh}(2021)}]{Gowling:2021gcy}%
  \BibitemOpen
  \bibfield  {author} {\bibinfo {author} {\bibfnamefont {Chloe}\ \bibnamefont
  {Gowling}}\ and\ \bibinfo {author} {\bibfnamefont {Mark}\ \bibnamefont
  {Hindmarsh}},\ }\bibfield  {title} {\enquote {\bibinfo {title}
  {{Observational prospects for phase transitions at LISA: Fisher matrix
  analysis}},}\ }\href {\doibase 10.1088/1475-7516/2021/10/039} {\bibfield
  {journal} {\bibinfo  {journal} {J. Cosmol. Astropart. Phys.}\ }\textbf
  {\bibinfo {volume} {10}},\ \bibinfo {pages} {039} (\bibinfo {year} {2021})},\
  \Eprint {http://arxiv.org/abs/2106.05984} {arXiv:2106.05984 [astro-ph.CO]}
  \BibitemShut {NoStop}%
\bibitem [{\citenamefont {Giese}\ \emph {et~al.}(2021)\citenamefont {Giese},
  \citenamefont {Konstandin},\ and\ \citenamefont {van~de
  Vis}}]{Giese:2021dnw}%
  \BibitemOpen
  \bibfield  {author} {\bibinfo {author} {\bibfnamefont {Felix}\ \bibnamefont
  {Giese}}, \bibinfo {author} {\bibfnamefont {Thomas}\ \bibnamefont
  {Konstandin}}, \ and\ \bibinfo {author} {\bibfnamefont {Jorinde}\
  \bibnamefont {van~de Vis}},\ }\bibfield  {title} {\enquote {\bibinfo {title}
  {{Finding sound shells in LISA mock data using likelihood sampling}},}\
  }\href {\doibase 10.1088/1475-7516/2021/11/002} {\bibfield  {journal}
  {\bibinfo  {journal} {J. Cosmol. Astropart. Phys.}\ }\textbf {\bibinfo
  {volume} {11}},\ \bibinfo {pages} {002} (\bibinfo {year} {2021})},\ \Eprint
  {http://arxiv.org/abs/2107.06275} {arXiv:2107.06275 [astro-ph.CO]}
  \BibitemShut {NoStop}%
\bibitem [{\citenamefont {Boileau}\ \emph {et~al.}(2023)\citenamefont
  {Boileau}, \citenamefont {Christensen}, \citenamefont {Gowling},
  \citenamefont {Hindmarsh},\ and\ \citenamefont {Meyer}}]{Boileau:2022ter}%
  \BibitemOpen
  \bibfield  {author} {\bibinfo {author} {\bibfnamefont {Guillaume}\
  \bibnamefont {Boileau}}, \bibinfo {author} {\bibfnamefont {Nelson}\
  \bibnamefont {Christensen}}, \bibinfo {author} {\bibfnamefont {Chloe}\
  \bibnamefont {Gowling}}, \bibinfo {author} {\bibfnamefont {Mark}\
  \bibnamefont {Hindmarsh}}, \ and\ \bibinfo {author} {\bibfnamefont {Renate}\
  \bibnamefont {Meyer}},\ }\bibfield  {title} {\enquote {\bibinfo {title}
  {{Prospects for LISA to detect a gravitational-wave background from first
  order phase transitions}},}\ }\href {\doibase 10.1088/1475-7516/2023/02/056}
  {\bibfield  {journal} {\bibinfo  {journal} {J. Cosmol. Astropart. Phys.}\
  }\textbf {\bibinfo {volume} {02}},\ \bibinfo {pages} {056} (\bibinfo {year}
  {2023})},\ \Eprint {http://arxiv.org/abs/2209.13277} {arXiv:2209.13277
  [gr-qc]} \BibitemShut {NoStop}%
\bibitem [{\citenamefont {Gowling}\ \emph {et~al.}(2023)\citenamefont
  {Gowling}, \citenamefont {Hindmarsh}, \citenamefont {Hooper},\ and\
  \citenamefont {Torrado}}]{Gowling:2022pzb}%
  \BibitemOpen
  \bibfield  {author} {\bibinfo {author} {\bibfnamefont {Chloe}\ \bibnamefont
  {Gowling}}, \bibinfo {author} {\bibfnamefont {Mark}\ \bibnamefont
  {Hindmarsh}}, \bibinfo {author} {\bibfnamefont {Deanna~C.}\ \bibnamefont
  {Hooper}}, \ and\ \bibinfo {author} {\bibfnamefont {Jes\'us}\ \bibnamefont
  {Torrado}},\ }\bibfield  {title} {\enquote {\bibinfo {title} {{Reconstructing
  physical parameters from template gravitational wave spectra at LISA: First
  order phase transitions}},}\ }\href {\doibase 10.1088/1475-7516/2023/04/061}
  {\bibfield  {journal} {\bibinfo  {journal} {J. Cosmol. Astropart. Phys.}\
  }\textbf {\bibinfo {volume} {04}},\ \bibinfo {pages} {061} (\bibinfo {year}
  {2023})},\ \Eprint {http://arxiv.org/abs/2209.13551} {arXiv:2209.13551
  [astro-ph.CO]} \BibitemShut {NoStop}%
\bibitem [{\citenamefont {Romero}\ \emph {et~al.}(2021)\citenamefont {Romero},
  \citenamefont {Martinovic}, \citenamefont {Callister}, \citenamefont {Guo},
  \citenamefont {Mart\'\i{}nez}, \citenamefont {Sakellariadou}, \citenamefont
  {Yang},\ and\ \citenamefont {Zhao}}]{Romero:2021kby}%
  \BibitemOpen
  \bibfield  {author} {\bibinfo {author} {\bibfnamefont {Alba}\ \bibnamefont
  {Romero}}, \bibinfo {author} {\bibfnamefont {Katarina}\ \bibnamefont
  {Martinovic}}, \bibinfo {author} {\bibfnamefont {Thomas~A.}\ \bibnamefont
  {Callister}}, \bibinfo {author} {\bibfnamefont {Huai-Ke}\ \bibnamefont
  {Guo}}, \bibinfo {author} {\bibfnamefont {Mario}\ \bibnamefont
  {Mart\'\i{}nez}}, \bibinfo {author} {\bibfnamefont {Mairi}\ \bibnamefont
  {Sakellariadou}}, \bibinfo {author} {\bibfnamefont {Feng-Wei}\ \bibnamefont
  {Yang}}, \ and\ \bibinfo {author} {\bibfnamefont {Yue}\ \bibnamefont
  {Zhao}},\ }\bibfield  {title} {\enquote {\bibinfo {title} {{Implications for
  first-order cosmological phase transitions from the third LIGO-Virgo
  observing run}},}\ }\href {\doibase 10.1103/PhysRevLett.126.151301}
  {\bibfield  {journal} {\bibinfo  {journal} {Phys. Rev. Lett.}\ }\textbf
  {\bibinfo {volume} {126}},\ \bibinfo {pages} {151301} (\bibinfo {year}
  {2021})},\ \Eprint {http://arxiv.org/abs/2102.01714} {arXiv:2102.01714
  [hep-ph]} \BibitemShut {NoStop}%
\bibitem [{\citenamefont {Roper~Pol}\ \emph {et~al.}(2024)\citenamefont
  {Roper~Pol}, \citenamefont {Procacci}, \citenamefont {Midiri},\ and\
  \citenamefont {Caprini}}]{kin_sp_SSM}%
  \BibitemOpen
  \bibfield  {author} {\bibinfo {author} {\bibfnamefont {Alberto}\ \bibnamefont
  {Roper~Pol}}, \bibinfo {author} {\bibfnamefont {Simona}\ \bibnamefont
  {Procacci}}, \bibinfo {author} {\bibfnamefont {Antonino~S.}\ \bibnamefont
  {Midiri}}, \ and\ \bibinfo {author} {\bibfnamefont {Chiara}\ \bibnamefont
  {Caprini}},\ }\bibfield  {title} {\enquote {\bibinfo {title} {{Irrotational
  fluid perturbations from first-order phase transitions}},}\ }\href@noop {}
  {\bibfield  {journal} {\bibinfo  {journal} {{to be published}}\ } (\bibinfo
  {year} {2024})}\BibitemShut {NoStop}%
\bibitem [{\citenamefont {Kolb}\ and\ \citenamefont
  {Turner}(1990)}]{Kolb:1990vq}%
  \BibitemOpen
  \bibfield  {author} {\bibinfo {author} {\bibfnamefont {Edward~W.}\
  \bibnamefont {Kolb}}\ and\ \bibinfo {author} {\bibfnamefont {Michael~S.}\
  \bibnamefont {Turner}},\ }\href {\doibase 10.1201/9780429492860} {\emph
  {\bibinfo {title} {{The early Universe}}}},\ Vol.~\bibinfo {volume} {69}\
  (\bibinfo {address} {CRC Press, Boca Raton, FL, USA},\ \bibinfo {year}
  {1990})\BibitemShut {NoStop}%
\bibitem [{\citenamefont {Fixsen}(2009)}]{Fixsen:2009ug}%
  \BibitemOpen
  \bibfield  {author} {\bibinfo {author} {\bibfnamefont {D.~J.}\ \bibnamefont
  {Fixsen}},\ }\bibfield  {title} {\enquote {\bibinfo {title} {{The temperature
  of the cosmic microwave background}},}\ }\href {\doibase
  10.1088/0004-637X/707/2/916} {\bibfield  {journal} {\bibinfo  {journal}
  {Astrophys. J.}\ }\textbf {\bibinfo {volume} {707}},\ \bibinfo {pages} {916}
  (\bibinfo {year} {2009})},\ \Eprint {http://arxiv.org/abs/0911.1955}
  {arXiv:0911.1955 [astro-ph.CO]} \BibitemShut {NoStop}%
\bibitem [{\citenamefont {Isserlis}(1916)}]{Isserlis:1916}%
  \BibitemOpen
  \bibfield  {author} {\bibinfo {author} {\bibfnamefont {L.}~\bibnamefont
  {Isserlis}},\ }\bibfield  {title} {\enquote {\bibinfo {title} {{On certain
  probable errors and correlation of multiple frequency distributions with skew
  regression}},}\ }\href {\doibase 10.1093/biomet/11.3.185} {\bibfield
  {journal} {\bibinfo  {journal} {Biometrika}\ }\textbf {\bibinfo {volume}
  {11}},\ \bibinfo {pages} {185} (\bibinfo {year} {1916})}\BibitemShut
  {NoStop}%
\bibitem [{\citenamefont {Monin}\ and\ \citenamefont {Yaglom}(1975)}]{MY75}%
  \BibitemOpen
  \bibfield  {author} {\bibinfo {author} {\bibfnamefont {A.~S.}\ \bibnamefont
  {Monin}}\ and\ \bibinfo {author} {\bibfnamefont {A.~M.}\ \bibnamefont
  {Yaglom}},\ }\href@noop {} {\emph {\bibinfo {title} {{Statistical fluid
  mechanics: Mechanics of turbulence}}}},\ Vol.~\bibinfo {volume} {2}\
  (\bibinfo {address} {MIT press, Cambridge, MA, USA.},\ \bibinfo {year}
  {1975})\BibitemShut {NoStop}%
\bibitem [{\citenamefont {Caprini}\ \emph {et~al.}(2004)\citenamefont
  {Caprini}, \citenamefont {Durrer},\ and\ \citenamefont
  {Kahniashvili}}]{Caprini:2003vc}%
  \BibitemOpen
  \bibfield  {author} {\bibinfo {author} {\bibfnamefont {Chiara}\ \bibnamefont
  {Caprini}}, \bibinfo {author} {\bibfnamefont {Ruth}\ \bibnamefont {Durrer}},
  \ and\ \bibinfo {author} {\bibfnamefont {Tina}\ \bibnamefont
  {Kahniashvili}},\ }\bibfield  {title} {\enquote {\bibinfo {title} {{The
  cosmic microwave background and helical magnetic fields: The tensor mode}},}\
  }\href {\doibase 10.1103/PhysRevD.69.063006} {\bibfield  {journal} {\bibinfo
  {journal} {Phys. Rev. D}\ }\textbf {\bibinfo {volume} {69}},\ \bibinfo
  {pages} {063006} (\bibinfo {year} {2004})},\ \Eprint
  {http://arxiv.org/abs/astro-ph/0304556} {arXiv:astro-ph/0304556} \BibitemShut
  {NoStop}%
\bibitem [{\citenamefont {Sharma}\ \emph {et~al.}(2023)\citenamefont {Sharma},
  \citenamefont {Dahl}, \citenamefont {Brandenburg},\ and\ \citenamefont
  {Hindmarsh}}]{Sharma:2023mao}%
  \BibitemOpen
  \bibfield  {author} {\bibinfo {author} {\bibfnamefont {Ramkishor}\
  \bibnamefont {Sharma}}, \bibinfo {author} {\bibfnamefont {Jani}\ \bibnamefont
  {Dahl}}, \bibinfo {author} {\bibfnamefont {Axel}\ \bibnamefont
  {Brandenburg}}, \ and\ \bibinfo {author} {\bibfnamefont {Mark}\ \bibnamefont
  {Hindmarsh}},\ }\bibfield  {title} {\enquote {\bibinfo {title} {{Shallow
  relic gravitational wave spectrum with acoustic peak}},}\ }\href {\doibase
  10.1088/1475-7516/2023/12/042} {\bibfield  {journal} {\bibinfo  {journal} {J.
  Cosmol. Astropart. Phys.}\ }\textbf {\bibinfo {volume} {12}},\ \bibinfo
  {pages} {042} (\bibinfo {year} {2023})},\ \Eprint
  {http://arxiv.org/abs/2308.12916} {arXiv:2308.12916 [gr-qc]} \BibitemShut
  {NoStop}%
\bibitem [{\citenamefont {Huber}\ and\ \citenamefont
  {Konstandin}(2008{\natexlab{b}})}]{Huber:2007vva}%
  \BibitemOpen
  \bibfield  {author} {\bibinfo {author} {\bibfnamefont {Stephan~J.}\
  \bibnamefont {Huber}}\ and\ \bibinfo {author} {\bibfnamefont {Thomas}\
  \bibnamefont {Konstandin}},\ }\bibfield  {title} {\enquote {\bibinfo {title}
  {{Production of gravitational waves in the nMSSM}},}\ }\href {\doibase
  10.1088/1475-7516/2008/05/017} {\bibfield  {journal} {\bibinfo  {journal} {J.
  Cosmol. Astropart. Phys.}\ }\textbf {\bibinfo {volume} {05}},\ \bibinfo
  {pages} {017} (\bibinfo {year} {2008}{\natexlab{b}})},\ \Eprint
  {http://arxiv.org/abs/0709.2091} {arXiv:0709.2091 [hep-ph]} \BibitemShut
  {NoStop}%
\bibitem [{\citenamefont {Weir}(2018)}]{Weir:2017wfa}%
  \BibitemOpen
  \bibfield  {author} {\bibinfo {author} {\bibfnamefont {David~J.}\
  \bibnamefont {Weir}},\ }\bibfield  {title} {\enquote {\bibinfo {title}
  {{Gravitational waves from a first order electroweak phase transition: A
  brief review}},}\ }\href {\doibase 10.1098/rsta.2017.0126} {\bibfield
  {journal} {\bibinfo  {journal} {Phil. Trans. Roy. Soc. Lond. A}\ }\textbf
  {\bibinfo {volume} {376}},\ \bibinfo {pages} {20170126} (\bibinfo {year}
  {2018})},\ \Eprint {http://arxiv.org/abs/1705.01783} {arXiv:1705.01783
  [hep-ph]} \BibitemShut {NoStop}%
\bibitem [{\citenamefont {Hindmarsh}\ \emph {et~al.}(2021)\citenamefont
  {Hindmarsh}, \citenamefont {L\"uben}, \citenamefont {Lumma},\ and\
  \citenamefont {Pauly}}]{Hindmarsh:2020hop}%
  \BibitemOpen
  \bibfield  {author} {\bibinfo {author} {\bibfnamefont {Mark~B.}\ \bibnamefont
  {Hindmarsh}}, \bibinfo {author} {\bibfnamefont {Marvin}\ \bibnamefont
  {L\"uben}}, \bibinfo {author} {\bibfnamefont {Johannes}\ \bibnamefont
  {Lumma}}, \ and\ \bibinfo {author} {\bibfnamefont {Martin}\ \bibnamefont
  {Pauly}},\ }\bibfield  {title} {\enquote {\bibinfo {title} {{Phase
  transitions in the early universe}},}\ }\href {\doibase
  10.21468/SciPostPhysLectNotes.24} {\bibfield  {journal} {\bibinfo  {journal}
  {SciPost Phys. Lect. Notes}\ }\textbf {\bibinfo {volume} {24}},\ \bibinfo
  {pages} {1} (\bibinfo {year} {2021})},\ \Eprint
  {http://arxiv.org/abs/2008.09136} {arXiv:2008.09136 [astro-ph.CO]}
  \BibitemShut {NoStop}%
\bibitem [{\citenamefont {Kraichnan}(1965)}]{Kraichnan:1965zz}%
  \BibitemOpen
  \bibfield  {author} {\bibinfo {author} {\bibfnamefont {Robert~H.}\
  \bibnamefont {Kraichnan}},\ }\bibfield  {title} {\enquote {\bibinfo {title}
  {{Inertial-range spectrum of hydromagnetic turbulence}},}\ }\href {\doibase
  10.1063/1.1761412} {\bibfield  {journal} {\bibinfo  {journal} {Phys. Fluids}\
  }\textbf {\bibinfo {volume} {8}},\ \bibinfo {pages} {1385} (\bibinfo {year}
  {1965})}\BibitemShut {NoStop}%
\bibitem [{\citenamefont {Pen}\ and\ \citenamefont
  {Turok}(2016)}]{Pen:2015qta}%
  \BibitemOpen
  \bibfield  {author} {\bibinfo {author} {\bibfnamefont {Ue-Li}\ \bibnamefont
  {Pen}}\ and\ \bibinfo {author} {\bibfnamefont {Neil}\ \bibnamefont {Turok}},\
  }\bibfield  {title} {\enquote {\bibinfo {title} {{Shocks in the early
  universe}},}\ }\href {\doibase 10.1103/PhysRevLett.117.131301} {\bibfield
  {journal} {\bibinfo  {journal} {Phys. Rev. Lett.}\ }\textbf {\bibinfo
  {volume} {117}},\ \bibinfo {pages} {131301} (\bibinfo {year} {2016})},\
  \Eprint {http://arxiv.org/abs/1510.02985} {arXiv:1510.02985 [astro-ph.CO]}
  \BibitemShut {NoStop}%
\bibitem [{\citenamefont {Roper~Pol}()}]{GH}%
  \BibitemOpen
  \bibfield  {author} {\bibinfo {author} {\bibfnamefont {Alberto}\ \bibnamefont
  {Roper~Pol}},\ }\bibfield  {title} {\enquote {\bibinfo {title} {{GitHub
  project \href{https://github.com/AlbertoRoper/cosmoGW}{\texttt{cosmoGW}}}},}\
  }\href {\doibase 10.5281/zenodo.6045844} {\
  10.5281/zenodo.6045844}\BibitemShut {NoStop}%
\end{thebibliography}%

\label{RealLastPage}

\end{document}